\documentclass[aps,pre,reprint,superscriptaddress,footinbib,floatfix]{revtex4-1}

\usepackage{graphicx}
\usepackage{bm}
\usepackage{amssymb}
\usepackage{amsbsy}
\usepackage{amsmath}
\usepackage{mathtools}
\usepackage{xcolor}
\usepackage{stmaryrd}
\usepackage{mathrsfs}
\usepackage[mathscr]{euscript}
\usepackage{tikz}
\usetikzlibrary{shapes}
\usepackage{cancel}
\usepackage[normalem]{ulem}

\begin{document}


\title{Wet and dry internal friction can be measured with the Jarzynski equality}


\author{R. Kailasham}
\affiliation{IITB-Monash Research Academy, Indian Institute of Technology Bombay, Mumbai, Maharashtra -  400076, India}
\affiliation{Department of Chemistry, Indian Institute of Technology Bombay, Mumbai, Maharashtra -  400076, India}
\affiliation{Department of Chemical Engineering, Monash University,
Melbourne, VIC 3800, Australia}
\author{Rajarshi Chakrabarti}
\email{rajarshi@chem.iitb.ac.in}
\affiliation{Department of Chemistry, Indian Institute of Technology Bombay, Mumbai, Maharashtra -  400076, India}
\author{J. Ravi Prakash}
\email{ravi.jagadeeshan@monash.edu}
\affiliation{Department of Chemical Engineering, Monash University,
Melbourne, VIC 3800, Australia}


\date{\today}

\begin{abstract}
The existence of two types of internal friction\textemdash wet and dry\textemdash is revisited, and a simple protocol is proposed for distinguishing between the two types and extracting the appropriate internal friction coefficient. The scheme requires repeatedly stretching a polymer molecule, and measuring the average work dissipated in the process by applying the Jarzynski equality. The internal friction coefficient is then estimated from the average dissipated work in the extrapolated limit of zero solvent viscosity. The validity of the protocol is established through analytical calculations on a one-dimensional free-draining Hookean spring-dashpot model for a polymer, and Brownian dynamics simulations of: (a) a single-mode nonlinear spring-dashpot model for a polymer, and (b) a finitely extensible bead-spring chain with cohesive intra-chain interactions, both of which incorporate fluctuating hydrodynamic interactions.  Well-established single-molecule manipulation techniques, such as optical tweezer-based pulling, can be used to implement the suggested protocol experimentally.
\end{abstract}


\maketitle


\section{\label{sec:intro}Introduction}

Conformational transitions in polymer molecules are impeded by solvent molecules, and sometimes additionally by intramolecular interactions. The dissipation caused by the latter are termed as \emph{internal friction}~\cite{kuhn1945,degennes,Manke1985,Sagnella2000,Hagen2010385,Hridya2018}, and examples of such dissipation include the damping of protein folding~\cite{Ansari1992,Qiu2004,Cellmer2008,Wensley2010,Soranno2017}, the modulation of stretching transitions in polysaccharides~\cite{Khatri20071825}, and the enhancement of dissipated work  in the stretch-relaxation of polymers~\cite{Murayama2007,Alexander-Katz2009,Schulz20154565}. While the microscopic origin of internal friction is manyfold~\cite{Khatri20071825,Murayama2007,Alexander-Katz2009,Schulz20154565,Echeverria2014,DeSancho2014,Sashi2016,Jain2016,Ameseder2018,Jas2001,Soranno2017}, it has been broadly classified as being either of the \emph{wet} or \emph{dry} kind~\cite{Soranno201217800,Hagen2010385}.

The transition of a protein from an unfolded to its native folded state is commonly interpreted as a diffusive search process over a rugged energy landscape~\cite{onuchic1997}, and the internal friction associated with landscape roughness is typically considered to be of the wet type~\cite{Wensley2010,DeSancho2014,Neupane2017}. An analytical solution for the diffusion coefficient in one dimension was derived by Zwanzig~\cite{Zwanzig1988} who observed that the effective friction $\gamma_{\text{eff}}$, is related purely multiplicatively to the solvent friction $\gamma_{\text{s}}$, as $\gamma_{\text{eff}}=\gamma_{\text{s}}\exp\left[\left(\delta/k_BT\right)^2\right]$, where $k_B$ is Boltzmann's constant, $T$ is the absolute temperature, and $\delta^2$ is the variance of the heights of the normally distributed undulations. Since $\gamma_{\text{s}}\sim\eta_{\text{s}}$, where $\eta_{\text{s}}$ denotes the solvent viscosity, the effective friction is higher than that expected due to solvent friction alone, at any finite value of $\eta_{\text{s}}$. It is also clear that the internal friction would vanish in the extrapolated limit of zero solvent viscosity ($\eta_{\textrm{s}}\to 0$), which is a characteristic feature of wet internal friction.

On the other hand, experimental measurements of the dependence of the reconfiguration time of small proteins on $\eta_{\text{s}}$ find a finite value in the extrapolated limit of $\eta_{\text{s}}\to0$, indicating the presence of a solvent-viscosity-independent resistance to folding~\cite{Qiu2004,Cellmer2008,Soranno201217800}. Hagen~\cite{Hagen2010385}  has proposed a heterogeneous reaction friction model with the unfolded and native states separated by two consecutive barriers, one for each mode of friction, as an explanation for the presence of dry internal friction. 

The commonly accepted operational definition for dry internal friction~\cite{Qiu2004,Avdoshenko2017,Soranno2017} as the reconfiguration time in the limit $\eta_{\text{s}}\to0$ is not a direct quantitative measure of the internal friction coefficient, nor does it apply to the case of wet internal friction. In principle, the effective friction could be calculated from experimental measurements of the diffusion coefficient of biomolecules hopping between native and unfolded states~\cite{Chung2013,Chung2015,Neupane2016,Neupane2018}. However, such measurements would not determine if the internal friction were of the wet or the dry type unless they were performed at multiple solvent viscosities, followed by extrapolation to the limit $\eta_{\text{s}}\to0$. There is clearly a need for a protocol that can directly estimate the internal friction coefficient and distinguish between the two types.

The internal friction coefficient can be obtained directly from dissipated work, as demonstrated by the experiments on condensed DNA by~\citet{Murayama2007}, and simulations of polypeptide-stretching by~\citet{Schulz20154565}. Indeed in the protocol proposed by Netz and coworkers~\cite{Alexander-Katz2009,Schulz20154565}, the work required to stretch a macromolecule is separated into two parts: reversible free energy increase due to the extension of the molecule, and irreversible work required to overcome rate-dependent restoring forces arising from solvent and internal friction. Within this framework, they show that the average dissipated work  in the limit of $\eta_{\text{s}}\to 0$ scales with the number of hydrogen bonds, which are considered to be responsible for internal friction~\cite{Schulz20154565}.

Here we propose a novel application of the Jarzynski equality (JE)~\cite{Jarzynski1997,JARZYNSKI2007495} and show that by focussing on measuring the dissipation associated with stretching a macromolecule rather than on obtaining the free-energy difference, a quantitative measure of the internal friction can be obtained. The JE has been routinely employed for reconstructing the free energy landscape of biomolecules from experiments~\cite{Liphardt2002,Harris2007,Gupta2011} and simulations~\cite{Hummer2010,Hodges2016}, while dissipation has largely been ignored (except for estimating the accuracy of the free-energy difference~\cite{Ritort2002,Jarzynski2006,YungerHalpern2016}). In the proposed protocol, multiple realizations of the pulling experiment are performed and the JE is used to extract both the free-energy difference and the average dissipated work at finite pulling rates, 
\begin{equation}
\label{eq:je_use}
\begin{split}
\left<\exp\left[-W/k_BT\right]\right>&=\exp\left[-\Delta A/k_BT\right] ; \\
\left<W_{\text{dis}}\right>&=\left<W\right>-\Delta A
\end{split}
\end{equation}
where the $\left<...\right>$ in Eq.~(\ref{eq:je_use}) represents an average with respect to the probability distribution of work values. Prior studies~\cite{Murayama2007,Alexander-Katz2009,Schulz20154565} estimate $\Delta A$ from the work done in the quasi-static limit~\cite{Callen1985} and calculate $\left<W_{\text{dis}}\right>$ at finite pulling rates by subtracting $\Delta A$ from the total work done, rather than estimating both components of work simultaneously, as is done here. 

In essence, the proposed protocol consists of calculating $\left<W_{\text{dis}}\right>$ at fixed values of both the pulling velocity $v$ and distance $d$ over which the molecule is stretched, but at various values of $\eta_{\text{s}}$. The value in the limit $\eta_{\text{s}}\to0$,  $\left<W_{\text{dis}}\right>_{\eta_{\!_{\, \text{s}}}\to\,0}$, is then obtained by extrapolation. By repeating this process for a number of values of $v$ and plotting the ratio $\left<W_{\text{dis}}\right>_{\eta_{\!_{\, \text{s}}}\to\,0}/d$ as a function of $v$, the internal friction coefficient can be determined from the slope of the linear region at sufficiently small velocities. Clearly, dry internal friction corresponds to cases where $\left<W_{\text{dis}}\right>_{\eta_{\!_{\, \text{s}}}\to\,0}$ is non-zero, while wet friction is indicated when it is zero. In the latter case, the protocol measures the enhancement in friction at any finite value of $\eta_{\text{s}}$.

The validity of the proposed protocol is established for both types of internal friction using coarse-grained polymer models. Additionally, since hydrodynamic interactions (HI) are known to affect the dynamic response of polymers~\cite{Kailasham2018,Prakash2019}, the effect of HI on dissipated work is also examined. 

For dry internal friction, a spring-dashpot model~\cite{kuhn1945,Manke1985,Khatri20076770,Samanta2016165} is considered where the molecule is represented as massless beads connected by a spring and dashpot in parallel with each other. The spring accounts for entropic elasticity, while dissipative effects due to internal friction are captured by the dashpot~\cite{Bird1987b}. The drag on the beads is responsible for solvent friction. By its very construction, this model describes dry internal friction, as the dashpot contributes to dissipation even in the limit of $\eta_{\text{s}}\to0$. Within this framework, two examples are considered. In the first case, the work distribution for a free-draining Hookean spring-dashpot model subjected to constant-velocity pulling is analytically calculated. In the second case, pulling simulations on a nonlinear-spring-dashpot model with fluctuating HI are performed using Brownian dynamics (BD). In both these cases, it is demonstrated that the internal friction coefficient estimated from $\left<W_{\text{dis}}\right>$ in the limit $\eta_{\text{s}}\to0$ is identical to the damping coefficient of the dashpot, which is a model input parameter, thereby establishing the validity of the proposed protocol. It is also shown that HI does not affect $\left<W_{\text{dis}}\right>$. 

For wet internal friction, a bead-spring chain with cohesive interactions between the beads is considered. A similar model was used by Netz and coworkers~\cite{Alexander-Katz2009} to compare simulated values of internal friction with experimental data on force-induced unraveling of collapsed DNA~\cite{Murayama2007}. By using Zwanzig's interpretation~\cite{Zwanzig1988} to estimate energy landscape roughness due to cohesive interactions, they implicitly assume wet internal friction. Using our protocol, it is established directly that the internal friction due to cohesive interactions in this coarse-grained polymer model is wet in nature.  Further, it is observed that while HI reduces the total resistance to pulling, the enhancement in the friction coefficient remains unaffected.

The rest of the paper is organized as follows. In Sec.~\ref{sec:dry_sd}, analytical calculations and Brownian dynamics simulations on a spring-dashpot model are presented. Sec.~\ref{sec:cg} covers the application of the protocol to pulling simulations on a single polymer chain with cohesive interactions between the beads. A discussion of the results and concluding remarks are provided in Sec.~\ref{sec:concl}. Appendix~\ref{sec:oned_gov_deriv} contains the derivation of the governing equation of motion for a one-dimensional Hookean spring-dashpot tethered at one end and pulled at the other. In Appendix~\ref{sec:add_dis_steps}, additional details pertaining to the derivation of the moments of the work distribution in the Hookean spring-dashpot model is presented. 
In Appendix~\ref{sec:gov_eq_db}, the derivation of the Fokker-Planck and stochastic differential equations for a spring-dashpot with a nonlinear force-extension relation and fluctuating hydrodynamic interactions is presented, along with the details of the solver algorithm. In Appendix~\ref{sec:valid_db}, the validation of the code for the dumbbell case is presented, by comparing the free energy difference obtained from the Jarzynski equality and that obtained from direct numerical integration. In Appendix~\ref{sec:valid_sc}, the probability of the work distribution determined numerically for a Rouse chain tethered at one end and subjected to constant velocity pulling is compared against the analytical results derived by Dhar~\cite{Dhar2005}, thereby validating the code for the single chain case.

\section{\label{sec:dry_sd}Dry internal friction}

\subsection{\label{sec:analytical}One-dimensional free draining Hookean spring-dashpot}

\begin{figure}[ptbh]
\centering
\includegraphics[width=0.8\linewidth]{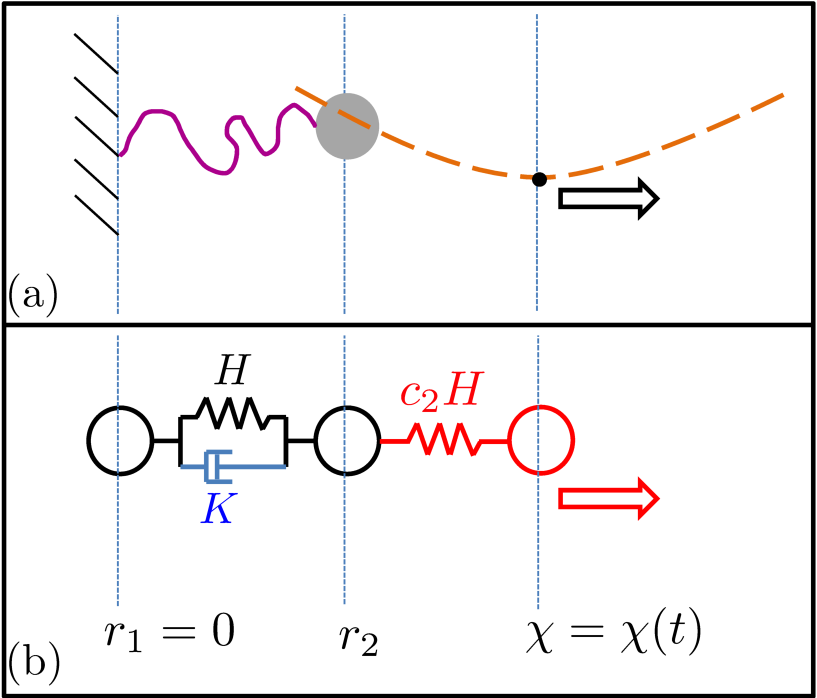}  
\caption{\small  (Color online) \textbf{Schematic of a one-dimensional polymer model subjected to pulling}. (a) A cartoon depicting a single polymer strand that is tethered to a surface at one end, and attached to a bead at the other end. The bead is under the influence of an optical tweezer whose position is varied in time according to a deterministic protocol. (b) Representation of the polymer as a single-mode spring-dashpot, connected to a bead that is manipulated by a predetermined protocol, $\chi(t)$. Internal friction is modeled using the dashpot, whose damping coefficient is $K$. The Hookean spring constant associated with the spring of the polymer is $H$, and that associated with the trap is $c_2H$.}
\label{fig:scheme_1d}
\end{figure}

The simple analytically tractable dumbbell model is shown in Fig.~\ref{fig:scheme_1d}, with one bead, at $r_1$, fixed at the origin ($r_1=0$), and the other bead, at $r_2$, connected to a bead at $\chi$ which is indicative of the cantilever of an atomic force microscope (AFM), or the location of the optical trap. The dumbbell is suspended in an incompressible,  Newtonian solvent of viscosity $\eta_{\text{s}}$. The bead radius is taken to be $a$, and its associated friction co-efficient given by $\zeta=6\pi\eta_{\text{s}}a$. The bead at $\chi$ is manipulated using a pre-determined protocol, given by $\chi=\chi(t)$. All the springs considered in the present model are Hookean: the spring in parallel with the dashpot has a spring constant of $H$, whereas the spring connecting the spring-dashpot setup to the driven bead has a spring constant of $c_2H$, where $c_2$ is an arbitrary positive constant. The damping coefficient of the dashpot is denoted by $K$. It is evident that the only degree-of-freedom in the system is $r_2$, which is allowed to execute stochastic motion. The Hamiltonian of the system is then written as
\begin{align}\label{eq:hamilt}
\mathcal{H}&=\dfrac{H}{2}r^2_2+\dfrac{c_2H}{2}\left[r_2-\chi(t)\right]^2
\end{align}
For ease of algebra, it is convenient to work with non-dimensional variables until the need for dimensional variables arises. Using $l_{\text{H}}\equiv\sqrt{k_BT/H}$ and $\lambda_{\text{H}}\equiv\zeta/4H$ as the length- and time-scales, respectively, and denoting dimensionless quantities using an asterisk as superscript, the governing stochastic differential equation can be derived to be
\begin{align}\label{eq:gov_eq}
\dfrac{dr^{*}_2}{dt^{*}}=-\dfrac{E \, r^{*}_2}{\theta}+\dfrac{c_2 \, \chi^{*}(t^{*})}{4\theta}+\dfrac{1}{\theta}\, \xi(t^{*})
\end{align}
where $E=\left[\left(c_2+1\right)/4\right]$, $\theta=\left[1+\left(K/\zeta\right)\right]$, and the noise term, $\xi(t^{*})$, obeys $\left<\xi(t^{*})\right>=0$ and $\left<\xi(t^{*})\xi(t_1^{*})\right>=\left({\theta}/{2}\right)\delta(t^{*}-t^{*}_1)$.
The steps for the derivation of Eq.~(\ref{eq:gov_eq}) have been detailed in Appendix~\ref{sec:oned_gov_deriv}. The solution to Eq.~(\ref{eq:gov_eq}) can be written as
\begin{align}\label{eq:traj_def}
r^{*}_2(t^{*})=r^{*}_2(0)G(t^{*})+\dfrac{1}{\theta}\int_{0}^{t^{*}}dt_1^{*}&G\left(t^{*}-t_1^{*}\right)\nonumber\\[5pt]\times\Biggl(\dfrac{c_2\chi^{*}(t_1^{*})}{4}
&+\xi(t_1^{*})\Biggr)
\end{align} 
where $G(t^{*})=e^{-Et^{*}/\theta}$.

The work done during one realization of the pulling performed in the interval $\left[0,\tau\right]$ is~\cite{Hodges2016,Chaki2018} 
\begin{align}
W&=\int_{0}^{\tau} \!  \dfrac{\partial \mathcal{H}}{\partial t} \, dt = \int_{0}^{\tau} \! \dfrac{\partial \mathcal{H}}{\partial \chi} \, \dot{\chi} \, dt \nonumber\\[5pt]
& = \int_{0}^{\tau} \!\! c_2 \, H\left(\chi(t)-r_2\right) \dot{\chi}(t) \, dt \nonumber\\[5pt]
&=k_BT \left[ c_2 \int_{0}^{\tau^{*}} \!\! \!\! \left(\chi^{*}(t^{*})-r^{*}_2\right)\dot{\chi}^{*}(t^{*}) \, dt^{*} \right]
\end{align}
and the dimensionless work, $W^{*}=W/k_BT$, is then 
\begin{align}\label{eq:dimless_work}
W^{*}=\dfrac{c_2}{2}\left[\chi^{2*}\left(\tau^{*}\right)-\chi^{2*}\left(0\right)\right]-c_2\int_{0}^{\tau^{*}} \!\! \!\! dt^{*}\dot{\chi}^{*}(t^{*}) \, r^{*}_2
\end{align}
Upon substituting the expression for $r^{*}_2$ from Eq.~(\ref{eq:traj_def}) into Eq.~(\ref{eq:dimless_work}), one obtains the complete expression for $W^*$. It is clear that the distribution of $W^*$ ought also be Gaussian, since $W^*$ is linear in $r^{*}_2(0)$ and $\xi(t^{*})$, both of which are Gaussian variables. It therefore suffices to evaluate the mean and variance of $W^*$ in order to completely determine the distribution. There is only one bead that is allowed to move freely in this problem, and as a result, the dimensionless free-energy, $A^* = -\ln Z^*$, can be obtained once the dimensionless partition function, $Z^{*}$ is known. Details of all the intermediate steps in the calculation of $\left<W^{*}\right>$, $\left<(W^*-\left<W^*\right>)^2\right>$ and $A^{*}$ are given in Appendix~\ref{sec:add_dis_steps}. For ease of exposition, only the salient results are reproduced below.

The dimensionless free-energy can be shown to be given by the expression,
\begin{equation}\label{eq:free_energy_def}
A^*\left(\chi^*\right) = \left[\dfrac{c_2}{2(c_2+1)}\right]\chi^{2*},
\end{equation}
while the expression for the average work is,
\begin{align}\label{eq:av_work6}
\left<W^{*}\right> =\Delta A^{*}+\dfrac{c_2^2}{c_2+1}\int_{0}^{\tau^{*}}dt^{*}&\int_{0}^{t^{*}}dt_1^{*}\Bigl[\dot{\chi}^{*}(t^{*})\\[5pt]
&\times G\left(t^{*}-t_1^{*}\right)\dot{\chi}^{*}(t_1^{*})\Bigr]\nonumber
\end{align}
where $\Delta A^{*}\equiv A^*\left[\chi^{*}(\tau^{*})\right]-A^*\left[\chi^{*}(0)\right]$. The variance of the work distribution, $\sigma^2 \equiv \left<(W^*-\left<W^*\right>)^2\right>$, can be shown to be
\begin{equation}\label{eq:sig_exp}
\sigma^2=\dfrac{2c_2^2}{\left(c_2+1\right)}\int_{0}^{\tau^{*}}dt^{*}\int_{0}^{t^{*}}dt_1^{*}\dot{\chi}^{*}(t^{*})G(t^{*}-t_1^{*})\dot{\chi}^{*}(t_1^{*})
\end{equation}
From Eqs.~(\ref{eq:av_work6}) and~(\ref{eq:sig_exp}), it is readily seen that
 \begin{equation}\label{eq:work_part}
 \left<W^{*}\right> = \Delta A^{*}+\dfrac{\sigma^2}{2}
 \end{equation}
 and the average dissipated work is given by
 \begin{equation}
 \left<W^{*}_{\text{dis}}\right> \equiv \left<W^{*}\right>-\Delta A^{*}=\dfrac{\sigma^2}{2}
 \end{equation}
It follows that the probability distribution of work is given by
\begin{equation}\label{eq:prob_gauss}
P^{*}(W^*)=\dfrac{1}{\sqrt{2\pi\sigma^2}}\exp\left[-\dfrac{\left(W^*-\left<W^*\right>\right)^2}{2\sigma^2}\right]
\end{equation}
The quantities $\left<W^{*}\right>,\,\Delta A^{*},\, \text{and} \left<W^{*}_{\text{dis}}\right>$ have been calculated analytically without explicit recourse to the Jarzynski equality. This is a consequence of the governing equation being linear in the position variable and the noise term, resulting in a Gaussian distribution of the work trajectories. The Jarzynski's equality is satisfied trivially for such systems, since
 \begin{align}\label{eq:jarz_auto}
 \left<\exp\left(-W^{*}\right)\right>&=\int_{-\infty}^{+\infty}\exp\left(-W^{*}\right)P^{*}(W^*)dW^{*}  \nonumber \\
 & = \exp\left(-\Delta A^{*}\right)
 \end{align}
 which has also been reported previously~\cite{Jarzynski1997,Speck2004}.
 
The development so far is applicable to any arbitrary pulling protocol, $\chi^{*}(t^{*})$. The particular value of the average dissipation is dependent on the protocol used to transition the system between its initial and final states. We focus attention on the special case of a constant-velocity pulling protocol commonly encountered in single-molecule force spectroscopy~\cite{Harris2007,Gupta2011}. Within this framework, $\chi^{*}(t^{*})=\chi^{\text{(i)}*}+\left(d^*t^{*}/\tau^{*}\right)$, implying that the last bead is moved across a distance $d^*$ over a time $\tau^{*}$, with a dimensionless pulling velocity given by $\dot{\chi}^{*}(t^{*})=v^*=d^*/\tau^{*}$. $\chi^{\text{(i)}*}$ and $\chi^{\text{(f)}*}$ represent the position of the last bead at $t^{*}=0$ and $t^{*}=\tau^*$ respectively. Under this protocol, the free energy difference, $\Delta A^*$, is given by
\begin{equation}\label{eq:delf_spec}
\Delta A^*  = \dfrac{c_2}{2\left(c_2+1\right)}\left[\left(\chi^{\text{(f)}*}\right)^2 - \left(\chi^{\text{(i)}*}\right)^2\right],
\end{equation}
and the average dissipated work can be evaluated to be
\begin{align}\label{eq:wdis_spec}
\left<W^{*}_{\text{dis}}\right> & = 4\left(\frac{c_2}{c_2+1}\right)^2\theta\,v^*d^* \nonumber\\[3pt] & + 16\left(\dfrac{c^2_2}{\left(c_2+1\right)^3}\right)\theta^2 v^{*2}\left[\exp\left(-\frac{(c_2+1)d^*}{4v^*\theta}\right)-1\right]
\end{align}
It is now appropriate to obtain the expression for the average dissipated work in the limit of zero solvent friction. Since the solvent friction is absorbed into the definition of the timescale, it is first necessary to convert all quantities in Eq.~(\ref{eq:wdis_spec}) to their dimensional form, before taking the limit. Using the non-dimensionalization scheme discussed previously, we obtain
\begin{flalign}\label{eq:dim_wdis_def}
\left<W\right>_{\text{dis}}&=\left(\dfrac{c_2}{c_2+1}\right)^2\left(\zeta+K\right)vd \nonumber\\[3pt]  +\dfrac{c_2^2}{\left(c_2+1\right)^3} & \dfrac{\left(\zeta+K\right)^2v^2}{H}\left[\exp\left(-\dfrac{H\left(c_2+1\right)d}{\left(\zeta+K\right)v}\right)-1\right]
\end{flalign} 
In the extrapolated limit of zero solvent friction, $\zeta\to 0$ by definition of the bead-friction coefficient. Upon taking this limit in Eq.~(\ref{eq:dim_wdis_def}), 
\begin{align}\label{eq:wdis_limit}
\left<W\right>_{\text{dis},\,\eta_{\text{s}}\to0}&=\left(\dfrac{c_2}{c_2+1}\right)^2{Kvd} \nonumber\\[3pt] +\dfrac{c_2^2}{\left(c_2+1\right)^3} & \dfrac{K^2v^2}{H}\left[\exp\left(-\dfrac{H\left(c_2+1\right)d}{Kv}\right)-1\right]
\end{align}
In the limit of high pulling trap stiffness ($c_2 \gg 1$), the second term on the RHS of Eq.~(\ref{eq:wdis_limit}) vanishes, and the parenthesized prefactor in the first term asymptotically tends to unity, leading to
\begin{align}
\dfrac{\left<W\right>_{\text{dis},\,\eta_{\text{s}}\to0}}{d}=Kv,
\end{align}
Clearly, the proposed protocol for determining the internal friction coefficient based on the Jarzynski equality recovers the damping coefficient $K$, establishing its validity in the case of the simple analytical model considered here.

\subsection{\label{sec:sp_dash_sim}Non-linear spring-dashpot with hydrodynamic interactions}

\subsubsection{\label{sec:mod_des_db}Model description}

\begin{figure}[ptbh]
\centering
\includegraphics[width=0.9\linewidth]{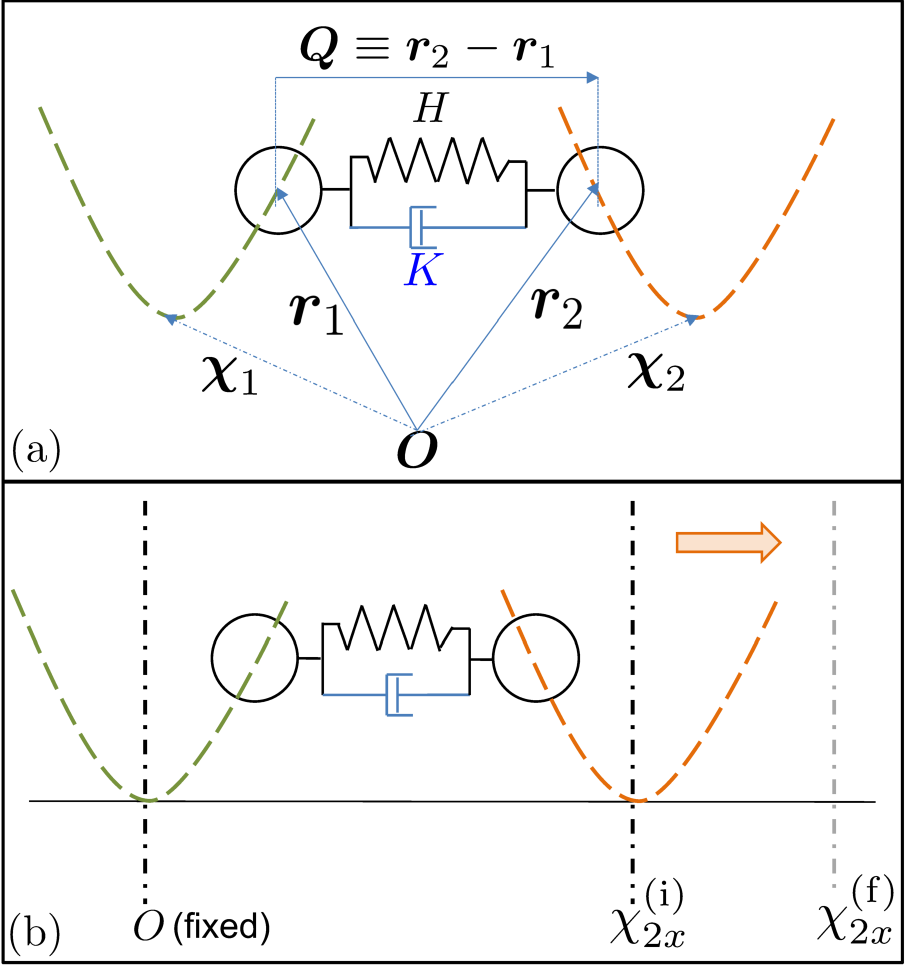} 
\caption{\small (Color online) \textbf{Schematic of the proposed simulation/experiment}. (a) Schematic diagram of the coarse-grained polymer model entrapped between two optical tweezers. The spring connecting the two beads is finitely extensible, upto a length $Q_0$. Internal friction is modelled using the dashpot, whose damping coefficient is $K$. The Hookean spring constant associated with the spring is $H$. The strengths of the two traps, modelled as Harmonic potential wells, are $H_1=c_1H$ and $H_2=c_2H$. (b) The one-dimensional pulling protocol : the position of the first trap is taken to be the origin, and remains stationary throughout the experiment. The second trap is moved from its initial position, $\chi^{\textrm{(i)}}_{2x}$ to its final position, $\chi^{\textrm{(f)}}_{2x}$, over a time-interval $\tau$, stretching the spring-dashpot setup in the process. The difference between the initial and the final positions of the mobile trap is $d$, and the velocity of pulling is $v_{x}$.}
\label{fig:scheme}
\end{figure}

In the more general case, a dumbbell model with fluctuating internal friction and hydrodynamic interactions is considered, as shown in Fig.~\ref{fig:scheme}~(a). The beads, each of radius $a$, are joined by a spring, with maximum stretchability $Q_0$ and a Hookean spring constant $H$, in parallel with a dashpot of damping coefficient $K$. The Marko-Siggia force expression~\cite{Marko1995}, widely employed to model the force-extension relationship in synthetic polymer molecules~\cite{Black2017}, as well as biopolymers~\cite{Latinwo2014,Raman2014,Sunthar2005,Sasmal2016}, is used to describe the entropic elasticity in the dumbbell. The positions of the two beads are $\bm{r}_1$ and $\bm{r}_2$, the connector vector joining the two beads is denoted by $\bm{Q}\equiv\bm{r}_2-\bm{r}_1$, and the position of the centre of mass by $\bm{R}\equiv(1/2)\left(\bm{r}_1+\bm{r}_2\right)$. Note that while the bead co-ordinates are allowed to sample the entirety of the three-dimensional coordinate space, the pulling is restricted to the $x$-axis alone. An alternative protocol in which the pulling direction is also in general three-dimensional space can be implemented, but this would not alter the analysis and arguments presented here. The positions of the two beads can be manipulated using optical traps, modelled here as harmonic potential wells. The trap stiffnesses are denoted by $H_1=c_1H$, and $H_2=c_2H$ (in units of the dumbbell spring constant), and the co-ordinates of the minimum of the wells are represented by $\bm{\chi}_1$ and $\bm{\chi}_2$, respectively. A temperature of $T=300\,K$ is considered in all our simulations, as a matter of convenience. The viscosity of the solvent at this temperature is taken to be $\eta_{\textrm{s},0}=0.001$ kg/m s, which is close to the viscosity of water at room temperature. In this protocol, values of solvent viscosity which are multiples of $\eta_{\textrm{s},0}$ will be considered.

The finite extensibility parameter, $b$, is defined as $b\equiv Q^2_0/l^2_{\textrm{H}}$. The internal friction parameter, $\varphi\,(\coloneqq 2K/\zeta)$, is defined as the ratio of the internal friction coefficient to the bead friction coefficient. The hydrodynamic interaction parameter is given by $h^*=a/\sqrt{\pi}l_{\textrm{H}}$, and $h^*=0$ corresponds to the free-draining case. 

In Fig.~\ref{fig:scheme}~(b), the pulling protocol employed in this study is depicted. Without any loss of generality, $\bm{\chi}_1$ is chosen as the origin of our frame of reference. In all pulling simulations throughout this work, the first trap is held stationary, and the second trap is moved from its initial position, $\bm{\chi}_2^{\text{(i)}}\equiv(\chi^{\textrm{(i)}}_{2x},0,0)$, to its final position, $\bm{\chi}_2^{\text{(f)}}\equiv(\chi^{\textrm{(f)}}_{2x},0,0)$. The notation ``$\chi^{\textrm{(i)}}_{2x}\to\chi^{\textrm{(f)}}_{2x}$'' represents such a pulling event. The stretching distance is denoted by $d\equiv \left[\chi^{\textrm{(f)}}_{2x}-\chi^{\textrm{(i)}}_{2x}\right]$, the time interval for stretching by $\tau$, and the pulling velocity by $\bm{v}\equiv(v_x,0,0)$, where $v_x=d/\tau$. 

In the more general case, the Hamiltonian, $\mathcal{H}$, of the system for any value of the trap position, $\bm{\chi}_2$,  is given by
\begin{align}\label{eq:ham_def}
\mathcal{H}&=U_{\text{MS}}(\bm{Q})+\frac{H_1}{2}\bm{r}_1^2+\frac{H_2}{2}\left(\bm{r}_2-\bm{\chi}_2\right)^2
\end{align}
where $U_{\text{MS}}(\bm{Q})$ represents the potential energy in the Marko-Siggia spring.
The generalized Jarzynski work corresponding to the pulling protocol discussed above is, in this case, given by
\begin{equation}\label{eq:jarz_work}
W = \int_{0}^{\tau}\left(\frac{\partial \mathcal{H}}{\partial \bm{\chi}_2}\right)\cdot \bm{v}\,dt
\end{equation}
where $\bm{v}=d\boldsymbol{\chi}_2/dt$.
The average dissipation associated with the stretching process is calculated using Jarzynski's equality as shown in Eq.~(\ref{eq:je_use}).

\subsubsection{\label{sec:sim_details_db}Simulation details}

The system of stochastic differential equations governing the time evolution of the non-dimensional connector vector $\bm{Q}^{*}$ and centre of mass coordinates $\bm{R}^{*}$, subjected to the pulling protocol depicted in Fig.~\ref{fig:scheme}, can be derived as shown in Appendix~\ref{sec:gov_eq_db}. Brownian dynamics simulations are used here to solve these equations. To begin with, the initial values of $\bm{Q^*}$ and $\bm{R^*}$ are picked from a Gaussian distribution of zero mean and unit variance. With the first trap held at the origin, and the second at $\bm{\chi}^{\text{(i)}*}_2=(\chi^{\text{(i)}*}_{2x},0,0)$, the dumbbell is equilibrated for a duration of fifty-five dimensionless times. Equilibration is ascertained by checking that $\left<Q^{*2}\right>$ has reached a steady value with respect to time. Then, the pulling is commenced (at $t^*=0$), by varying the position of the second trap linearly, as $\chi^{*}_{2x}=\chi^{\text{(i)}*}_{2x}+v_x^*t^*$, till $t^*=\tau^*$. The window $[0,\tau^*]$ is uniformly divided into $N_{\text{t}}$ intervals, such that $\Delta t_j^*\equiv t^*_{j}-t^*_{j-1}=\tau^*/N_{\text{t}}$, where $j=1,2,..., (N_{\text{t}}+1)$. 

The dimensionless equivalent of the work done by the mobile trap during one realization of the pulling event is calculated using a simple rectangular quadrature as follows,
\begin{equation}\label{eq:numwork}
W^*=c_2\sum_{j=1}^{N_{\text{t}}}\left(\chi^*_{2x}-r^{*}_{2x}\right)_jv_{x}^*\Delta t^*_{j}
\end{equation}
where the subscript $j$ on the first term indicates that it is evaluated at time, $t_j^*$, and $r_{2x}^{*}$ refers to the $x$-coordinate of the position of the dumbbell bead subjected to pulling. For representative values of the molecular and control parameters, which are discussed in more detail in the next subsection, the average dissipated work is computed using the time-step widths $\Delta t^*=\left\{10^{-5},10^{-4},10^{-3}\right\}$. The values of the average work calculated at all the time-steps concur within statistical error bars of the simulation, and the largest of the three time-step widths, i.e; $\Delta t^*=1\times 10^{-3}$, is used for all the cases where $c_2=1000$. For $c_2=100$, $\Delta t^*=1\times 10^{-2}$ is found to suffice, whereas $c_2=10000$ requires $\Delta t^*=1\times 10^{-4}$. The choice of the time-step width is also affected by the values of the internal friction and the finite extensibility parameter as higher values of these parameters necessitate the use of smaller time-steps.

The protocol proposed here involves pulling the molecule over a pre-determined distance at the same dimensional velocity but different solvent viscosities. In this context, it is essential to note that the timescale varies linearly with the solvent viscosity, $\lambda_{\text{H}}\propto \eta_{\text{s}}$. In order to maintain the same dimensional pulling time ($\tau=\tau^*\lambda_{\text{H}}$) across simulations with differing solvent viscosity, the dimensionless pulling time ($\tau^*$) is scaled by $1/\eta_{\text{s}}$ as the solvent viscosity is increased. 

\subsubsection{\label{sec:par_spec}Molecular and control parameters}

The parameters used in the present work are broadly classified into molecular and control parameters. Molecular parameters pertain to the polymer that is being stretched, whereas control parameters are set by the experiments or the simulations used in the study of stretching the molecule. 


The choice of molecular parameters is based on the $\lambda$-phage DNA (48.5 kbp) used in Murayama et al.'s~\cite{Murayama2007} work, which has a contour length, $L_{\text{c}}$, of $16.5\,\mu\text{m}$, and Kuhn segment length, $b_{\text{K}}$, of approximately 88 nm. In order to model this molecule as a dumbbell, the model parameters, $b$, and $l_{\text{H}}$, are chosen such that the contour length and the radius of gyration of the model and the DNA molecule are the same. Following the procedure for parameter selection described in detail in ref.~\citenum{Sunthar2005}, we obtain $b=811.25$ and $l_{\text{H}}=580.95\,\text{nm}$. We round down both these values, and use $b=800$ and $l_{\text{H}}=500\,\text{nm}$ as the parameters to model $\lambda$-phage DNA. 

The Hookean spring constant of the model, $H$, is then found using
\begin{align*}
&H=\frac{k_BT}{l^2_{\text{H}}}=\frac{4.142\,\,\text{pN nm}}{\left(500\right)^2\left(\text{nm}\right)^2}=1.657\times10^{-5}\,\text{pN/nm}
\end{align*}

The choice of the bead radius, $a$, is motivated by Alexander-Katz \textit{et al}.'s~\cite{Alexander-Katz2009} work, where it is suggested that the monomeric radius may be taken as the persistence length of the molecule. We choose $a=30\,\text{nm}$, which is identical to the choice made by Alexander-Katz \textit{et al}.~\cite{Alexander-Katz2009} for comparing the results of BD simulations against experiments on DNA.   

Other values of $b$ and $l_{\text{H}}$, of the same order-of-magnitude as obtained for the $\lambda$-phage DNA case, have been used in this study and in addition to $a=30\,\text{nm}$, bead radii of $80\,\text{nm}$ and $100\,\text{nm}$ have also been used.


\begin{figure}[tb]
\centering
\includegraphics[width=0.99\linewidth]{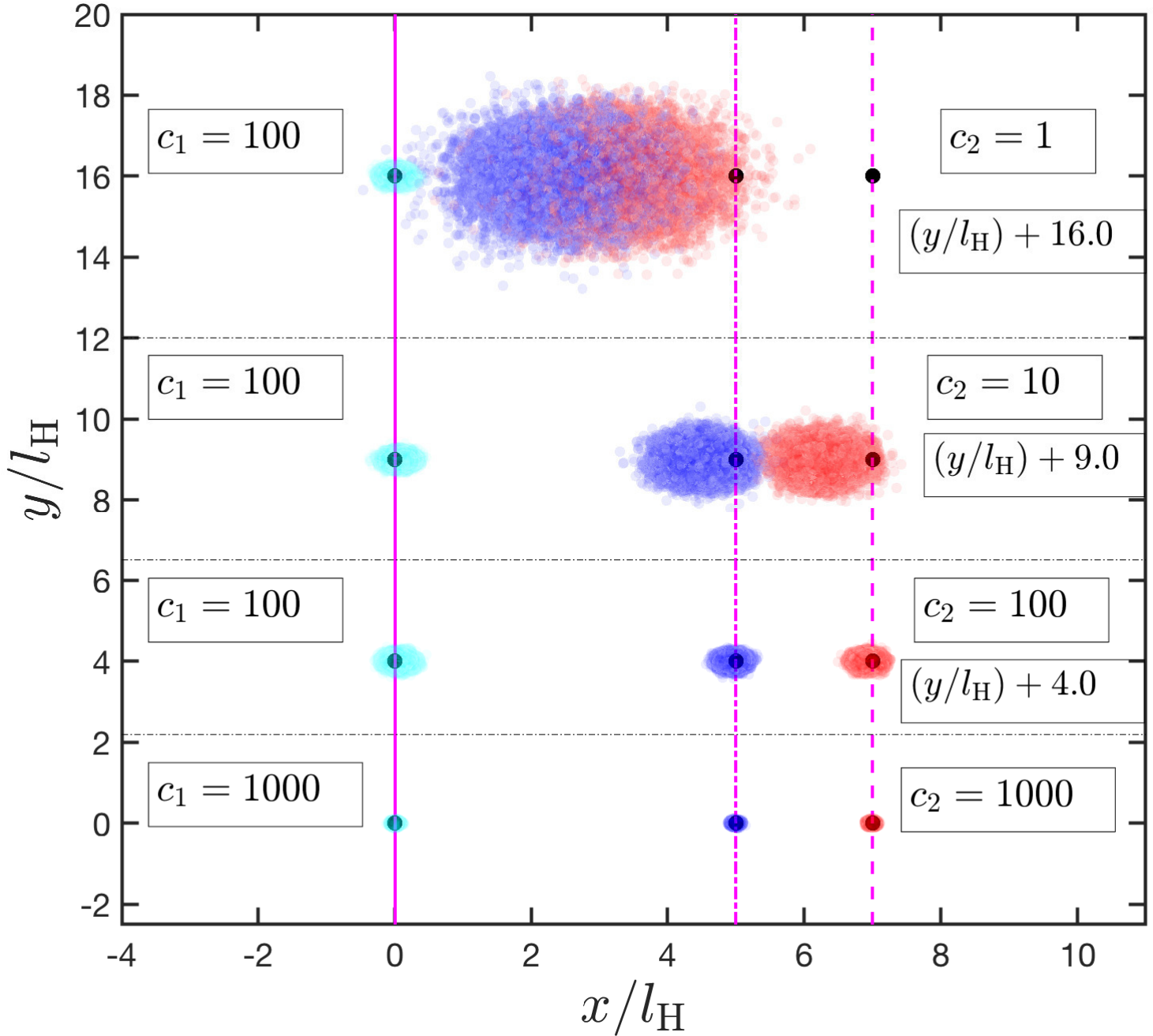}
\caption{\small (Color online) \textbf{Trap stiffness determines the distribution of bead positions.} An $x-y$ projection of the equilibrated positions of the beads, for an ensemble of free-draining dumbbells ($N=1\times10^4$), for different values of the trap stiffness. From left to right, the solid vertical line, the dash-dotted vertical line, and the dashed vertical line represent the positions of the first trap $(\bm{\chi}^{*}_1)$, the initial position of the second trap $(\bm{\chi}^{\text{(i)}*}_2)$, and the final position of the second trap $(\bm{\chi}^{\text{(f)}*}_2)$, respectively. From left to right, data points in cyan, close to the solid vertical line, correspond to the positions of the first bead with the corresponding trap at the origin, those in blue denote the positions of the second bead when the trap position is at $\bm{\chi}^{\text{(i)}}_2=\left(5\,l_{\text{H}},0,0\right)$, and those in red represent the positions of the second bead when the trap is located at $\bm{\chi}^{\text{(f)}}_2=\left(7\,l_{\text{H}},0,0\right)$. The data sets have been shifted vertically for clarity, and the offsets for each of the shifted cases are indicated in the figure.}
\label{fig:bead_pos}
\end{figure}

Fig.~\ref{fig:bead_pos} provides a snapshot of the $x$-$y$ projection of the positions of an ensemble of beads of the dumbbell, obtained after an equilibration of fifty-five dimensionless times at the initial trap positions $\bm{\chi}^{*}_1=\left(0,0,0\right), \,\bm{\chi}^{\text{(i)}*}_2=\left(5,0,0\right)$ and final trap states $\bm{\chi}^{*}_1=\left(0,0,0\right) , \,\bm{\chi}^{\text{(f)}*}_2=\left(7,0,0\right)$, as a function of the optical trap stiffness. 
It is clearly seen that the strength of the trap ($c_1$ or $c_2$) determines its ability to confine the bead near the position of its minimum. Since it is intended to hold the position of the first bead fixed at the origin, a trap strength of $c_1=1000$ is used throughout our simulations.

\begin{figure}[tb]
\begin{center}
\includegraphics[width=0.8\linewidth]{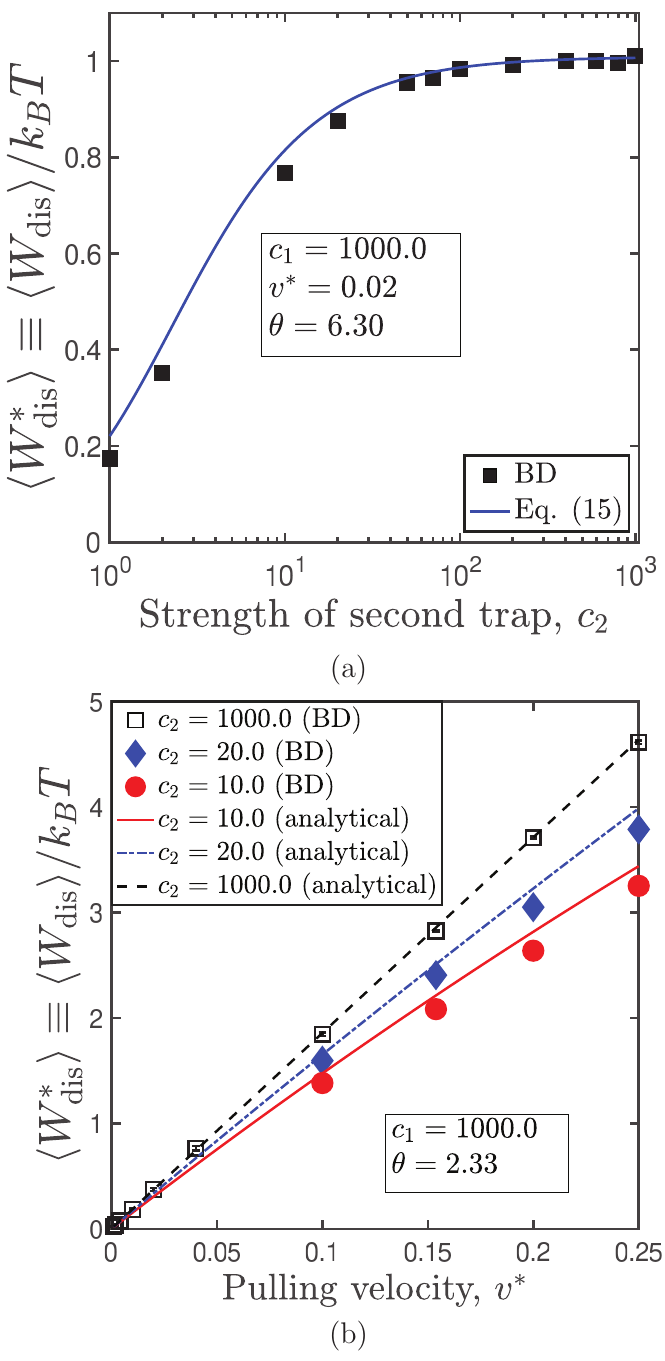}
\end{center}
\caption{\small (Color online) \textbf{Comparison of the average dissipation from BD simulations against analytical results} Plots of the average dissipation for the pulling protocol $5\,l_{\mathrm{H}}\to7\,l_{\mathrm{H}}$, as a function of (a) mobile trap stiffness, for a fixed value of the pulling velocity, $v^*=0.02$, and (b) pulling velocity, at three different values of the mobile trap stiffness. The lines indicate analytical value of the average dissipation for a freely draining Hookean spring-dashpot [Eq.~(\ref{eq:wdis_spec})] plotted for the parameter values indicated in the figure. The symbols are results from BD simulations on a freely draining spring-dashpot with Marko-Siggia force law, and $b=800$.}
\label{fig:trp_stiff}
\end{figure}

In Figs.~\ref{fig:trp_stiff}, the effect of the mobile trap stiffness ($c_2$) on the dissipation is shown for a fixed value of the stationary trap stiffness, for freely-draining dumbbells. In Fig.~\ref{fig:trp_stiff}~(a), the average dissipation as the dumbbell is pulled at a constant velocity and a fixed pulled distance, is plotted as a function of the mobile trap stiffness. It is seen that the dissipation reaches a plateau at $c_2\approx100$. This behavior is in agreement with the dissipation calculated for the analytical model [Eq.~(\ref{eq:wdis_spec})], which reaches $\sim 98\%$ of its asymptotic value at $c_2\approx100$. In all our simulations, we set $c_1=c_2=1000$ (unless specified otherwise), in order to operate in a regime where the dissipation is independent of the trap stiffness. 

In Fig.~\ref{fig:trp_stiff}~(b), the average dissipation is plotted as a function of the pulling velocity, over the same fixed distance, for three different values of the mobile trap stiffness. At lower values of the trap stiffness, the dissipation grows linearly before scaling sub-linearly with the velocity. The onset of the non-linear regime is pushed to higher velocities as the trap stiffness is increased. In the asymptotic limit of high trap stiffness, the non-linear regime vanishes, and a linear scaling of the dissipation with the pulling velocity is observed for over two orders of magnitude in the velocity. There is a good agreement between the trends predicted by the simple analytical model using a Hookean spring-dashpot discussed in Sec.~\ref{sec:analytical} and BD simulations on a model with a nonlinear force law. Remarkably, the effect of the non-linear force law is more perceptible at lower trap stiffnesses, vanishing as the mobile trap stiffness is increased, until a quantitative agreement is obtained between the simulation and the analytical results.

The Marko-Siggia force expression is linear at low values of the extension of the molecule, and diverges as the fractional extension approaches unity. For initial extensions in the linear regime of the force-extension profile, a trap stiffness of $c_2=1000$ is sufficient to make the bead track the position of the trap, as shown in Fig.~{\ref{fig:bead_pos}}. Higher trap stiffnesses are found to be required for operating in the non-linear regime of the force-extension curve.

The Jarzynski equality is strictly \textit{exact} only in the limit of an infinite number of work trajectories, $N\to\infty$. In applications of the JE, the number of trajectories required to accurately recover the free-energy difference increases with average dissipated work in the process, as discussed in refs.~\citenum{Ritort2002,Jarzynski2006,YungerHalpern2016}. 

\begin{figure}[ptbh]
\begin{center}
\begin{tabular}{c}
\includegraphics[width=0.99\linewidth]{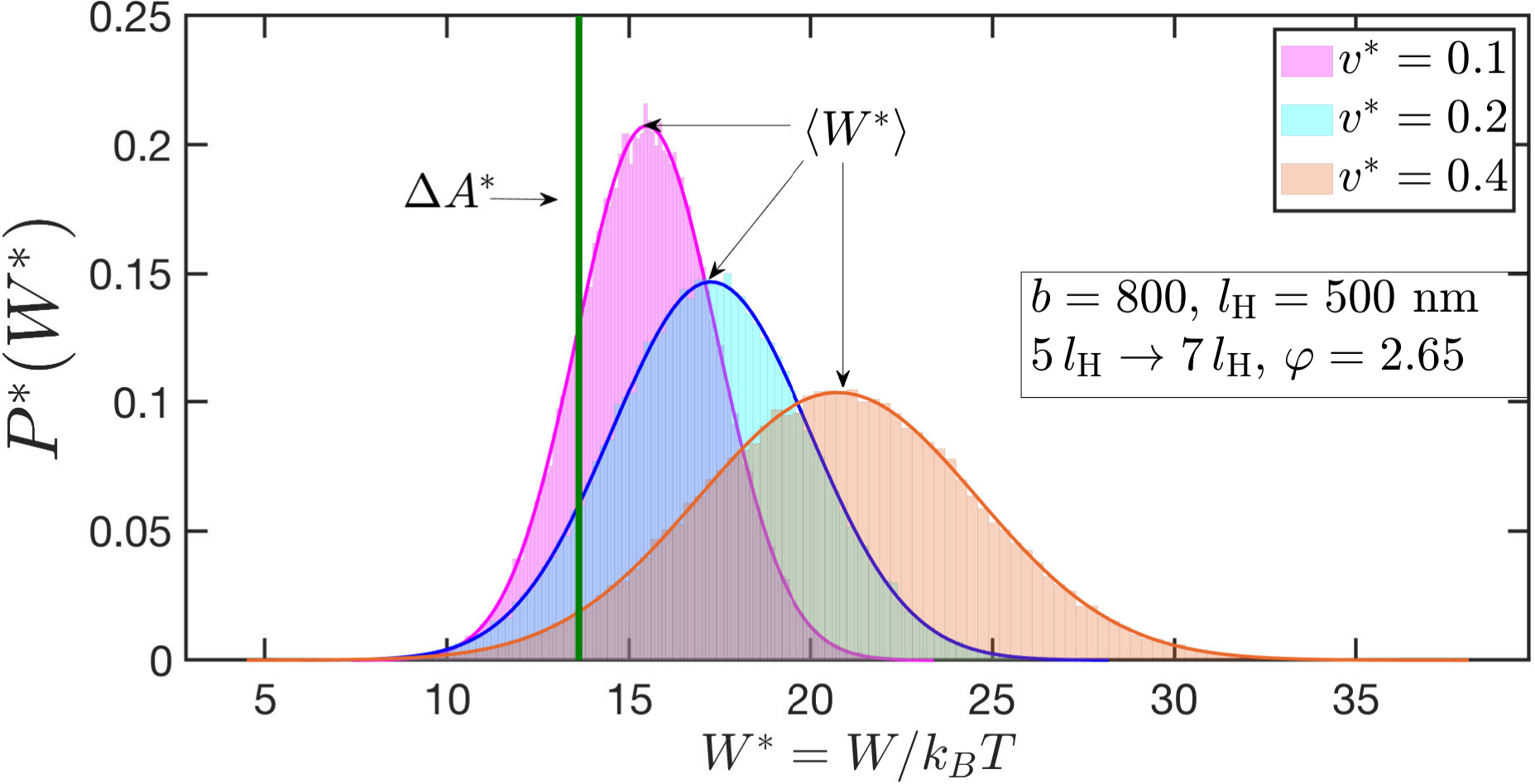}\\
(a) \\
\includegraphics[width=0.99\linewidth]{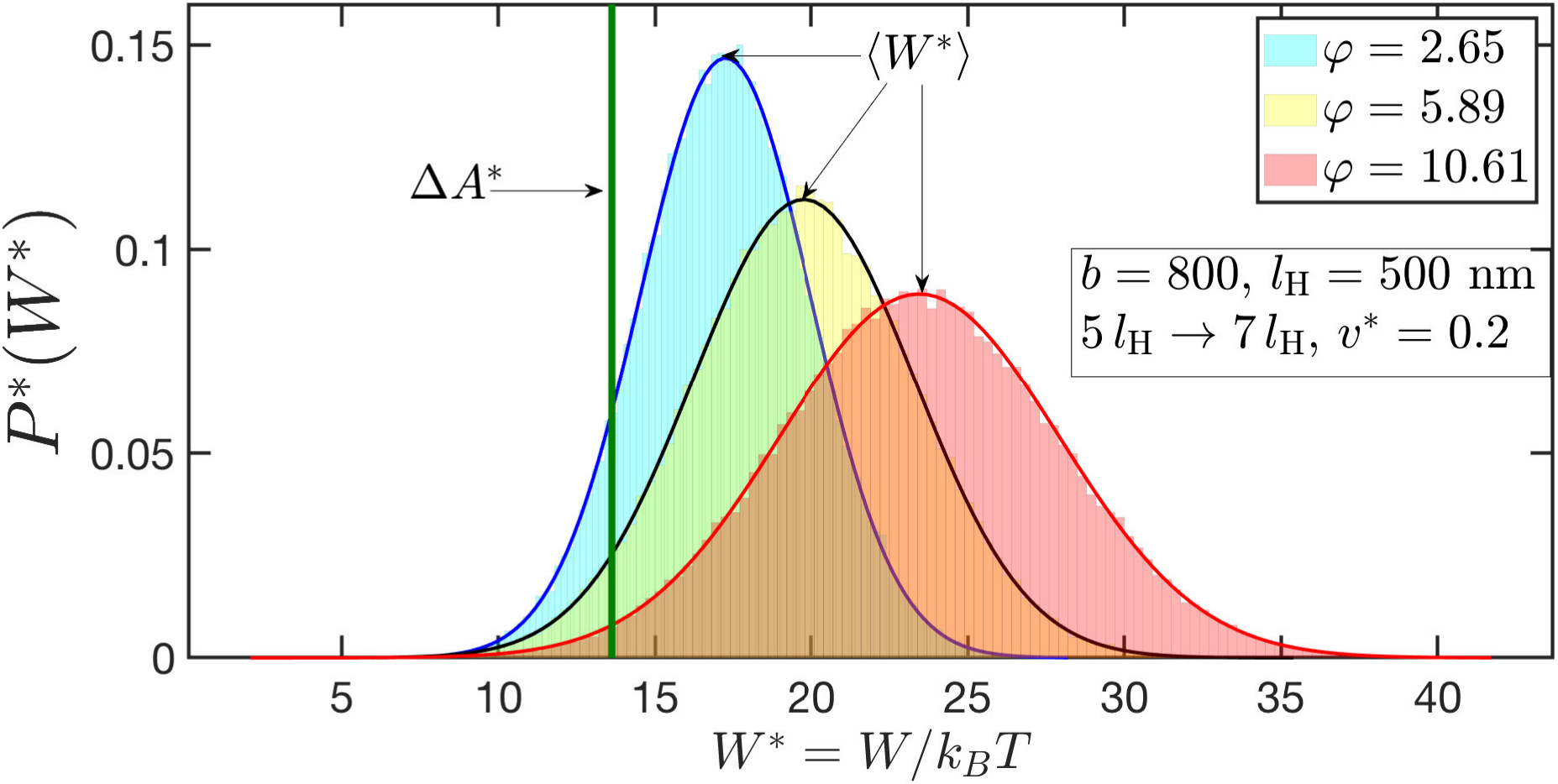}\\
(b)  \\
\end{tabular}
\end{center}
\caption{\small (Color online) \textbf{Pulling velocity and internal friction broaden the the probability density of work.} Probability densities of the work done over $10^5$ realizations of the pulling protocol, for: (a) a fixed value of the internal friction parameter, and three different values of the pulling velocity, and (b) a fixed pulling velocity, and three different values of the internal friction parameter. The green vertical line represents the free-energy difference obtained by taking an error-weighted mean of the values of the free energy difference obtained at pulling velocities $v^* \leq0.02$ using Jarzynski's equality. The solid lines are Gaussian fits to the data.}
\label{fig:dist_width}
\end{figure}

In Figs.~\ref{fig:dist_width}, the effect of internal friction and pulling velocity on the probability distribution of the work trajectories is plotted for freely-draining dumbbells. The vertical green lines in the figures indicate the free-energy difference, $\Delta A^*$, obtained by taking an error-weighted mean of the values of the free energy difference obtained at dimensionless pulling velocities $v^* \leq 0.02$.  Interestingly the work-distributions are normally distributed, as seen by the good agreement between the histogram data and the Gaussian fit. That it is so, despite the equations of motion for the system being non-linear, is an observation previously made by Speck and Seifert~\cite{Speck2004}. The Gaussianity of the distribution is attributed to the slow rate of the driving protocol ($1/\tau$) with respect to the molecular relaxation rate ($1/\lambda_{\text{H}}$). For all the data-sets plotted in Figs.~\ref{fig:dist_width}, the driving rate is at least five times slower than the molecular relaxation rate.

\begin{figure}[ptbh]
\centering
\includegraphics[width=0.95\linewidth]{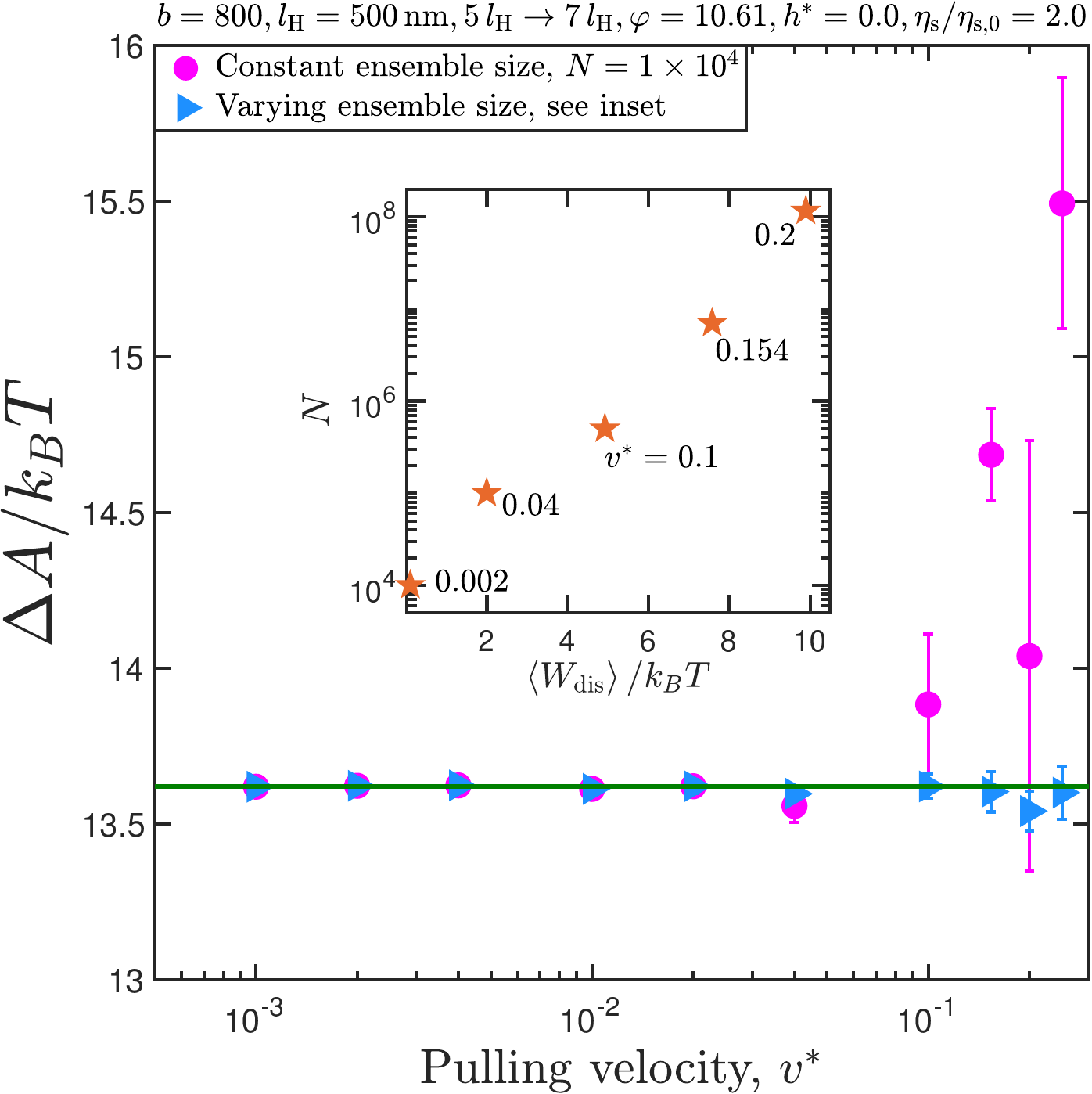}
\caption{\small (Color online) \textbf{Dissipation dictates ensemble size for the application of Jarzynski's equality.} Free energy difference as a function of pulling velocity, for a representative case. Inset shows the empirically chosen ensemble size as a function of the average dissipated work. The numbers next to the data points indicate the corresponding values of the dimensionless pulling velocity. Trap stiffness used is $c_1=c_2=1000$.}
\label{fig:ensemble_size}
\end{figure}

From Fig.~\ref{fig:dist_width}~(a), it is seen that increasing the pulling velocity at a fixed value of the internal friction parameter increases the average dissipated work, and the width of the distribution. An identical trend is observed in Fig.~\ref{fig:dist_width}~(b), where an increase in the internal friction parameter at a fixed pulling velocity causes the work distribution to shift rightwards, and results in an increased dissipation. Thus, the dissipation in our model is directly correlated with the pulling velocity, and the internal friction in the system. Under such conditions of high dissipation, the estimates for $\Delta A$ are dominated by rare realizations that occur near the tail of the work distribution, necessitating the use of a larger number of trajectories to obtain an accurate estimate of the free energy difference.

In Fig.~\ref{fig:ensemble_size}, we illustrate the above point using an alternative representation. Without prior knowledge of the ensemble size required for the simulations, an initial guess of $N=1\times10^4$ was chosen. The green horizontal line represents the free-energy difference obtained by taking an error-weighted mean of the values of the free energy difference obtained at pulling velocities $v^* \leq 0.02$ using Jarzynski's equality. An ensemble size of $N=1\times10^4$ is sufficient for an accurate estimation of the free energy difference at lower velocities (dissipation), but is found to become inadequate at $v^{*}\geq0.04$. Upon increasing the ensemble size for the higher velocity cases \textit{empirically}, the accuracy of the estimated free energy difference improves. The ensemble size is plotted as a function of the average dissipated work in the inset of Fig.~\ref{fig:ensemble_size}, which shows that the choice of the pulling velocity and the ensemble size are mutually related.

\begin{figure}[ptbh]
\centering
\includegraphics[width=0.99\linewidth]{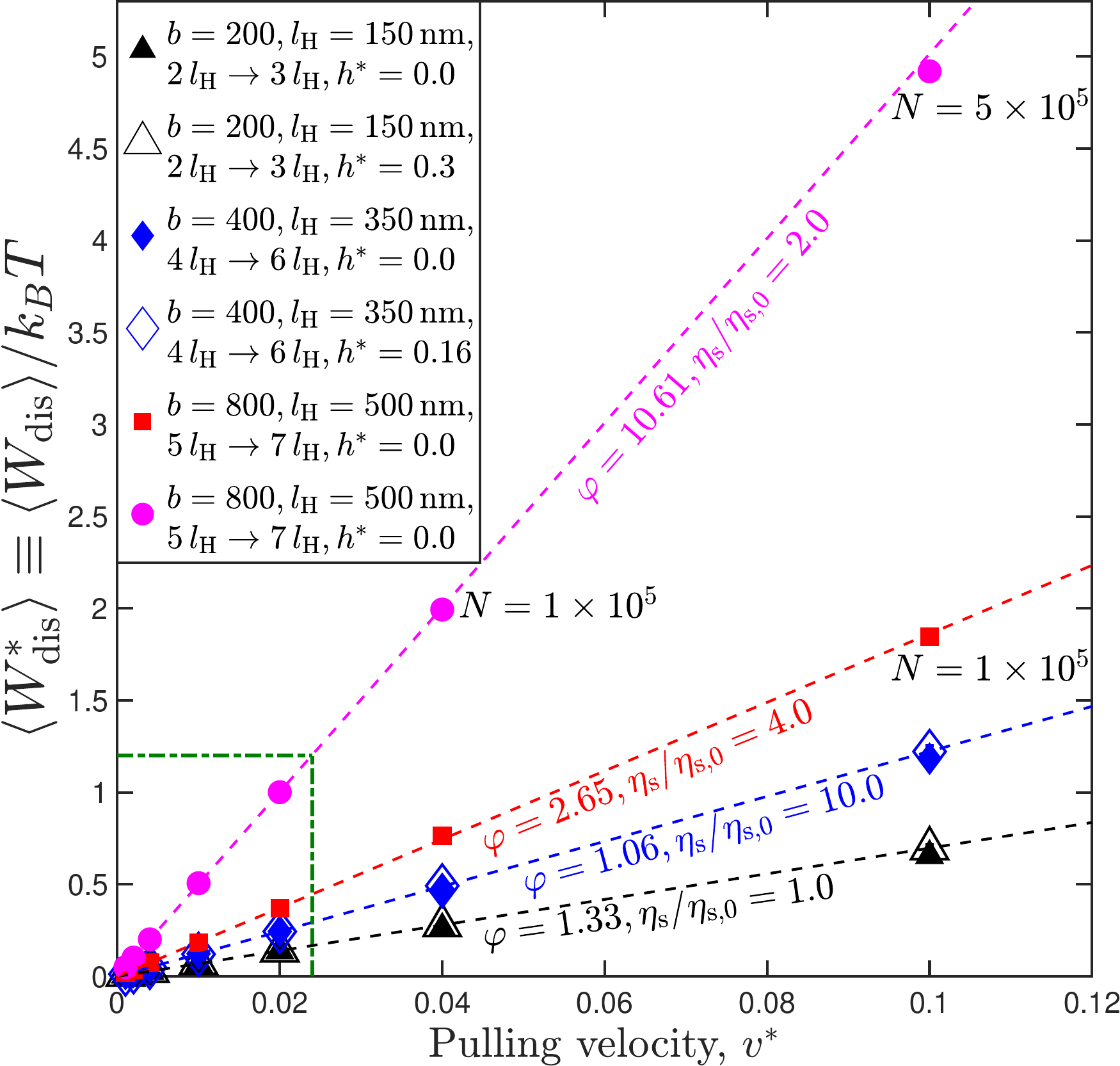}  
\caption{\textbf{Linearity between dissipation and pulling velocity determines regime of operation}. Average dissipated work as a function of the dimensionless pulling velocity, for various molecular and control parameters. Except when mentioned otherwise, an ensemble size of $N=1\times10^{4}$ is used for all the data points. Symbols indicating datasets with fluctuating hydrodynamic interactions have been enlarged for the sake of clarity. The boxed region indicates the regime of operation for the simulation results reported in this section.}
\label{fig:vel_linear}
\end{figure}

In Fig.~\ref{fig:vel_linear}, the average dissipated work (scaled by the thermal energy $k_BT$) calculated for a variety of molecular and control parameters is plotted against the magnitude of the dimensionless pulling velocity, $v^*$.  It is seen that the average dissipated work varies linearly over the entire range of the pulling velocity, $v^*=0.001-0.1$. The velocity range in dimensional units would depend on the molecular parameters.

\begin{table}[t]
\setlength{\tabcolsep}{4pt}
\centering
\caption{\label{exp_param} Typically observed lower and upper bounds on optical tweezer parameters.}
\begin{center}
\begin{ruledtabular}
\begin{tabular}{c c c}
Parameter & Lower bound & Upper bound\\
\hline
Trap stiffness  (pN/nm) & $0.0002$ [ref.~\citenum{Black2017}]& $0.9$ [ref.~\citenum{Gupta2011}]\\[5pt]
Pulling velocity (nm/s) & $10$ [ref.~\citenum{Gupta2011}]& $13560$ [ref.~\citenum{Trepagnier2004}] \\[5pt]
Stretching distance (nm) & $10$ [ref.~\citenum{Gupta2011}]& $8000$ [ref.~\citenum{Murayama2007}] \\[5pt]
\end{tabular}
\end{ruledtabular}
\end{center}
\end{table}

\begin{figure}[tb]
\begin{center}
\begin{tabular}{c}
\includegraphics[width=0.8\linewidth]{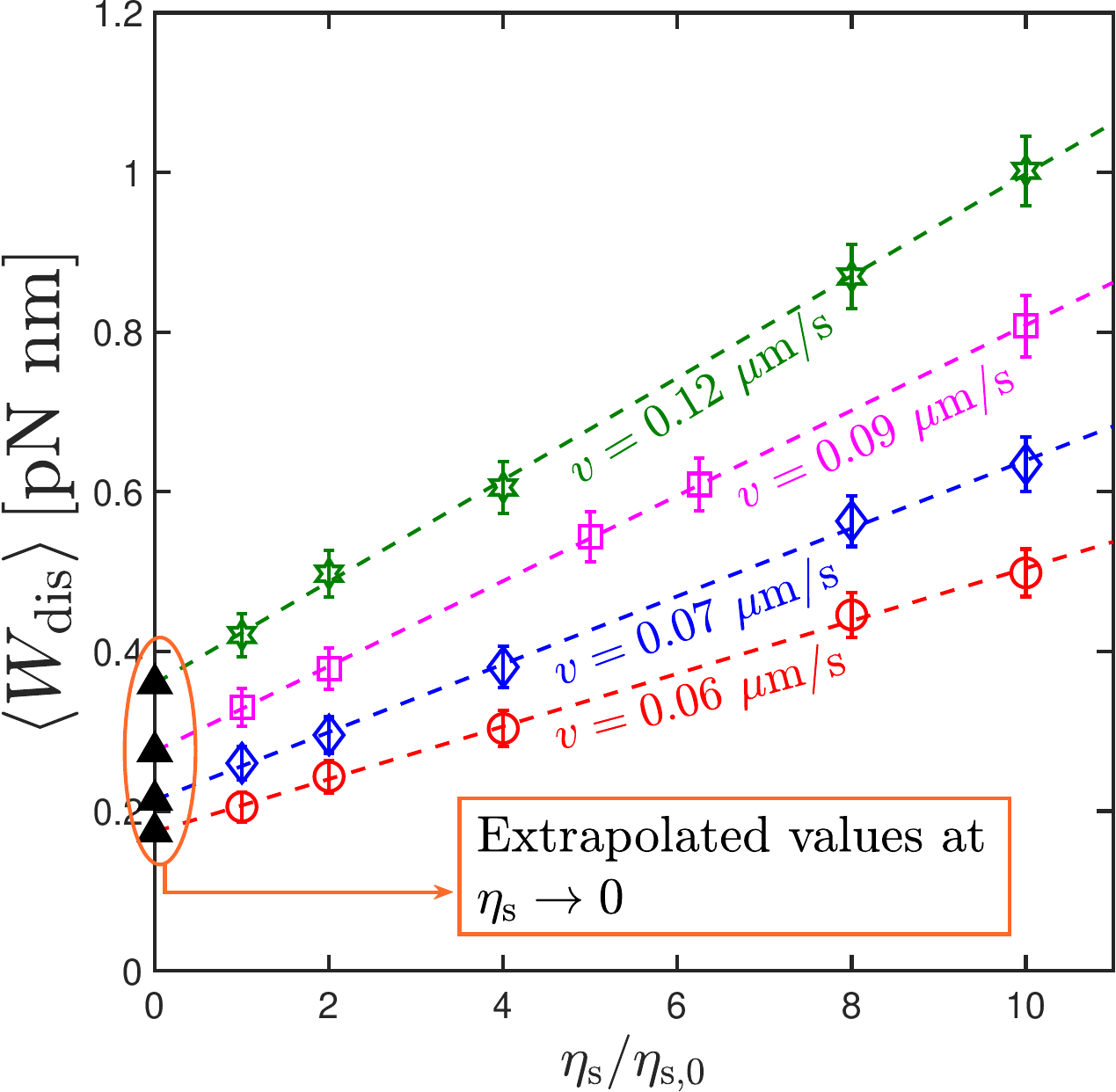}\\[5pt]
(a)\\[5pt]
\includegraphics[width=0.8\linewidth]{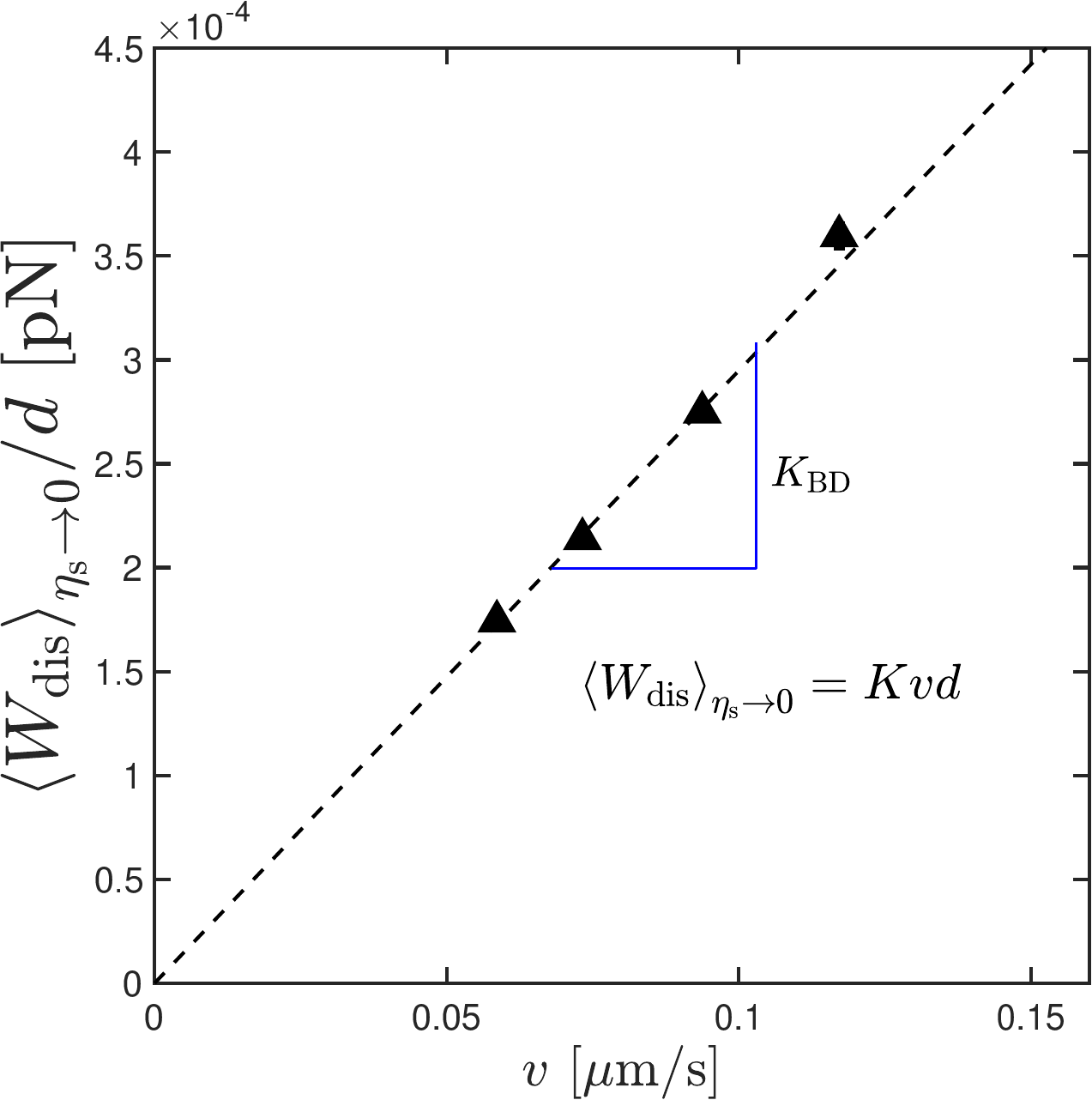}\\
(b)
\end{tabular}
\end{center}
\caption{\textbf{Protocol for the extraction of the internal friction coefficient.} (a) Average dissipated work as a function of the solvent viscosity, for molecules with the parameters: $\left\{b=800, l_{\mathrm{H}}=500\,\mathrm{nm}, K=3.0\times10^{-9}\,\text{kg/s}\right\}$, subjected to pulling denoted by $5 \, l_{\textrm{H}}\to7 \, l_{\textrm{H}}$ at various values of the pulling velocity, $v$, for an ensemble size, $N=1\times10^{4}$. (b) The extrapolated values of $\left<W_{\textrm{dis}}\right>$ in the hypothetical limit of zero solvent viscosity, divided by the stretching distance, as a function of the pulling velocity. The slope of the graph, $K_{\textrm{BD}}$, is an estimate of the internal friction coefficient.}
\label{fig:extract_iv}
\end{figure}

The experimental feasibility of the proposed protocol can be discussed in the context of the molecular parameters used for the dataset represented by filled circles in Fig.~\ref{fig:vel_linear}. For these set of parameters, the pulling velocities explored in Fig.~\ref{fig:vel_linear} vary from $v=29.3\,\textrm{nm/s}\,(v^*=0.001)$ to $v=2.93\,\mu\textrm{m/s}\,(v^*=0.1)$. The molecule is stretched over a distance of $1\mu\textrm{m}$. The stiffness of this molecule is $H=1.657\times10^{-5}$ pN/nm. In order to operate in a regime where the dissipated work is independent of the trap strength, as discussed earlier, the stiffness of the trap must be at least a hundred times that of the molecule, which implies $H_{\textrm{trap,min}}=1.657\times10^{-3}$ pN/nm. 

In Table~\ref{exp_param}, based on a survey of the literature, the range of trap stiffnesses, pulling velocities, and stretching distances typically accessible by optical tweezers is given.  Additionally, the position and force resolution limits of optical traps are inversely correlated: stiffer traps improve the spatial resolution but also introduce large fluctuations in the measured force. A rough estimate of these resolution limits may be obtained using the equipartition theorem, as explained in refs.~\citenum{Smith2007,Neuman2008}. Most commercial optical tweezer setups are equipped with filtering mechanisms that aid in improving the precision in the measurements, by reducing the resolution limits~\cite{Gupta2011}. A detailed discussion of the resolution offered by optical tweezers can be found in refs.~\citenum{Smith2007} and~\citenum{Neuman2008}. 

From Table~\ref{exp_param}, it is clear that the values of $v$, $d$, and $H_{\textrm{trap,min}}$ for the representative case lie well within the range of values explored experimentally.

Ritort et al.~\cite{Ritort2002} have established from computer simulations of mechanical unfolding that the number of trajectories required to obtain estimates for free energy difference within an error of $\mathcal{O}\left(k_BT\right)$ increases exponentially with the average dissipation associated with the unfolding process. They predict that for dissipation less than $4k_BT$, around 100 trajectories would suffice, and for a dissipation of $5k_BT$, about 1000 trajectories would be required. These predictions agree well with the average dissipation and ensemble sizes encountered in optical-tweezer-based pulling experiments. For example, Liphardt et al.~\cite{Liphardt2002} stretch RNA hairpins using optical tweezers, and estimate $\left<W_{\text{dis}}\right> = 2\,\text{-}\,3\,k_BT$ with $N=47$. Similarly, for pulling experiments on DNA hairpins performed by Gupta et al.~\cite{Gupta2011}, $\left<W_{\text{dis}}\right> = 1.1\pm0.7 k_BT$ for $N=99$, and $\left<W_{\text{dis}}\right> = 4.9\pm0.3 k_BT$ for $N=1293$.

For simulations on the single-mode spring dashpot, the statistical error in the free-energy difference is maintained to be $\sim \mathcal{O}\left(0.01k_BT\right)$, in order to obtain a sufficiently accurate estimate of the average dissipated work that enables the internal friction coefficient to be extracted reliably. By restricting the regime of operation to the boxed region in Fig.~\ref{fig:vel_linear}, with $v^*\leq0.02$ and $\left<W_{\textrm{dis}}\right>\sim k_BT$, it is found that $N=1\times10^{4}$ trajectories are sufficient to obtain the free energy difference within the desired error limits. It is possible to operate at higher values of dissipation, outside the boxed regime, provided that the ensemble size is suitably increased, as shown in the inset of Fig.~\ref{fig:ensemble_size}.

All the results prior to Fig.~\ref{fig:vel_linear} have been presented for the case of freely-draining dumbbells. In Fig.~\ref{fig:vel_linear}, it is observed that the inclusion of fluctuating HI does not affect the dissipated work values in a single-mode spring dashpot. Speck and coworkers~\cite{Speck2017} have shown in the context of colloidal suspensions that the inclusion of hydrodynamic interactions does not alter the dissipation along a single trajectory. The effect of fluctuating hydrodynamic interactions on the dissipation is markedly different in the case of wet internal friction, as will be discussed in greater detail in Sec.~\ref{sec:sc_res}. 

\subsubsection{\label{sec:results}Dry friction in equals dry friction out}

\renewcommand{\thefootnote}{\alph{footnote}}
\begin{table*}[t]
\centering
\caption{\label{k_bd} Internal friction coefficients estimated using the protocol described in Fig.~\ref{fig:extract_iv}, for various values of the molecular, and control parameters. The error associated with the protocol is calculated as, $\%\,\textrm{error}=100\times\left\vert\left(K_{\textrm{BD}}-K\right)/K\right\vert$. An ensemble size of $N=1\times10^4$ and a time-step width of $\Delta t^*=1\times10^{-3}$ was used to obtain the results. For the $K=100.0\times 10^{-9}$ kg/s case (Marko-Siggia), a smaller time-step, $\Delta t^*=1\times10^{-4}$, was used. For the  $K=4.0\times 10^{-9}$ kg/s case (FENE), a larger ensemble, $N=1\times10^5$, was used.}
\begin{center}
\begin{ruledtabular}

\begin{tabular}{c c  c c c c c}
& {Spring force law: }& {Marko-Siggia} & &  &\\[2pt]
\hline\hline
& Input, & $\qquad \qquad \qquad h^*=0.0$ & & & $h^*>0.0$\\ \cline{3-4}\cline{5-7} Parameters & $K\,[\times 10^9$ kg/s]& $K_{\text{BD}} [\times 10^9$ kg/s] & $\%$ error& $h^*$ & $K_{\text{BD}} [\times 10^9$ kg/s] & $\%$ error\\
\hline
$b=200,\,l_{\mathrm{H}}=150\,\mathrm{nm},\,2\,l_{\mathrm{H}}\to3\,l_{\mathrm{H}}$ & $1.0$& $1.04\pm0.01$&$4.42$&$0.3$& $0.97\pm0.02$ & $3.11$\\
$b=400,\,l_{\mathrm{H}}=350\,\mathrm{nm},\,4\,l_{\mathrm{H}}\to6\,l_{\mathrm{H}}$ & $1.0$& $0.99\pm0.03$&$1.36$& $0.16$&$1.04\pm0.02$ & $4.29$\\
$b=200,\,l_{\mathrm{H}}=150\,\mathrm{nm},\,2\,l_{\mathrm{H}}\to3\,l_{\mathrm{H}}$ & $10.0$& $10.04\pm0.05$&$0.47$&$0.3$& $9.87\pm0.09$ & $1.24$\\
$b=400,\,l_{\mathrm{H}}=350\,\mathrm{nm},\,4\,l_{\mathrm{H}}\to6\,l_{\mathrm{H}}$ & $10.0$& $9.922\pm0.008$&$0.78$& $0.16$& $9.95\pm0.09$ & $0.52$\\
$b=800,\,l_{\mathrm{H}}=500\,\mathrm{nm},\,5\,l_{\mathrm{H}}\to7\,l_{\mathrm{H}}$ & $3.0$& $2.95\pm0.02$&$1.74$& $-$&$-$&$-$\\
$b=800,\,l_{\mathrm{H}}=500\,\mathrm{nm},\,5\,l_{\mathrm{H}}\to7\,l_{\mathrm{H}}$ & $6.0$& $6.00\pm0.07$&$0.11$& $-$ & $-$ & $-$\\
{$b=800,\,l_{\mathrm{H}}=500\,\mathrm{nm},\,5\,l_{\mathrm{H}}\to7\,l_{\mathrm{H}}$} &{$100.0$}&{$97.4\pm1.4$}&{$2.56$}&{$-$} &{$-$} &{$-$}\\
\hline \hline
& {Spring force law: }& {FENE} & &  &\\[2pt]
\hline\hline
& Input, & $\qquad \qquad \qquad h^*=0.0$ & & & $h^*>0.0$\\ \cline{3-4}\cline{5-7} Parameters & $K\,[\times 10^9$ kg/s]& $K_{\text{BD}} [\times 10^9$ kg/s] & $\%$ error& $h^*$ & $K_{\text{BD}} [\times 10^9$ kg/s] & $\%$ error\\
\hline
$b=400,\,l_{\mathrm{H}}=350\,\mathrm{nm},\,2\,l_{\mathrm{H}}\to3\,l_{\mathrm{H}}$ & $2.0$& $-$&$-$& $0.23$& $1.993\pm0.009$ & $0.35$\\
$b=600,\,l_{\mathrm{H}}=250\,\mathrm{nm},\,3\,l_{\mathrm{H}}\to4\,l_{\mathrm{H}}$ & $3.0$& $-$&$-$& $0.09$& $2.985\pm0.009$ & $0.50$\\
$b=800,\,l_{\mathrm{H}}=500\,\mathrm{nm},\,3\,l_{\mathrm{H}}\to5\,l_{\mathrm{H}}$ & $4.0$& $3.992\pm0.007$&$0.18$& $-$&$-$&$-$\\
$b=800,\,l_{\mathrm{H}}=500\,\mathrm{nm},\,3\,l_{\mathrm{H}}\to5\,l_{\mathrm{H}}$ & $6.0$& $6.02\pm0.02$&$0.23$& $-$&$-$&$-$\\
\end{tabular}
\end{ruledtabular}
\end{center}
\end{table*}

The methodology to extract the internal friction coefficient is illustrated using a molecule with parameters: $\left\{b=800, l_{\mathrm{H}}=500\,\mathrm{nm}, K=3.0\times10^{-9}\,\text{kg/s}, h^*=0.0\right\}$ as an example. An ensemble of such molecules is pulled from an initial trap position of ${\chi}_{2x}^{\text{(i)}}=5\,l_{\text{H}}$ to a final trap position of ${\chi}_{2x}^{\text{(f)}}=7\,l_{\text{H}}$ at different pulling velocities. At each value of the pulling velocity, the average dissipated work is calculated at several values of the solvent viscosity in the range, $\eta_{\textrm{s}}=\eta_{\textrm{s},0}$ to $\eta_{\textrm{s}}=10\,\eta_{\textrm{s},0}$. In an experimental setting with water as the solvent, suitable viscogens, such as glucose or sucrose, may be added to the solvent in order to realize an approximately four-fold increase in its viscosity~\cite{Jas2001,Qiu2004}. In experiments that study the kinetics of intrachain contact formation in polypeptides~\cite{Bieri1999} suspended in a solvent mixture of ethanol and glycerol, the solvent viscosity was varied over two orders of magnitude by adjusting the proportion of glycerol in the mixture. 

As shown in Fig.~\ref{fig:extract_iv}~(a), for each value of the pulling velocity used, the average dissipated work in the hypothetical limit of zero solvent viscosity, $\left<W_{\text{dis}}\right>_{\eta_{\!_{\, \text{s}}}\to\,0}$, is obtained from a linear fit to the average dissipated work at finite solvent viscosities. Since the extrapolated value is finite, it is clear signature of the presence of dry internal friction. In Fig.~\ref{fig:extract_iv}~(b), the extrapolated values of the average dissipated work in the limit of zero solvent viscosity (divided by the stretching distance $d$), is plotted against the pulling velocity. The slope of the graph $(K_{\textrm{BD}})$ represents the internal friction coefficient extracted from simulations. 

Table~\ref{k_bd} shows a comparison between the value of the internal friction coefficient used as an input parameter in the Brownian dynamics simulations, and the corresponding value extracted from the dissipated work using the protocol proposed here, for various molecular and control parameters. Our protocol recovers the input internal friction coefficient to within $5\%$ accuracy, and is insensitive to the choice of the spring-force law, as shown for the Marko-Siggia and the FENE force laws in Table~\ref{k_bd}. Further, values of $K_{\textrm{BD}}$, for models with and without HI, lie close to each other, indicating that HI does not affect the dissipated work due to dry internal friction.

The validity of the proposed protocol for the extraction of the dry internal friction coefficient has thus been established. In the next section, the application of the protocol to a model with wet internal friction is discussed, in the context of a coarse-grained polymer model with cohesive intra-chain interactions.  

\section{\label{sec:cg}Wet internal friction}
In this section, the case of wet internal friction is investigated in the context of the force-induced unraveling of a coiled globule which has previously been studied experimentally by~\citet{Murayama2007} and with simulations by~\citet{Alexander-Katz2009}. The problem is revisited here with the goal of understanding the role of solvent viscosity and HI, both of which have not been considered previously.
  
\subsection{\label{sec:sc_particulars}Model description}
A bead-spring model with $N_{\text{b}}$ beads connected by FENE springs, each stretchable up to a maximum length of $Q_0$, is considered. The excluded volume interactions between beads are modelled using the Soddemann-D\"{u}nweg-Kremer (SDK) potential~\cite{Soddemann2001}, whose functional form is given by
\begin{align}\label{eq:pot_sdk}
\begin{scriptsize}
\dfrac{U_{\text{SDK}}(r)}{k_BT}= \left\{
\begin{array}{ll}
      4\left[\left(\dfrac{\sigma_{\text{{s}}}}{r}\right)^{12}-\left(\dfrac{\sigma_{\text{s}}}{r}\right)^{6}+\dfrac{1}{4}\right]-\epsilon; & r\leq 2^{1/6}\sigma_{\text{s}}\\[15pt]
      \dfrac{\epsilon}{2}\left[\cos\left(m_1 \left(\dfrac{r}{\sigma_{\text{s}}}\right)^2+m_2\right)-1\right]; &2^{1/6}\sigma_{\text{s}} \leq r \leq r_{\text{c}} \\[15pt]
       0; & r \geq r_{\text{c}} \\
\end{array} 
\right. 
\end{scriptsize}
\end{align}

The minimum of the potential occurs at at $r=2^{1/6}\sigma_{\text{s}}$, and the value of the potential at this location is $U_{\text{SDK}}=-\epsilon \, k_BT$. As seen from Eq.~(\ref{eq:pot_sdk}), the repulsive part of the pair-potential is modeled after the Weeks-Chandler-Andersen (WCA) potential, while the attractive part is constructed using a cosine function which smoothly approaches zero at the cut-off distance, $r_{\text{c}}$. A detailed comparison of the SDK potential with the Lennard-Jones and the WCA potential has been performed recently~\cite{Santra2019}. The parameter values for $m_1$ and $m_2$ depend on the choice of $r_{\text{c}}$. They have been refined in ref.~\citenum{Santra2019} in comparison to the values in the paper by \citet{Soddemann2001} in order to be more applicable to BD simulations, as discussed below and in Sec.~\ref{sec:sc_par_space}. Setting $\epsilon=0$ in Eq.~(\ref{eq:pot_sdk}) leads to purely repulsive inter-bead interactions, and corresponds to the athermal limit of solvent quality. Increasing the value of $\epsilon$ beyond zero results in a decrease in the solvent quality. A special feature of the SDK potential~\cite{Santra2019} is that modifying the value of $\epsilon$ allows one to tune the attractive interactions selectively, without affecting the repulsive branch of the pair-potential. This is in stark contrast to the more commonly used Lennard-Jones (LJ) potential for which changing the well-depth affects both the attractive and the repulsive branches. Furthermore, the exact truncation of the SDK potential at the cutoff distance results in an increased computational efficiency~\cite{Soddemann2001} in comparison to the LJ potential whose long attractive tail approaches zero only asymptotically, at large inter-bead separations.

The force on the $\mu$th bead due to bonded and non-bonded excluded volume interactions is denoted by $\bm{F}^{\text{(c)}}_{\mu}$, and $\bm{F}^{\text{(E)}}_{\mu}$, respectively, and the notation $\bm{F}_{\mu}^{\phi}\equiv\bm{F}^{\text{(c)}}_{\mu}+\bm{F}^{\text{(E)}}_{\mu}$ is used to denote the total force on a bead due to the FENE and SDK potentials.

As discussed previously, the terminal beads of the chain are subjected to harmonic trap potentials. One of the traps is held fixed at $\boldsymbol{\chi}_1=(0,0,0)$, and the other is moved from $\bm{\chi}_2^{\text{(i)}}\equiv(\chi^{\textrm{(i)}}_{2x},0,0)$, to $\bm{\chi}_2^{\text{(f)}}\equiv(\chi^{\textrm{(f)}}_{2x},0,0)$ at a constant velocity, $\bm{v}\equiv\left(v_{x},0,0\right)$. 

The hydrodynamic interaction between any pair of beads $\mu$ and $\nu$ is accounted for by defining the diffusion tensor $\bm{\Upsilon}_{\mu\nu}=\delta_{\mu\nu}\boldsymbol{\delta}+\zeta\boldsymbol{\Omega}_{\mu\nu}$, where $\delta_{\mu\nu}$ is the Kroenecker delta, and the hydrodynamic interaction tensor, $\boldsymbol{\Omega}_{\mu\nu}$, is approximated using the Rotne-Prager-Yamakawa expression, as shown in Eqs.~(\ref{eq:def_hi}) \textemdash ~(\ref{eq:less_Q}). For notational convenience, we define $\bm{\mathcal{D}}$, a block matrix of size $N_{\text{b}}\times N_{\text{b}}$, whose each element is the $3\times3$ matrix, $\bm{\Upsilon}_{\mu\nu}$. Additionally, the block matrix $\bm{\mathcal{B}}$ is defined as $\bm{\mathcal{B}}\cdot\bm{\mathcal{B}}^{T}=\bm{\mathcal{D}}$. 

The stochastic differential equation describing the time-evolution of the position of the $\mu$th bead is given by 
\begin{widetext}
\begin{equation}\label{eq:sc_gov_eq}
\bm{{r}^{*}}_{\mu}\left(t^{*}+\Delta t^{*}\right) = \bm{{r}^{*}}_{\mu}\left(t^{*}\right) + \dfrac{1}{4}\sum_{\nu=1}^{N_{\text{b}}}\Biggl[\bm{\Upsilon^*}_{\mu\nu}\cdot\bm{{F}}_{\nu}^{\bm{*}(\phi)}-\boldsymbol{\Upsilon}^{\bm{*}}_{\!_{1\nu}}\cdot\left[c_1\left(\bm{r^{*}}_1-\bm{\chi}^{*}_1\right)\right] - \boldsymbol{\Upsilon}^{\bm{*}}_{\!_{N_{\text{b}}}\nu}\cdot\left[c_2\left(\bm{r^{*}}_{N_{\text{b}}}-\bm{\chi}^{*}_2\right)\right]\Biggr]\Delta t^{*}
+\dfrac{1}{\sqrt{2}}\sum_{\nu=1}^{N_{\text{b}}}\bm{\mathscr{B}^{*}}_{\mu\nu}\cdot \Delta \bm{\mathscr{W}}^{\bm{*}}_{\nu}
\end{equation}
\end{widetext}
where $\bm{\mathscr{B}^{*}}_{\mu\nu}$ is the $\left(\mu,\nu\right)-$th element of $\bm{\mathcal{B}^{*}}$, and $\Delta \bm{\mathscr{W}^{*}}$ is a dimensionless Wiener process of zero mean and variance $\Delta t^*$.

\begin{figure}[tb]
\begin{center}
\includegraphics[width=0.85\linewidth]{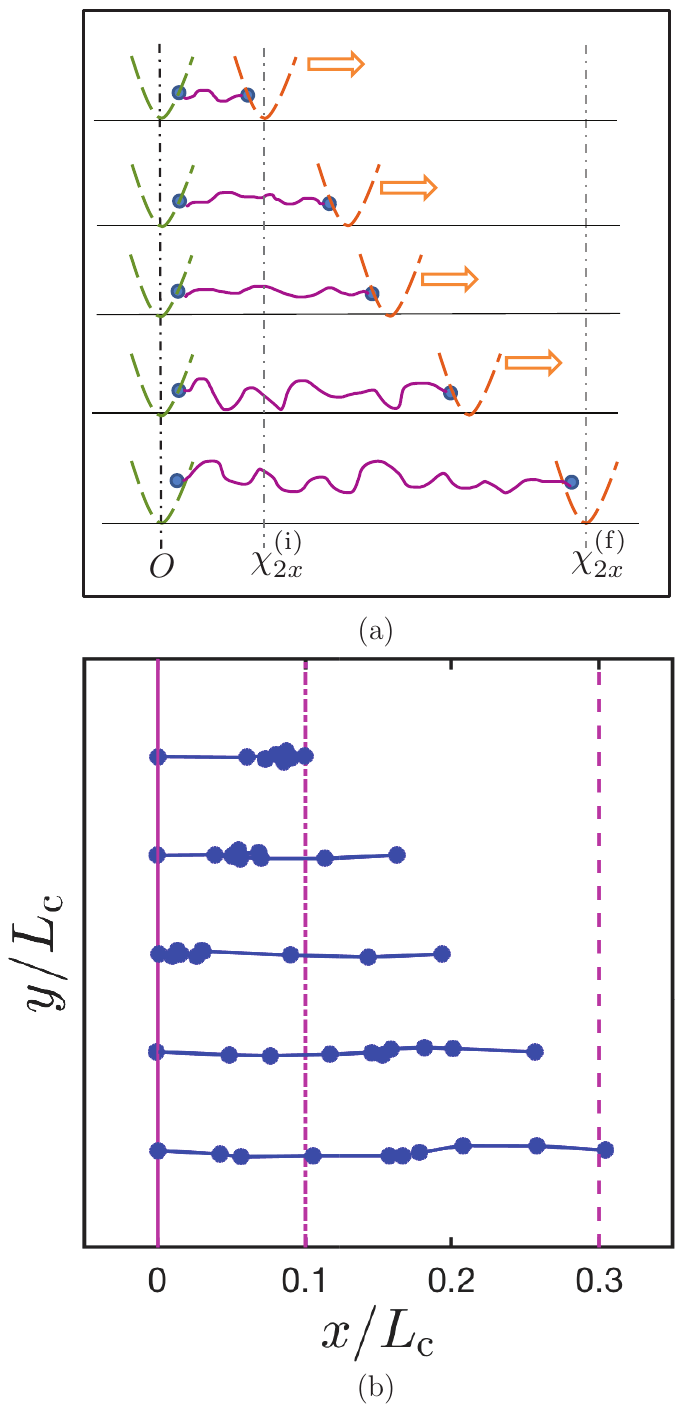}
\end{center}
\caption{\small (Color online) \textbf{Schematic and snapshot of single chain polymer model subjected to pulling} (a) Schematic representation of a polymer chain subjected to pulling, (b) snapshots from the BD simulations of a chain with $N_{\text{b}}=10$ and $\epsilon=3.45$, pulled with a constant velocity of $v^*=0.01$. From left to right, the solid vertical line, the dash-dotted vertical line, and the dashed vertical line represent the positions of the first trap $(\bm{\chi}^{*}_1)$, the initial position of the second trap $(\bm{\chi}^{\text{(i)}*}_2)$, and the final position of the second trap $(\bm{\chi}^{\text{(f)}*}_2)$, respectively.}
\label{fig:sc_pull_rep}
\end{figure}

In Fig.~\ref{fig:sc_pull_rep}~(a), a schematic representation of the pulling is shown, and in Fig.~\ref{fig:sc_pull_rep}~(b), snapshots from BD simulations on a ten-bead chain is presented.

\subsection{\label{sec:sc_sim_details}Simulation details}

The stochastic differential equation governing the pulling of a single polymer chain [Eq.~(\ref{eq:sc_gov_eq})] is solved numerically using Brownian dynamics simulations. The initial bead positions are picked from the equilibrium distribution function corresponding to the FENE force law. The chain is then equilibrated at the initial state, with $\chi^{*}_{2x}=\chi^{*\textrm{(i)}}_{2x}$, for fifty Rouse times. Equilibration is ascertained by checking that the mean-squared value of the dimensionless radius of gyration, $\left<R^{2*}_{\text{g}}\right>$, has reached a steady value with respect to time. Pulling is then commenced at $t^*=0$, by changing the position of the mobile trap linearly in time, as $\chi^{*}_{2x}=\chi^{*\textrm{(i)}}_{2x}+v^{*}_xt^{*}$, till $t^*=\tau^*$. 
The work done in one realization of the pulling event is given by Eq.~(\ref{eq:numwork}), with $r^{*}_{2x}$ replaced with $r^{*}_{N_{\text{b}}x}$, where $r^{*}_{N_{\text{b}}x}$ refers to the $x$-coordinate of the last bead in the chain, and the remaining symbols retain their original meaning as defined in Sec.~\ref{sec:sim_details_db}. A timestep width of $\Delta t^*=1.0\times 10^{-4}$ is used after ascertaining, for a ten-bead chain with representative parameter values, that the average work obtained for the $\Delta t^*=1.0\times 10^{-4}$ and $\Delta t^*=1.0\times 10^{-5}$ cases agree within error bars. The calculated work is then used to estimate the free-energy difference and the average dissipation as shown in Eq.~(\ref{eq:je_use}).

\subsection{\label{sec:sc_par_space}Parameter space specification}

A value of  $b=50$ is used throughout this section for the dimensionless FENE parameter. For a free chain (without confining potentials acting on the terminal beads), with this particular value of $b$, the choice of $r_c=1.82\sigma$, $m_1=1.5306333121$, and $m_2=1.213115524$ have been shown to lead to the correct scaling predictions in poor, $\theta$, and good solvent conditions for the radius of gyration with the number of beads in the chain, as described in detail in ref.~\citenum{Santra2019}, and hence have been used for all simulations on single chains discussed in this section. The $\theta$-temperature for this system is observed to be at $\epsilon\approx0.45$, where $\left<R^2_{\text{g}}\right>\sim\left(N_{\text{b}}-1\right)$ and the second virial coefficient, $B_2=0$~\cite{Santra2019}. This value of cohesive strength is denoted as $\epsilon_{\theta}$. Poor solvent scaling, namely,  $\left<R^2_{\text{g}}\right>\sim\left(N_{\text{b}}-1\right)^{0.67}$, is observed for $\epsilon\geq0.55$. The bead radius, $a$, is defined on the basis of $\sigma_{\text{s}}$, as $a=0.5\sigma_{\text{s}}$. We set $\sigma_{\text{s}}=l_{\text{H}}$ in all our simulations. 

A trap stiffness of $c_1=c_2=1000$ is used for both the stationary and the mobile traps. The initial and final positions of the mobile traps are chosen as $\chi^{\textrm{(i)}}_{2x}=0.1\,L_{\text{c}}$ and $\chi^{\textrm{(f)}}_{2x}=0.3\,L_{\text{c}}$, respectively, where $L_{\text{c}}\equiv\left(N_{\text{b}}-1\right)Q_0$ is the contour length of the chain. A chain size of $N_{\text{b}}=10$ has been used for all the simulation results reported in this section.

\subsection{\label{sec:sc_res}Globule unraveling is wet}

\begin{figure}[tb]
\centering
\includegraphics[width=0.95\linewidth]{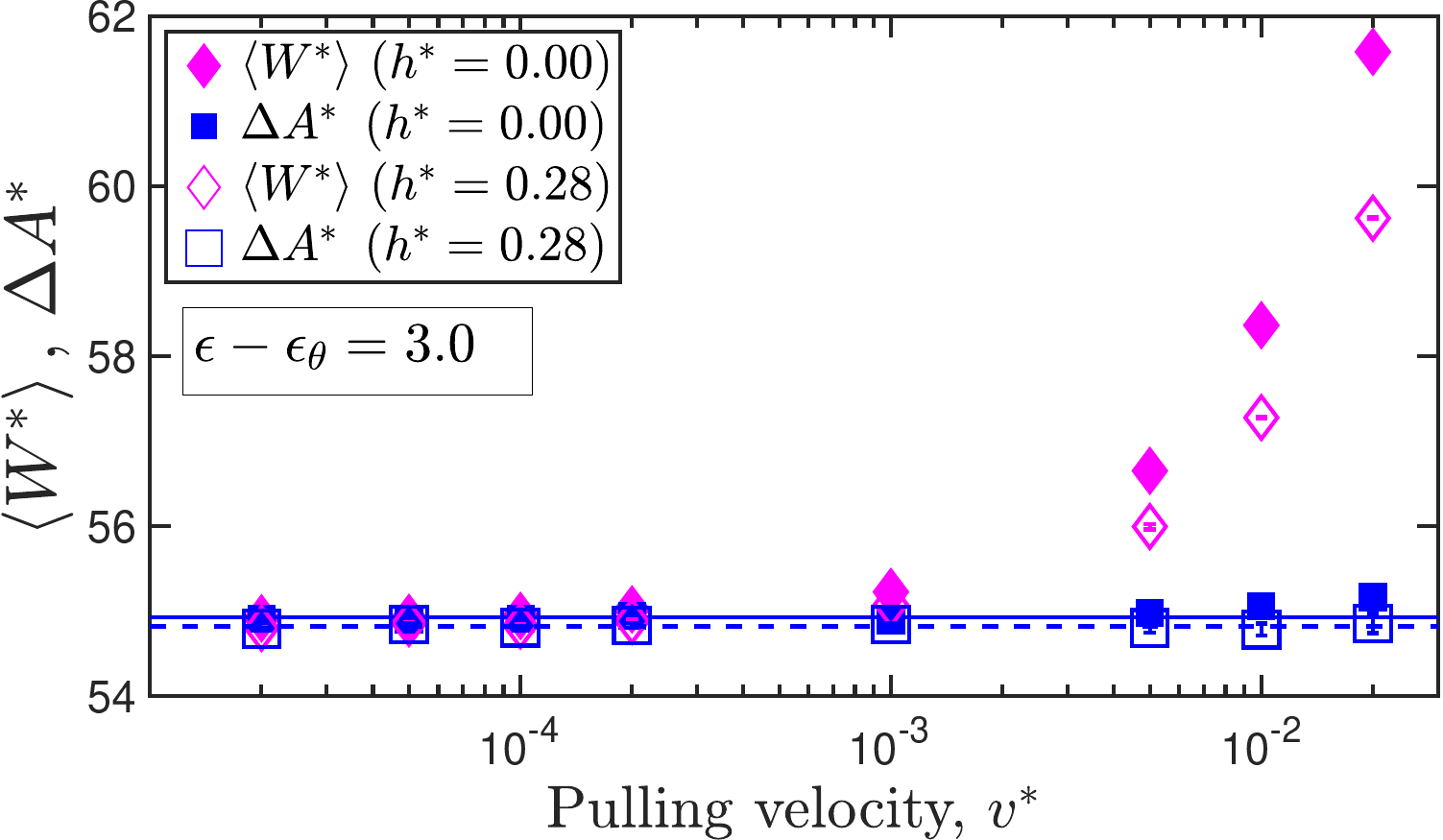}  
\caption{\small (Color online) \textbf{Equivalence between the free energy difference estimated using the Jarzynski equality and the classical definition}. Total work done as a function of the pulling velocity, for a well-depth of $\epsilon=3.45$ and a chain size of ten beads. The open and filled square symbols indicate the Jarzynski estimate of the free-energy difference evaluated for models with and without the inclusion of fluctuating hydrodynamic interactions, respectively.}
\label{fig:netz_jz_compare}
\end{figure}

Netz and coworkers~\cite{Alexander-Katz2009} have measured the internal friction associated with collapsed homopolymers by measuring the work dissipated in the force-induced unfolding of a single polymer chain. As mentioned previously, they estimate the free energy difference, denoted by $W_{\text{eq}}$, as the work done in the quasi-static pulling limit, that is, $W_{\text{eq}}\left(\epsilon\right)\equiv W\left(\epsilon,v\to0\right)$. The dissipated work at any finite pulling velocity is then calculated as the difference between the average work done at that velocity, and the reversible work. This is exactly the definition of dissipation that we have adopted throughout this paper, as indicated in Eq.~(\ref{eq:je_use}). However, rather than using the work done in the quasi-static limit to estimate the free-energy difference, we use the Jarzynski equality to evaluate the same quantity.

In Fig.~\ref{fig:netz_jz_compare}, the total dimensionless work done in pulling a ten-bead chain with a representative set of parameters is plotted as a function of the dimensionless pulling velocity, for cases with and without hydrodynamic interactions. Horizontal lines represent the error-weighted mean of the total work at the lowest four values of the pulling velocity, and is therefore a measure of the free-energy difference in the classical sense. Solid lines correspond to the freely-draining case, while dashed lines indicate the case with hydrodynamic interactions. It is seen that the free-energy differences for cases with and without HI concur within $~10^{-1}\,k_BT$, in agreement with the expectation that the free-energy difference, a static equilibrium property, remains unaffected by hydrodynamic fluctuations. Furthermore, it is also seen from the figure that the work done in the quasi-static limit agrees, within error bars, with the free-energy difference estimated using the Jarzynski equality (square symbols), thus establishing the validity of our approach for the estimation of the free-energy difference and the dissipation.

\begin{figure}[tb]
\centering
\includegraphics[width=0.99\linewidth]{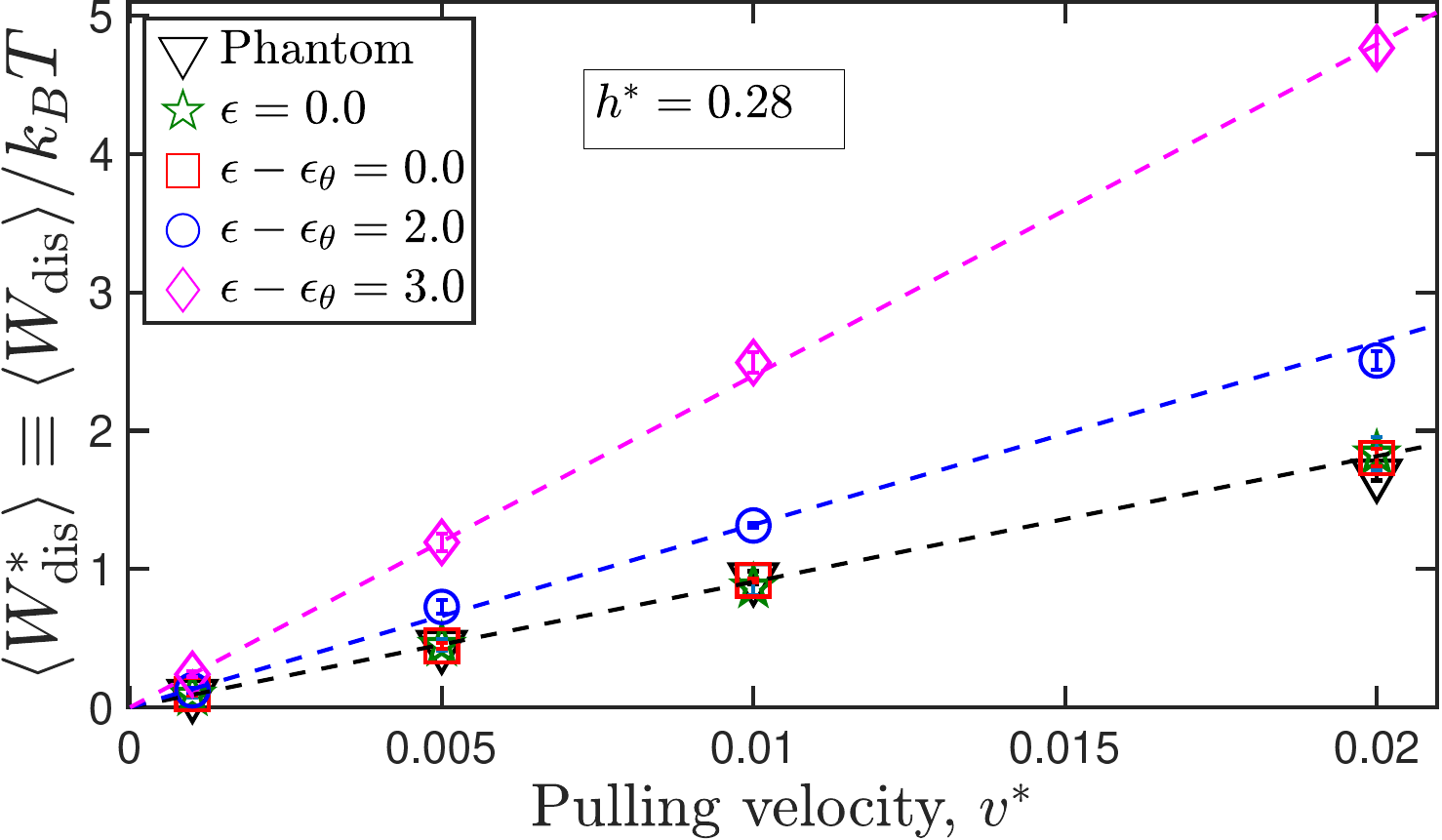}  
\caption{\small (Color online) \textbf{Regime of linear dependence between average dissipation and pulling velocity determined by strength of cohesive interactions. } Average dimensionless dissipation, as a function of the dimensionless pulling velocity. The dashed lines are linear fits to the data.}
\label{fig:sc_wdis_nb_vel}
\end{figure}

\begin{figure}[ptbh]
\begin{center}
\begin{tabular}{c}
\includegraphics[width=0.95\linewidth]{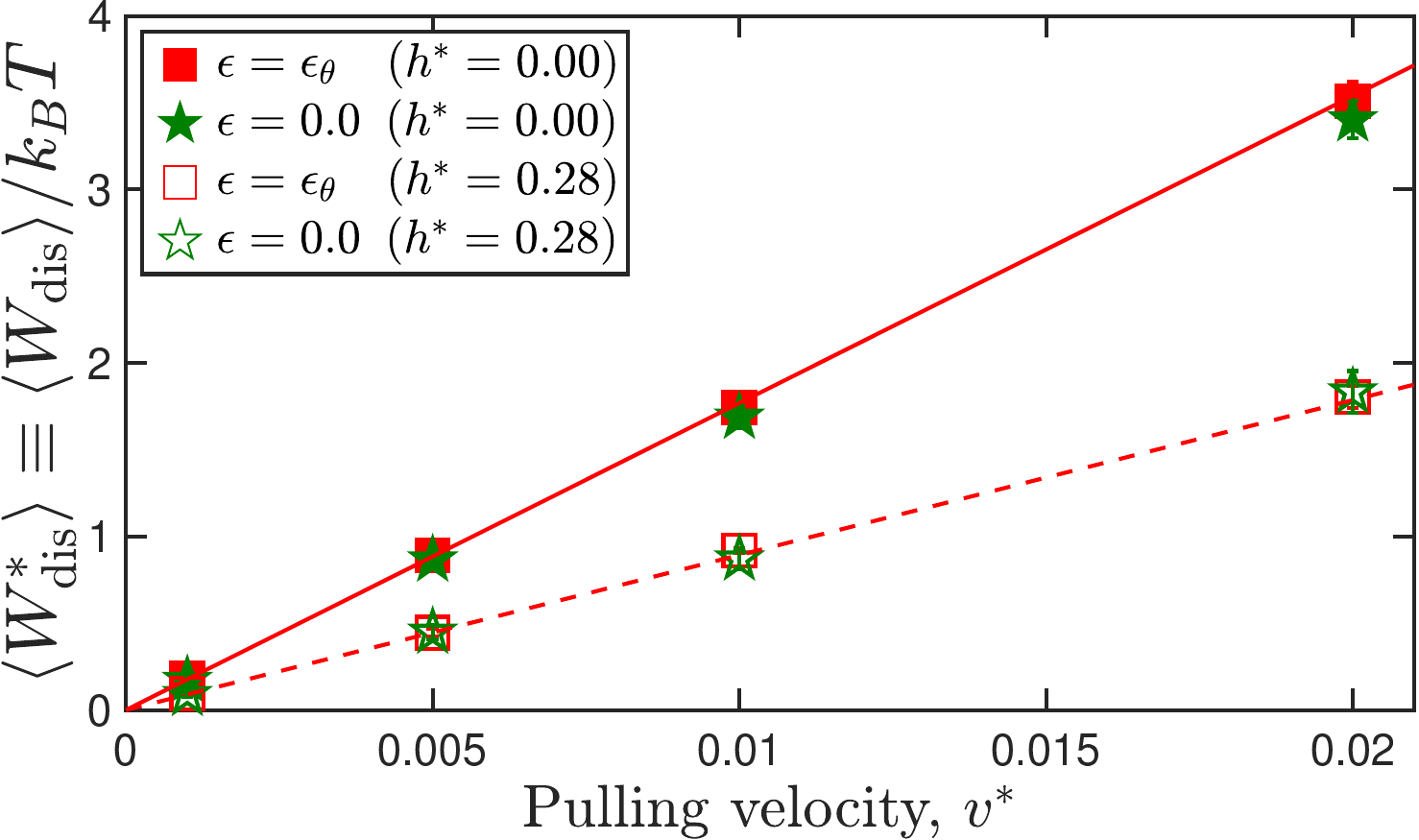}\\
(a) \\
\includegraphics[width=0.95\linewidth]{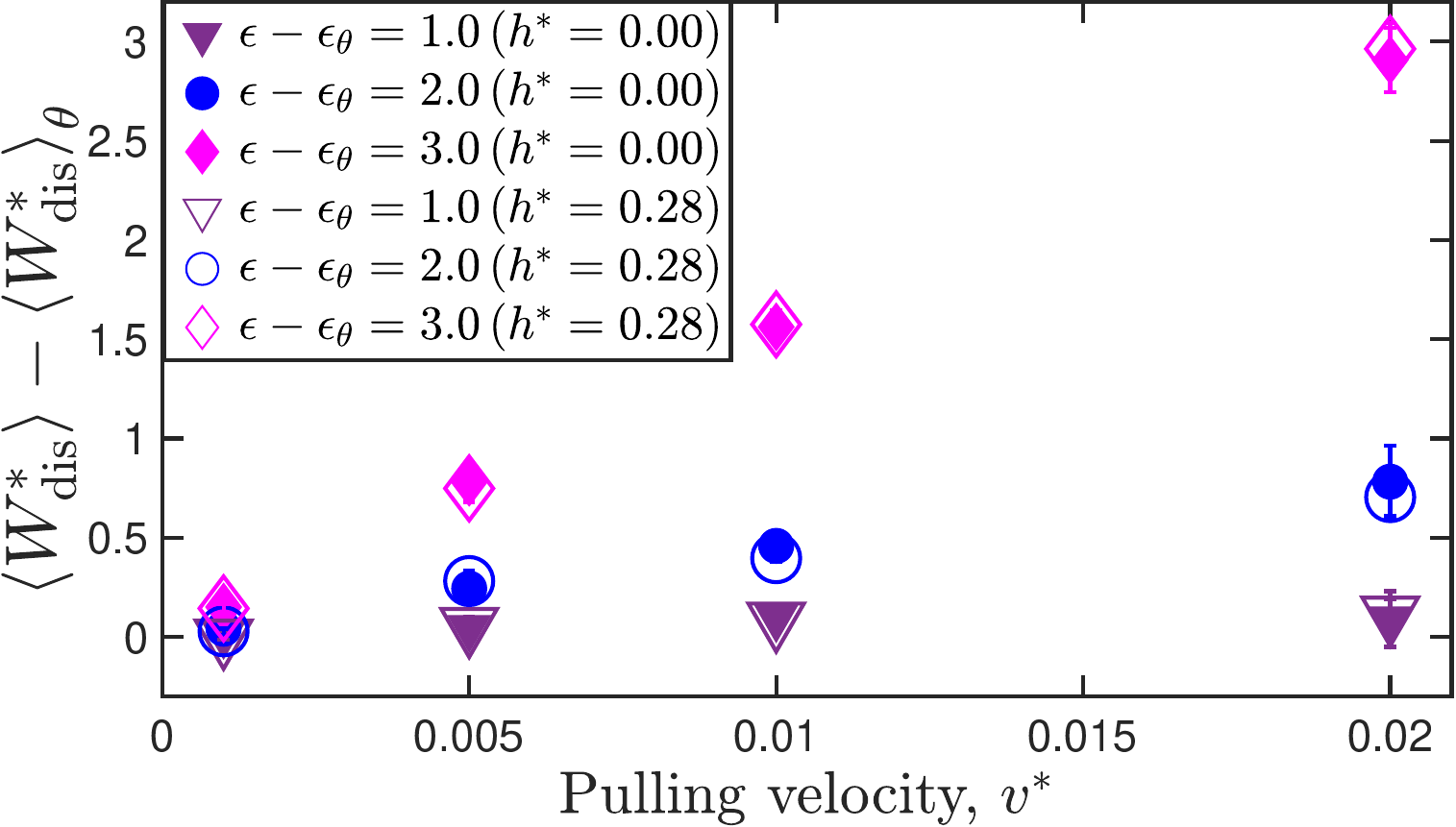}\\
(b)  \\
\end{tabular}
\end{center}
\caption{\small (Color online) \textbf{Effect of hydrodynamic interactions on the dissipation.} (a) the average dissipation, and (b) the enhancement in the average dissipation (with respect to the dissipation due to solvent) due to cohesive interactions, plotted as a function of the dimensionless pulling velocity. Symbols representing the enhancement values for cases with HI have been enlarged for clarity.}
\label{fig:hi_eff_dis}
\end{figure}

\begin{figure}[ptbh]
\centering
\includegraphics[width=0.8\linewidth]{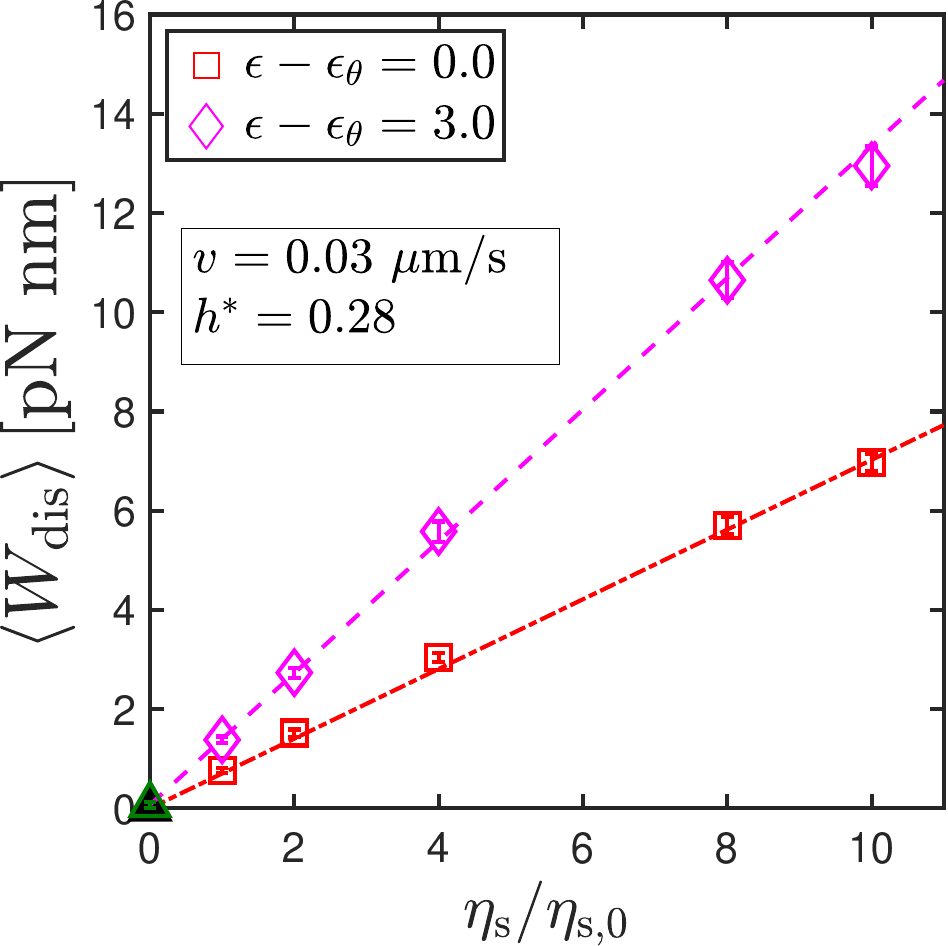}  
\caption{\small (Color online) \textbf{Protocol establishes the presence of wet friction.} Average dissipation as a function of the solvent viscosity for two different well-depths. The extrapolated value in the limit of zero solvent viscosity is indicated by an upright triangle, and the extrapolated values for both the cases are found to coincide within error bars of the simulation.}
\label{fig:sc_wdis_eta}
\end{figure}

\begin{figure}[ptbh]
\begin{center}
\begin{tabular}{c}
\includegraphics[width=0.95\linewidth]{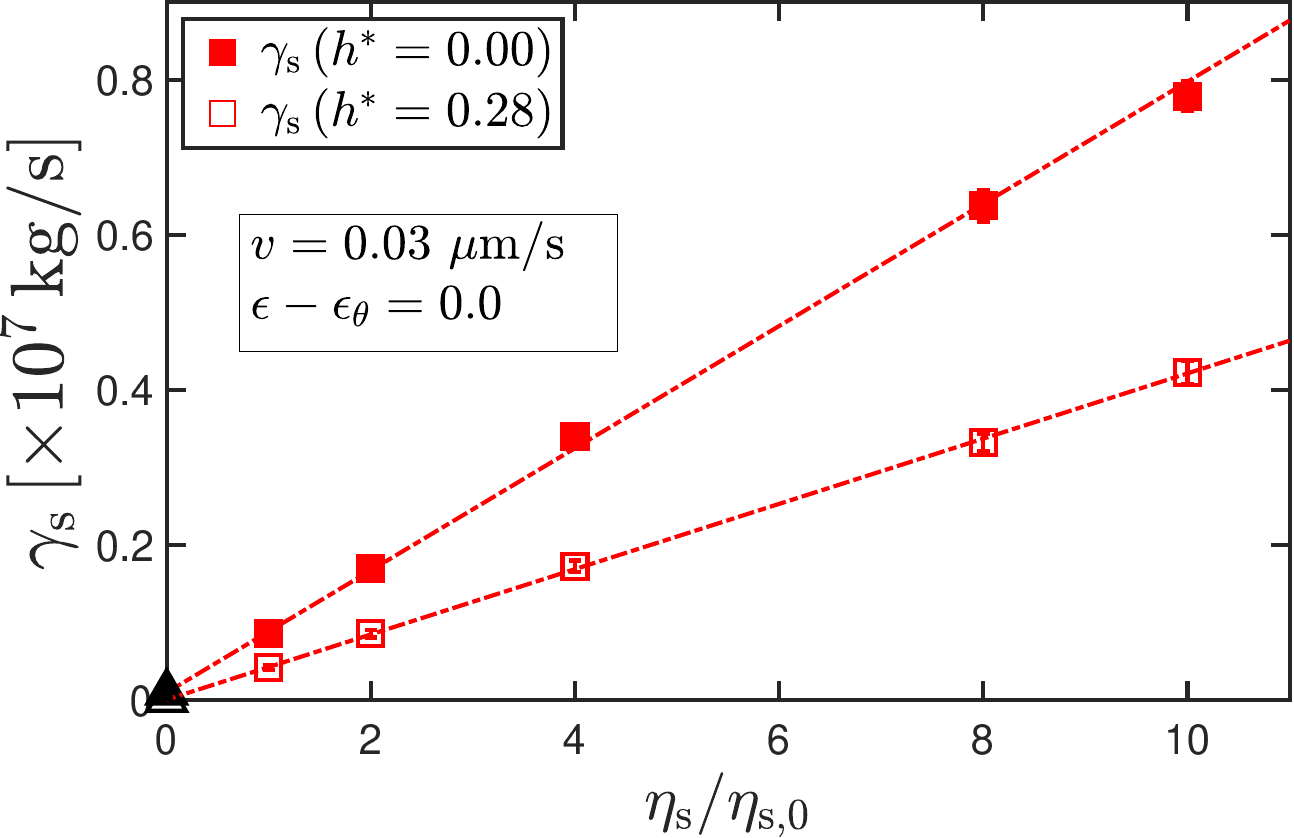}\\
(a) \\
\includegraphics[width=0.95\linewidth]{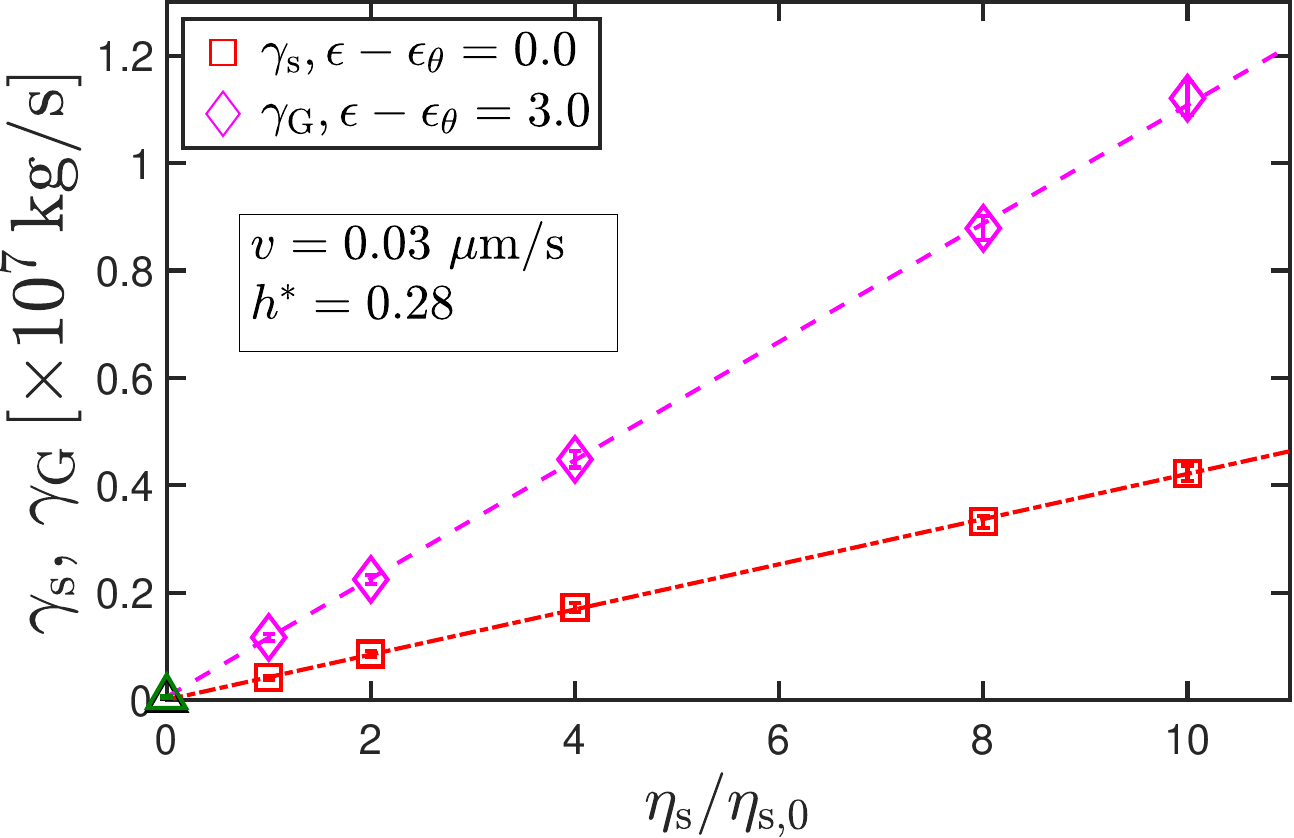}\\
(b)  \\
\end{tabular}
\end{center}
\caption{\small (Color online) \textbf{Effect of hydrodynamic interactions on the friction coefficient.} (a) Solvent friction, with and without hydrodynamic interactions and (b) a comparison between the solvent and globule friction coefficient, with hydrodynamic interactions, plotted as a function of solvent viscosity.}
\label{fig:hi_fric}
\end{figure}

\begin{figure}[ptbh]
\centering
\includegraphics[width=0.99\linewidth]{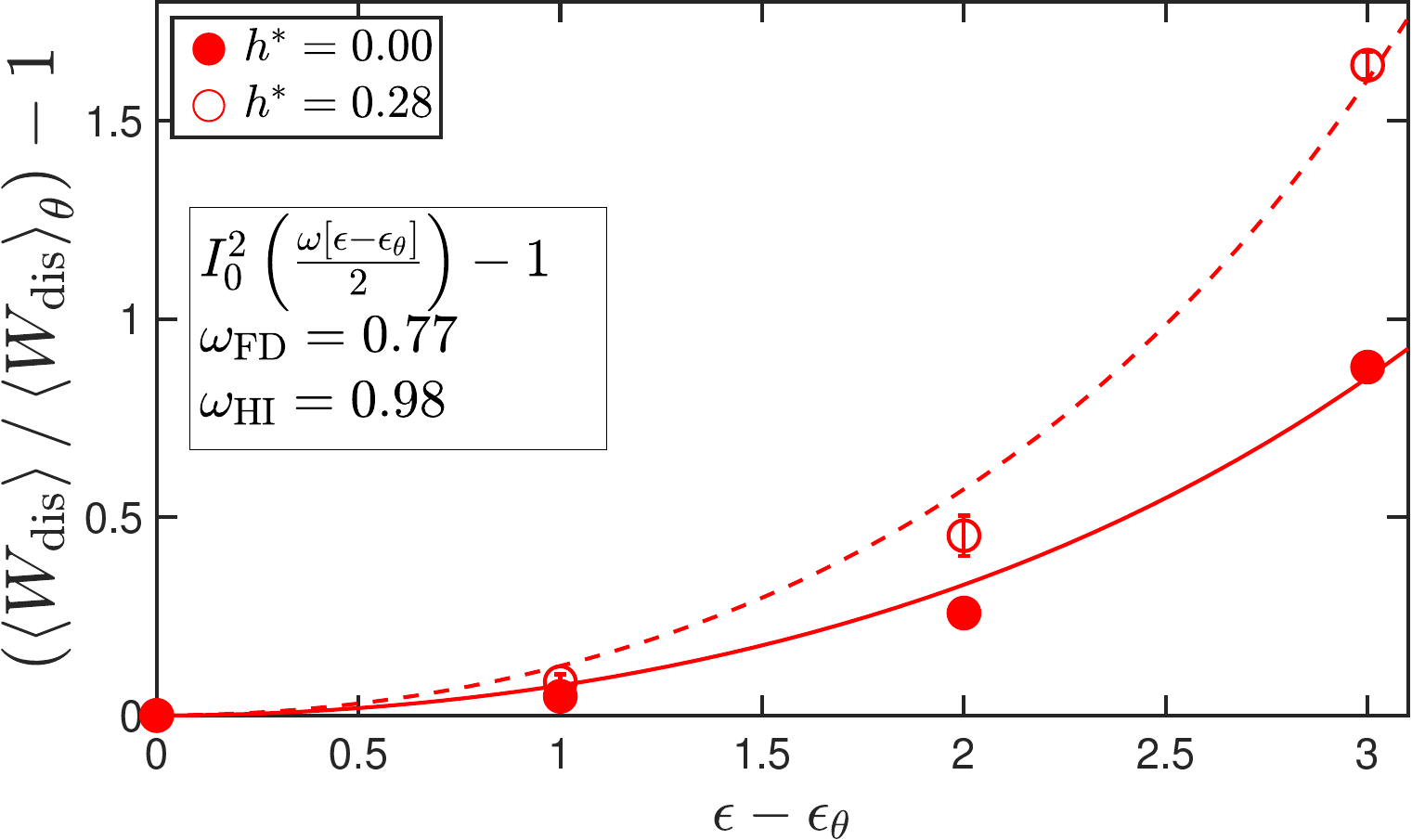}  
\caption{\small (Color online) \textbf{Magnification of internal friction due to cohesive interactions.} Rescaled excess contribution to dissipation as a function of the effective well-depth. Solid and dashed lines are used to fit the data points obtained for cases without and with HI, respectively, and correspond to Eq.~(\ref{eq:zw_fit}) with fitting parameters $\omega_{\text{FD}}=0.77$ and $\omega_{\text{HI}}=0.98$.}
\label{fig:sc_slope_rat}
\end{figure}

In Fig.~\ref{fig:sc_wdis_nb_vel}, the dimensionless average dissipation is plotted as a function of the dimensionless pulling velocity. It is seen that the data for the phantom chain (no excluded volume interactions), athermal chain (purely repulsive interactions) and the chain under $\theta-$conditions superimpose. This indicates that for uncollapsed chains, the work dissipated during pulling is expended entirely against the solvent friction, and there is no internal friction associated with these cases. Furthermore, increasing the well-depth beyond the $\theta-$point results in an increased dissipation at a fixed value of the pulling velocity, which is due to the additional work needed to unravel the globule as the polymer chain is stretched. The average dissipation scales linearly with the pulling velocity over the entire range of the latter quantity.

In the linear regime, for a collapsed globule ($\epsilon>\epsilon_{\theta}$), the following relationship for the dissipation can be written,
\begin{align}\label{eq:lin_def}
\left<W_{\text{dis}}\right>&=\gamma_{\text{G}} v d
\end{align}
where $\gamma_{\text{G}}$ is the globule friction coefficient. The same expression is valid in the uncollapsed case, with $\gamma_{\text{G}}$ replaced with $\gamma_{\text{s}}$, the solvent friction coefficient. As $\epsilon\to\epsilon_{\theta}$, the dissipation is entirely due to the solvent, and $\gamma_{\text{G}}\to\gamma_{\text{s}}$. 

Fig.~\ref{fig:hi_eff_dis} elucidates the effect of hydrodynamic interactions on the dissipation more clearly. As shown in Fig.~\ref{fig:hi_eff_dis}~(a), for chains under both $\theta$ and athermal conditions, the dissipation due to the solvent decreases identically upon the inclusion of fluctuating hydrodynamic interactions. In Fig.~\ref{fig:hi_eff_dis}~(b), it is seen that the enhancement in the dissipation due to cohesive interactions, measured as the difference between the total dissipation and the dissipation due to solvent alone, remains practically unaffected by hydrodynamic interactions. 

In Fig.~\ref{fig:sc_wdis_eta}, the average dissipated work is plotted for two values of the well-depth, for one value of the dimensional pulling velocity, as a function of the solvent viscosity. At any finite value of the solvent viscosity, the work dissipated for the collapsed globule case is greater than that for that in $\theta-$condition. However, in the extrapolated limit $\eta_{\text{s}} \to 0$, the dissipated work goes to zero. Clearly, this suggest that the additional dissipation due to cohesive interactions between the beads corresponds to the case of wet internal friction, which is not clear a priori, and can only be established following the protocol proposed here. 

In Figs.~\ref{fig:hi_fric}, the friction coefficients calculated using Eq.~(\ref{eq:lin_def}) have been plotted as a function of the solvent viscosity. From Fig.~\ref{fig:hi_fric}~(a), it is seen that friction coefficient due to the solvent scales linearly with the solvent viscosity, and that hydrodynamic interactions reduce the friction coefficient in comparison to the freely draining case. From Fig.~\ref{fig:hi_fric}~(b), it is seen that the inclusion of cohesive interactions results in an enhancement in the friction coefficient at all finite values of the solvent viscosity. In the extrapolated limit of zero solvent viscosity, however, the friction coefficient is also seen to tend to zero, as is typical of wet friction.

By taking a ratio of the slopes of the dissipation versus pulling velocity for the collapsed and the uncollpased states, for identical pulling distances in the linear regime, one gets
\begin{align}\label{eq:excess_dis}
\dfrac{\left<W_{\text{dis}}\right>}{\left<W_{\text{dis}}\right>_{\theta}}-1=\dfrac{\gamma_{\text{G}}-\gamma_{\text{s}}}{\gamma_{\text{s}}}
\end{align}
This quantity represents the rescaled excess contribution to the dissipation due to internal friction, and enables an investigation of the relationship between $\gamma_{\text{G}}$ and $\gamma_{\text{s}}$, and comparison with Zwanzig's prediction~\cite{Zwanzig1988}, as was done previously by~\citet{Alexander-Katz2009} in the absence of hydrodynamic interactions. For a Brownian particle moving in a corrugated one-dimensional potential of the form $U(x)=\left(\omega/2\right)\left(\epsilon-\epsilon_{\theta}\right)\sin\left(\pi x/a\right)$, the effective friction and the solvent friction are related by~\cite{Alexander-Katz2009,Zwanzig1988}
\begin{align}\label{eq:zw_fit}
\dfrac{\gamma_{\text{G}}-\gamma_{\text{s}}}{\gamma_{\text{s}}}=I^{2}_{0}\left(\dfrac{\omega\left(\epsilon-\epsilon_{\theta}\right)}{2}\right)-1
\end{align}
where $I_0(..)$ is the modified Bessel function of zeroth order, and $\omega$ is a fitting parameter.

In Fig.~\ref{fig:sc_slope_rat}, the rescaled excess dissipation due to internal friction, for models with and without fluctuating hydrodynamic interactions, is plotted as a function of the well-depth relative to the $\theta-$condition. It is seen from the figure that a good qualitative agreement is observed between the simulation results and Eq.~(\ref{eq:zw_fit}), but with a fitting parameter that depends on whether hydrodynamic interactions are incorporated in the simulations. This agreement between simulations and theory, also observed by ~\citet{Alexander-Katz2009}, suggests that Zwanzig's formulation, albeit based on a one-dimensional energy landscape, satisfactorily captures the scaling of internal friction with the strength of cohesive interactions in force-spectroscopy simulations on single molecules. The subject of diffusion on rugged energy landscapes of dimension higher than one has been treated rigorously in refs.~\citenum{Seki2015} and~\citenum{Seki2016}.

\section{\label{sec:concl}Conclusions}

In summary, we have introduced a simple and novel protocol based on the Jarzynski equality for determining both dry and wet internal friction coefficients of macromolecules that can be implemented experimentally using optical tweezers. Using Brownian dynamics simulations on a spring-dashpot model for a polymer, we establish proof-of-principle by recovering the dry internal friction coefficient which is used as a model input, and show that a bead-spring chain with cohesive interactions is an example of wet friction. It is conceivable that some real polymer chains might possess both wet and dry internal friction, and modeling such molecules would require the use of multi-bead-spring-dashpots with cohesive interactions. We envisage that the scheme proposed here may be applicable to a variety of macromolecules, and would enable a succinct characterization of the dissipative properties of the molecule.

\section*{\label{sec:acknwl}Acknowledgements}
We thank Burkhard D\"unweg, Subhashish Chaki, Ranjith Padinhateeri and Gareth McKinley for enlightening discussions. The work was supported by the MonARCH and SpaceTime-2 computational facilities of Monash University and IIT Bombay, respectively. CPU time grants from the National Computational Infrastructure (NCI) facility hosted by Australian National University is also acknowledged. R.C. acknowledges IRCC-IIT Bombay for funding (Project No. RD/0518-IRCCAW0-001). We also acknowledge the funding and general support received from the IITB-Monash Research Academy.

\appendix

\section{\label{sec:oned_gov_deriv} Governing equation for the one-dimensional spring-dashpot}

The Fokker-Planck equation for the configurational distribution function $\Psi\left(r_2,t\right)$ can be derived by following the procedure commonly used in polymer kinetic theory, i.e., by combining a force balance on the beads with an equation of continuity in probability space~\cite{Bird1987b}. The force balance essentially states that the (i) the internal friction force due to the dashpot, (ii) the restoring force due to the finitely extensible spring, (iii) external forces (like the force due to the optical traps, in the present case),  (iv) the random Brownian force due to bombardment by solvent molecules, and (v) the hydrodynamic force responsible for the solvent-mediated propagation of momentum on each bead, must sum up to zero. 

The force balance over the free bead may be written as
\begin{align}
0=-\zeta\llbracket\dot{r}_2\rrbracket-\dfrac{\partial \mathcal{H}}{\partial r_2}-K\llbracket\dot{r}_2\rrbracket-k_BT\left(\dfrac{\partial \ln\,\Psi}{\partial r_2}\right)
\end{align}
Upon simplification, 
\begin{align}
\zeta\llbracket\dot{r}_2\rrbracket&=-\left(c_2+1\right)Hr_2+c_2H\chi(t)\nonumber\\[5pt]
&-K\llbracket\dot{r}_2\rrbracket-k_BT\left(\dfrac{\partial \ln\,\Psi}{\partial r_2}\right)
\end{align}
Grouping together the terms containing $\llbracket\dot{r}_2\rrbracket$, and defining $\theta=\left[1+\left(K/\zeta\right)\right]$, where $\varphi\coloneqq 2K/\zeta$ is the internal friction parameter,
\begin{equation}\label{eq:force_balance}
\llbracket\dot{r}_2\rrbracket = -\dfrac{\left(c_2+1\right)H}{\zeta\theta}r_2+\dfrac{c_2H}{\zeta\theta}\chi(t)
 -\dfrac{k_BT}{\zeta\theta}\left(\dfrac{\partial \ln\,\Psi}{\partial r_2}\right)
\end{equation}
The equation of continuity for the probability density, $\Psi\left(r_2,t\right)$, is written as
\begin{equation}\label{eq:continuity}
\dfrac{\partial \Psi}{\partial t}=-\dfrac{\partial}{\partial r_2}\Biggl(\llbracket\dot{r}_2\rrbracket\Psi\Biggr)
\end{equation}
Substituting the expression for $\llbracket\dot{r}_2\rrbracket$ from Eq.~(\ref{eq:force_balance}) into Eq.~(\ref{eq:continuity}), one obtains
\begin{align}
\dfrac{\partial \Psi}{\partial t}&=-\dfrac{\partial}{\partial r_2}\left\{-\dfrac{\left(c_2+1\right)H}{\zeta\theta}r_2\Psi +\dfrac{c_2H}{\zeta\theta}\chi(t)\Psi\right\}\nonumber\\[5pt]
&+\dfrac{k_BT}{\zeta\theta}\dfrac{\partial}{\partial r_2}\left[\left(\dfrac{\partial \ln\,\Psi}{\partial r_2}\right)\Psi\right],
\end{align}
which can be rewritten as
\begin{align}\label{eq:fp}
\dfrac{\partial \Psi}{\partial t}&=-\dfrac{\partial}{\partial r_2}\left\{\left[\dfrac{-\left(c_2+1\right)H}{\zeta\theta}r_2+\dfrac{c_2H}{\zeta\theta}\chi(t)\right]\Psi\right\}\\[5pt]
&+\dfrac{1}{2}\left(\dfrac{2k_BT}{\zeta\theta}\right)\dfrac{\partial^2 \Psi}{\partial r^2_2}\nonumber
\end{align}
The stochastic differential equation corresponding to Eq.~(\ref{eq:fp}) is given by~\cite{Ottinger1996}
\begin{align}\label{eq:sde_deriv}
dr_2=\left[\dfrac{-\left(c_2+1\right)H}{\zeta\theta}r_2+\dfrac{c_2H}{\zeta\theta}\chi(t)\right]dt+\sqrt{\dfrac{2k_BT}{\zeta\theta}}dW_t
\end{align}
where $W_t$ represents a Wiener process, and has dimensions of $[\text{time}]^{1/2}$. Eq.~(\ref{eq:sde_deriv}) can be recast in the Langevin form as
\begin{align}\label{eq:langevin_deriv}
\dfrac{dr_2}{dt}=-\dfrac{\left(c_2+1\right)H}{\zeta\theta}r_2+\dfrac{c_2H}{\zeta\theta}\chi(t)+\sqrt{\dfrac{2k_BT}{\zeta\theta}}f_{\text{B}}(t)
\end{align}
where $\left<f_{\text{B}}(t)\right>=0$ and $\left<f_{\text{B}}(t)f_{\text{B}}(t')\right>=\delta(t-t')$, and $f_{\text{B}}(t)$ has dimensions of $[\text{time}]^{-1/2}$.
Setting $f^{*}_{\text{B}}(t^{*})=f_{\text{B}}(t)\sqrt{\lambda_{\text{H}}}$, where $\left<f^{*}_{\text{B}}(t^{*})\right>=0$ and $\left<f^{*}_{\text{B}}(t^{*})f^{*}_{\text{B}}(t_1^{*})\right>=\delta(t^{*}-t_1^{*})$,
the variables in Eq.~(\ref{eq:langevin_deriv}) are cast into their dimensionless form as,
\begin{align}\label{eq:intermed_gov}
\left(\dfrac{l_{\text{H}}}{\lambda_{\text{H}}}\right)\dfrac{dr^{*}_2}{dt^{*}}&=-\dfrac{\left(c_2+1\right)H}{\zeta\theta}r^{*}_2\,l_{\text{H}}+\dfrac{c_2H}{\zeta\theta}\chi^{*}(t^{*})l_{\text{H}}\nonumber\\[5pt]
&+\sqrt{\dfrac{2k_BT}{\zeta\theta}}\left(\dfrac{f^{*}_{\text{B}}(t^{*})}{\sqrt{\lambda_{\text{H}}}}\right)
\end{align}
Multiplying Eq.~(\ref{eq:intermed_gov}) throughout by $\left(\lambda_{\text{H}}/l_{\text{H}}\right)$, and simplifying, the dimensionless governing equation is obtained as 
\begin{align}\label{eq:dimless_lang}
\dfrac{dr^{*}_2}{dt^{*}}=-\dfrac{\left(c_2+1\right)H}{4\theta}r^{*}_2+\dfrac{c_2\chi^{*}(t^{*})}{4\theta}+\dfrac{1}{\theta}\sqrt{\dfrac{\theta}{2}}f^{*}_{\text{B}}(t^{*})
\end{align}
Eq.~(\ref{eq:dimless_lang}) is rewritten as Eq.~(\ref{eq:gov_eq}), after further simplification.

\begin{widetext}
\section{\label{sec:add_dis_steps} Work distribution for freely draining one-dimensional Hookean spring-dashpot}
The intermediate steps in the derivation of the mean and variance of the probability of work distribution is provided in this section. 

Firstly, the complete expression for the work done during one realization of pulling is obtained by substituting the expression for $r^{*}_2$ from Eq.~(\ref{eq:traj_def}) into Eq.~(\ref{eq:dimless_work}), as   
\begin{equation}\label{eq:work_def}
W^{*}=\dfrac{c_2}{2}\left[\chi^{2*}\left(\tau^{*}\right)-\chi^{2*}\left(0\right)\right]-c_2\int_{0}^{\tau^{*}}dt^{*}\dot{\chi}^{*}(t^{*})\Biggl(r^{*}_2(0)G(t^{*})+\dfrac{c_2}{4\theta}\int_{0}^{t^{*}}dt_1^{*}G\left(t^{*}-t_1^{*}\right)\chi^{*}(t_1^{*})
+\dfrac{1}{\theta}\int_{0}^{t^{*}}dt_1^{*}G\left(t^{*}-t_1^{*}\right)\xi(t_1^{*})\Biggr)
\end{equation}
where the noise term, $\xi(t^{*})$, obeys $\left<\xi(t^{*})\right>=0$ and
\begin{align}\label{eq:noise_strength}
\left<\xi(t^{*})\xi(t_1^{*})\right>=\dfrac{\theta}{2}\delta(t^{*}-t^{*}_1)
\end{align}

The dimensionless partition function of the system can be derived to be,
\begin{align}
\label{eq:dimless_part_func}
Z^{*}&\equiv\int^{+\infty}_{-\infty}\exp\left[-\dfrac{\mathcal{H}}{k_BT}\right]dr^{*}_2=\sqrt{\dfrac{2\pi}{\left(c_2+1\right)}}\exp\left(-\dfrac{c_2\chi^{*2}}{2\left(c_2+1\right)}\right)
\end{align}
and the dimensionless free-energy, $A^*=-\ln Z^*$, is then simply
\begin{align}\label{eq:free_energy_def_suppl}
A^*\left(\chi^*\right)&=\left[\dfrac{c_2}{2(c_2+1)}\right]\chi^{2*},
\end{align}
after ignoring constant prefactors. This equation is reproduced as Eq.~(\ref{eq:free_energy_def}) in Sec.~\ref{sec:dry_sd}. 
The probability distribution function for the position of the bead is given by
\begin{equation*}\label{eq:prob_dist}
\Psi^*\left(r^{*}_2\right)\equiv\dfrac{1}{Z^*}\exp\left[-\dfrac{\mathcal{H}}{k_BT}\right]=\left(\dfrac{c_2+1}{2\pi}\right)^{1/2}\exp\left\{-\dfrac{1}{2}\left(c_2+1\right)\left[r^{*}_2-\left(\dfrac{c_2\chi^*}{c_2+1}\right)\right]^2\right\}
\end{equation*}
and the moments of the distribution are
\begin{align}\label{eq:mean_work}
\left<r^{*}_2\right> & =\dfrac{c_2\chi^*}{c_2+1} \\
\left<\left(r^{*}_2-\left<r^{*}_2\right>\right)^2\right> & =\dfrac{1}{c_2+1} \label{eq:var_work}
\end{align}
The expression for the average work is obtained by taking an ensemble average of Eq.~(\ref{eq:work_def}), 
\begin{align}\label{eq:av_work11}
\left<W^{*}\right>&=\dfrac{c_2}{2}\left[\chi^{2*}\left(\tau^{*}\right)-\chi^{2*}\left(0\right)\right]-c_2\int_{0}^{\tau^{*}}dt^{*}\dot{\chi}^{*}(t^{*})\Biggl[G(t^{*})\left<r^{*}_2(0)\right>+\dfrac{c_2}{4\theta}\int_{0}^{t^{*}}dt_1^{*}G\left(t^{*}-t_1^{*}\right)\chi^*(t_1^{*})\Biggr]
\end{align}
Substituting Eq.~(\ref{eq:mean_work}) into the second term on the RHS of Eq.~(\ref{eq:av_work11}),
\begin{align}\label{eq:av_work12}
\left<W^{*}\right>&=\dfrac{c_2}{2}\left[\chi^{2*}\left(\tau^{*}\right)-\chi^{2*}\left(0\right)\right]-\dfrac{c^2_2}{c_2+1}\chi^*(0)\int_{0}^{\tau^{*}}dt^{*}\dot{\chi}^{*}(t^{*})G(t^{*})-\dfrac{c^{2}_2}{4\theta}\int_{0}^{\tau^{*}}dt^{*}\dot{\chi}^{*}(t^{*})\underline{\int_{0}^{t^{*}}dt_1^{*}G\left(t^{*}-t_1^{*}\right)\chi^*(t_1^{*})}
\end{align}
 The underlined term is simplified as 
 \begin{align}\label{eq:byparts1}
 \int_{0}^{t^{*}}dt_1^{*}G\left(t^{*}-t_1^{*}\right)\chi^*(t_1^{*})&= \int_{0}^{t^{*}}dt_1^{*}\exp\left[-\dfrac{E\left(t^{*}-t_1^{*}\right)}{\theta}\right]\chi^*(t_1^{*})=\int_{0}^{t^{*}}dt_1^{*}\chi^*(t_1^{*})\dfrac{d}{dt_1^{*}}\left(\theta E^{-1}\exp\left[-\dfrac{E\left(t^{*}-t_1^{*}\right)}{\theta}\right]\right)
 \end{align} 
 Integrating expression on the RHS of Eq.~(\ref{eq:byparts1}) by parts, 
   \begin{align}\label{eq:byparts2}
 \int_{0}^{t^{*}}dt_1^{*}G\left(t^{*}-t_1^{*}\right)\chi^*(t_1^{*})&=\Biggl[\chi^*(t_1^{*})\theta E^{-1}\exp\left[-\dfrac{E\left(t^{*}-t_1^{*}\right)}{\theta}\right]\Biggr]_{0}^{t^*}- \int_{0}^{t^{*}}dt_1^{*}\dot{\chi}^{*}(t_1^{*})\theta E^{-1}\exp\left[-\dfrac{E\left(t^{*}-t_1^{*}\right)}{\theta}\right]
 \end{align}  
 one obtains, 
  \begin{align}\label{eq:byparts3}
 \int_{0}^{t^{*}}dt_1^{*}G\left(t^{*}-t_1^{*}\right)\chi^*(t_1^{*})&=\Bigl[\chi^*(t^{*})\theta E^{-1}\Bigr]-\Biggl[\chi^*\left(0\right)\theta E^{-1}\exp\left[-\dfrac{E t^*}{\theta}\right]\Biggr]- \int_{0}^{t^{*}}dt_1^{*}\dot{\chi}^{*}(t_1^{*})\theta E^{-1}\exp\left[-\dfrac{E\left(t^{*}-t_1^{*}\right)}{\theta}\right]
 \end{align}  
which can then be written as 
 \begin{align}\label{eq:byparts4}
 \int_{0}^{t^{*}}dt_1^{*}G\left(t^{*}-t_1^{*}\right)\chi^*(t_1^{*})&=\theta E^{-1} \Biggl\{ \chi^*(t^{*}) - \chi^*\left(0\right)G\left(t^*\right) - \int_{0}^{t^{*}}dt_1^{*}\dot{\chi}^{*}(t_1^{*})G\left(t^{*}-t_1^{*}\right)\Biggr\} 
 \end{align}  
 Substituting Eq.~(\ref{eq:byparts4}) into Eq.~(\ref{eq:av_work12}) and simplifying, 
  \begin{align}\label{eq:av_work_simp}
\left<W^{*}\right>&=\dfrac{c_2}{2}\left[\chi^{2*}\left(\tau^{*}\right)-\chi^{2*}\left(0\right)\right]-\cancel{\dfrac{c^2_2}{c_2+1}\chi^*(0)\int_{0}^{\tau^{*}}dt^{*}\dot{\chi}^{*}(t^{*})G(t^{*})}-\dfrac{c^{2}_2}{c_2+1}\int_{0}^{\tau^{*}}dt^{*} \dot{\chi}^{*}(t^{*}) \chi^*(t^{*}) \\[5pt]
&+\cancel{\dfrac{c^2_2}{c_2+1}\chi^*(0)\int_{0}^{\tau^{*}}dt^{*}\dot{\chi}^{*}(t^{*})G(t^{*})} + \dfrac{c^2_2}{c_2+1} \int_{0}^{\tau^{*}}dt^{*}\int_{0}^{t^{*}}dt_1^{*}\dot{\chi}^{*}(t^{*})G\left(t^{*}-t_1^{*}\right)\dot{\chi}^{*}(t_1^{*}) \nonumber
 \end{align}
 one obtains
 \begin{align}\label{eq:av_work_simp2}
\left<W^{*}\right>&=\left(\dfrac{c_2}{2\left(c_2+1\right)}\right) \left[\chi^{2*}\left(\tau^{*}\right)-\chi^{2*}\left(0\right)\right]+ \dfrac{c^2_2}{c_2+1} \int_{0}^{\tau^{*}}dt^{*}\int_{0}^{t^{*}}dt_1^{*}\dot{\chi}^{*}(t^{*})G\left(t^{*}-t_1^{*}\right)\dot{\chi}^{*}(t_1^{*})
 \end{align}
Recognizing that the first term on the RHS of Eq.~(\ref{eq:av_work_simp2}) is the free-energy difference, $\Delta A^{*}\equiv A^*\left[\chi^{*}(\tau^{*})\right]-A^*\left[\chi^{*}(0)\right]$, with $A$ given by Eq.~(\ref{eq:free_energy_def_suppl}), the expression for the average work can be rewritten as
 \begin{align}\label{eq:av_work_suppl}
\left<W^{*}\right>=\Delta A^{*}+\dfrac{c_2^2}{c_2+1}\int_{0}^{\tau^{*}}dt^{*}&\int_{0}^{t^{*}}dt_1^{*}\Bigl[\dot{\chi}^{*}(t^{*})G\left(t^{*}-t_1^{*}\right)\dot{\chi}^{*}(t_1^{*})\Bigr]
\end{align}
This equation is reproduced as Eq.~(\ref{eq:av_work6}) in Sec.~\ref{sec:dry_sd}.
Using Eqs.~(\ref{eq:work_def}) and (\ref{eq:av_work11}), the variance of the work distribution, $\sigma^2=\left<\left(W^*-\left<W^*\right>\right)^2\right>$, is written as,
\begin{align}\label{eq:sig_derv2}
\sigma^2&=c_2^2\int_{0}^{\tau^{*}}dt^{*}\int_{0}^{\tau^{*}}dt_1^{*}\dot{\chi}^{*}(t^{*})G(t^{*})\underline{\left[\left<\left(r^{*}_2(0)-\left<r^{*}_2(0)\right>\right)^2\right>\right]}G(t_1^{*})\dot{\chi}^{*}(t_1^{*})\\[5pt]
&+ \dfrac{c_2^2}{\theta^2}\int_{0}^{\tau^{*}}dt^{*}_1\int_{0}^{t^{*}_1}d\tilde{t}^{*}_1\int_{0}^{\tau^{*}}dt^{*}_2\int_{0}^{t^{*}_2}d\tilde{t}_2^{*}\dot{\chi}^{*}(t^{*}_1) G(t^{*}_1-\tilde{t}^{*}_1)\underline{\left<\xi(\tilde{t}^{*}_1)\xi(\tilde{t}^{*}_2)\right>}\dot{\chi}^{*}(t^{*}_2)G(t^{*}_2-\tilde{t}^{*}_2), \nonumber   
\end{align}
Eqs.~(\ref{eq:var_work}) and (\ref{eq:noise_strength}) can be used to simplify the underlined terms in Eq.~(\ref{eq:sig_derv2}), and we obtain, 
\begin{align}\label{eq:sig_derv3}
\sigma^2&=\dfrac{c_2^2}{c_1+1}\int_{0}^{\tau^{*}}dt^{*}\int_{0}^{\tau^{*}}dt_1^{*}\dot{\chi}^{*}(t^{*})G(t^{*})G(t_1^{*})\dot{\chi}^{*}(t_1^{*})\\[5pt]
&+ \dfrac{c_2^2}{2\theta}\int_{0}^{\tau^{*}}dt^{*}_1\int_{0}^{t^{*}_1}d\tilde{t}^{*}_1\int_{0}^{\tau^{*}}dt^{*}_2\int_{0}^{t^{*}_2}d\tilde{t}_2^{*}\dot{\chi}^{*}(t^{*}_1) G(t^{*}_1-\tilde{t}^{*}_1)\delta\left(\tilde{t}^{*}_1-\tilde{t}^{*}_2\right)\dot{\chi}^{*}(t^{*}_2)G(t^{*}_2-\tilde{t}^{*}_2) \nonumber   
\end{align}
Subsequent integration over $\tilde{t}^{*}_1$ in the second integral yields
\begin{equation}\label{eq:sig_intermed}
\sigma^2 = \dfrac{c_2^2}{c_2+1}\int_{0}^{\tau^{*}}dt^{*}\int_{0}^{\tau^{*}}dt_1^{*}\dot{\chi}^{*}(t^{*})G(t^{*})G(t_1^{*})\dot{\chi}^{*}(t_1^{*}) + \dfrac{c_2^2}{2\theta}\int_{0}^{\tau^{*}}dt^{*}_1\int_{0}^{\tau^{*}}dt^{*}_2\int_{0}^{t^{*}_2}d\tilde{t}^{*}_2\dot{\chi}^{*}(t^{*}_1)\dot{\chi}^{*}(t^{*}_2)G(t^{*}_1-\tilde{t}^{*}_2)G(t^{*}_2-\tilde{t}^{*}_2)   
\end{equation}

The following identity
\begin{equation}
\label{eq:g_simp}
\int_{0}^{t^{*}_2}d\tilde{t}^{*}_2\,G(t^{*}_1-\tilde{t}^{*}_2)G(t^{*}_2-\tilde{t}^{*}_2) =
\left(\dfrac{2\theta}{c_2+1}\right)\Biggl[G(t^{*}_1-t^{*}_2)-G(t^{*}_1)G(t^{*}_2)\Biggr]
\end{equation}
\end{widetext}
can be used used to simplify the second term on the RHS of Eq.~(\ref{eq:sig_intermed}), resulting in 
\begin{align}\label{eq:sig_exp_app}
&\sigma^2=\dfrac{2c_2^2}{\left(c_2+1\right)}\int_{0}^{\tau^{*}}dt^{*}\int_{0}^{t^{*}}dt_1^{*}\dot{\chi}^{*}(t^{*})G(t^{*}-t_1^{*})\dot{\chi}^{*}(t_1^{*})
\end{align}
This equation is reproduced as Eq.~(\ref{eq:sig_exp}) in Sec.~\ref{sec:dry_sd}.

\section{\label{sec:gov_eq_db}Governing equation and simulation details for the dumbbell obeying the Marko-siggia force law}

The dumbbell model described in Sec.~\ref{sec:mod_des_db} is suspended in a fluid where the velocity field at any location $\bm{r}_{\text{f}}$ is given by $\bm{v}_{\text{f}}(\bm{r},t)\equiv\bm{v}_0+\boldsymbol{\kappa}(t)\cdot\bm{r}_{\text{f}}$, where $\bm{v}_0$ is a constant vector, and $\boldsymbol{\kappa}\equiv\left(\nabla\bm{v}_{\text{f}}\right)^{\bm{T}}$ is the transpose of the velocity gradient tensor. In the present work, both $\bm{v}_0$ and $\boldsymbol{\kappa}$ are set to $\bm{0}$ as the pulling experiments are simulated in a quiescent fluid. However, these terms have been included in the governing equations for the sake of generality. The configurational distribution function, $\Psi\left(\bm{Q},\bm{R},t\right)$, denotes the probability of finding the dumbbell at a position between $\bm{R}$ and $\bm{R}+d\bm{R}$, with an extension that lies between $\bm{Q}$ and $\bm{Q} + d\bm{Q}$, at any time $t$. The total restoring force in the connector vector, $\bm{F}^{\text{(c)}}$, has contributions from both the spring, $\bm{F}^{\text{(s)}}\equiv\partial U_{\text{MS}}/\partial \bm{Q}$, and the dashpot, $\bm{F}^{\text{(d)}}$
\begin{equation}
\bm{F}^{\text{(c)}}=\bm{F}^{\text{(s)}}+\bm{F}^{\text{(d)}},
\end{equation}
with
\begin{equation}
\bm{F}^{\text{(s)}}=\frac{HQ_0}{3}\left[\frac{1}{2\left(1-Q/Q_0\right)^2}-\frac{1}{2}+2\left(\frac{Q}{Q_0}\right)\right]\frac{\bm{Q}}{Q},
\end{equation}
where the potential energy of the spring, $U_{\text{MS}}$, is given by
\begin{align*}\label{eq:ms_dimen}
U_{\text{MS}}&=\frac{HQ_0^2}{3}\Biggl[\frac{1}{2\left(1-Q/Q_0\right)}-\frac{1}{2}\left(\frac{Q}{Q_0}\right)+\left(\frac{Q}{Q_0}\right)^2\Biggr], 
\end{align*}
and
\begin{equation}
\bm{F}^{\text{(d)}}=K\frac{\bm{QQ}}{\bm{Q}^2}\cdot\llbracket{\dot {\bm{Q}}}\rrbracket
\end{equation}
where $\llbracket{\dot {\bm{Q}}}\rrbracket$ is the momentum-averaged rate-of-change of the connector vector, $\bm{Q}$.

The force balance on the beads can be solved to obtain the following equations of motion for the position vectors of the beads~\cite{Kailasham2018},
\begin{align}\label{eq:rsim_1}
\llbracket{\dot {\bm{r}}}_{1}\rrbracket & = \left[\boldsymbol{\delta}-\frac{\varphi\beta}{\varphi\beta+2}\frac{\bm{Q}\bm{Q}}{Q^2}\right]\cdot\Biggl(\bm{v}_{0}+\boldsymbol{\kappa}\cdot\bm{r}_1-\frac{1}{2\zeta}\bm{M}\cdot\frac{\partial U_{\text{MS}}}{\partial \bm{r}_{1}}\nonumber\\[5pt]
&-\frac{k_BT}{2\zeta}\bm{M}\cdot\frac{\partial \ln \Psi}{\partial \bm{r}_{1}}+\frac{\varphi}{4}\bm{M}\cdot\frac{\bm{Q}\bm{Q}}{Q^2}\cdot\llbracket{\dot {\bm{r}}}_{2}\rrbracket\nonumber\\[5pt]
&-\frac{{H_{1}}}{\zeta}\left(\bm{r}_{1}-\bm{\chi}_{1}\right)-\boldsymbol{\Omega}\cdot {H_{2}}\left(\bm{r}_{2}-\bm{\chi}_{2}\right)\Biggr)
\end{align}
and
\begin{align}\label{eq:rsim_2}
\llbracket{\dot {\bm{r}}}_{2}\rrbracket & = \left[\boldsymbol{\delta}-\frac{\varphi\beta}{\varphi\beta+2}\frac{\bm{Q}\bm{Q}}{Q^2}\right]\cdot\Biggl(\bm{v}_{0}+\boldsymbol{\kappa}\cdot\bm{r}_2-\frac{1}{2\zeta}\bm{M}\cdot\frac{\partial U_{\text{MS}}}{\partial \bm{r}_{2}}\nonumber\\[5pt]
&-\frac{k_BT}{2\zeta}\bm{M}\cdot\frac{\partial \ln \Psi}{\partial \bm{r}_{2}}+\frac{\varphi}{4}\bm{M}\cdot\frac{\bm{Q}\bm{Q}}{Q^2}\cdot\llbracket{\dot {\bm{r}}}_{1}\rrbracket\nonumber\\[5pt]
&-\frac{{H_{2}}}{\zeta}\left(\bm{r}_{2}-\bm{\chi}_{2}\right)-\boldsymbol{\Omega}\cdot {H_{1}}\left(\bm{r}_{1}-\bm{\chi}_{1}\right)\Biggr)
\end{align}
where $\bm{\Omega}$ is the hydrodynamic interaction tensor, defined as 
\begin{equation}\label{eq:def_hi}
\bm{\Omega}\left(\bm{Q}\right)=\frac{h}{\zeta Q}\left(A\boldsymbol{\delta} + B\frac{\bm{QQ}}{Q^2}\right)
\end{equation}
with $h=\left(3/4\right)a$ and $\bm{M}=2\left(\boldsymbol{\delta}-\zeta\boldsymbol{\Omega}\right)$. The terms $A$ and $B$ depend on the choice of the expression for the HI tensor, as shown in ref.~\citenum{Kailasham2018}. Here we choose the Rotne-Prager-Yamakawa (RPY) expression for the HI tensor~\cite{Rotne1969,Yamakawa}, in which the variables $A$ and $B$ are defined as follows
\begin{align}\label{eq:more_Q}
A & = 1 + \frac{2}{3}\left(\frac{a}{Q}\right)^2; B = 1 - 2\left(\frac{a}{Q}\right)^2 \text{for} \hspace{3pt} Q\geq2a \\[10pt]
\label{eq:less_Q}
A & = \frac{4}{3}\left(\frac{Q}{a}\right) - \frac{3}{8}\left(\frac{Q}{a}\right)^2; B = \frac{1}{8}\left(\frac{Q}{a}\right)^2 \text{for} \hspace{3pt} Q<2a
\end{align}

The quantity $\beta$ that appears in Eqs.~(\ref{eq:rsim_1}) and (\ref{eq:rsim_2}) is defined as 
\begin{equation}
\beta=1-\frac{h}{Q}\left(A+B\right)
\end{equation}
Using $\bm{r}_1=\bm{R}-(1/2)\bm{Q}\,;\,\bm{r}_2=\bm{R}+(1/2)\bm{Q}$, and the chain-rule for partial differentiation to operate on $\partial \Psi/\partial \bm{r}_1$ and $\partial \Psi/\partial \bm{r}_2$, leads to,
\begin{widetext}
\begin{align}\label{eq:rd1}
\llbracket{\dot {\bm{r}}}_{1}\rrbracket & = \bm{v}_{0}+\boldsymbol{\kappa}\cdot\left(\bm{R}-\frac{1}{2}\bm{Q}\right)+\boldsymbol{\Omega}\cdot\Biggl(-{k_BT}\left[\frac{1}{2}\frac{\partial \ln \Psi}{\partial \bm{R}}+\frac{\partial \ln\Psi}{\partial \bm{Q}}\right]-\frac{\partial U_{\text{MS}}}{\partial \bm{r}_{2}}-K\frac{\bm{QQ}}{Q^2}\cdot\llbracket{\dot {\bm{Q}}}\rrbracket-{H_{2}}\left(\bm{R}+\frac{1}{2}\bm{Q}-\bm{\chi}_{2}\right)\Biggr)\nonumber\\[5pt]
&- \frac{k_BT}{\zeta}\Biggl[\frac{1}{2}\frac{\partial \ln \Psi}{\partial \bm{R}}-\frac{\partial \ln\Psi}{\partial \bm{Q}}\Biggr]-\frac{1}{\zeta}\frac{\partial U_{\text{MS}}}{\partial \bm{r}_{1}}+\frac{\varphi}{2}\frac{\bm{QQ}}{Q^2}\cdot\llbracket{\dot {\bm{Q}}}\rrbracket-\frac{{H_{1}}}{\zeta}\left(\bm{R}-\frac{1}{2}\bm{Q}-\bm{\chi}_{1}\right)
\end{align}
and
\begin{align}\label{eq:rd2}
\llbracket{\dot {\bm{r}}}_{2}\rrbracket & = \bm{v}_{0}+\boldsymbol{\kappa}\cdot\left(\bm{R}+\frac{1}{2}\bm{Q}\right)+\boldsymbol{\Omega}\cdot\Biggl(-{k_BT}\left[\frac{1}{2}\frac{\partial \ln \Psi}{\partial \bm{R}}-\frac{\partial \ln\Psi}{\partial \bm{Q}}\right]-\frac{\partial U_{\text{MS}}}{\partial \bm{r}_{1}}+K\frac{\bm{QQ}}{Q^2}\cdot\llbracket{\dot {\bm{Q}}}\rrbracket-{H_{1}}\left(\bm{R}-\frac{1}{2}\bm{Q}-\bm{\chi}_{1}\right)\Biggr)\nonumber\\[5pt]
&- \frac{k_BT}{\zeta}\Biggl[\frac{1}{2}\frac{\partial \ln \Psi}{\partial \bm{R}}+\frac{\partial \ln\Psi}{\partial \bm{Q}}\Biggr]-\frac{1}{\zeta}\frac{\partial U_{\text{MS}}}{\partial \bm{r}_{2}}+\frac{\varphi}{2}\frac{\bm{QQ}}{Q^2}\cdot\llbracket{\dot {\bm{Q}}}\rrbracket-\frac{{H_{2}}}{\zeta}\left(\bm{R}+\frac{1}{2}\bm{Q}-\bm{\chi}_{2}\right)
\end{align}
\end{widetext}
By adding and subtracting Eqs.~(\ref{eq:rd1}) and (\ref{eq:rd2}) suitably, we obtain
\begin{align}\label{eq:r_dot}
\llbracket{\dot {\bm{R}}}\rrbracket & = \bm{v}_{0}+\boldsymbol{\kappa}\cdot\bm{R}-\frac{k_BT}{2\zeta}\left(\boldsymbol{\delta}+\zeta\boldsymbol{\Omega}\right)\cdot\frac{\partial \ln \Psi}{\partial \bm{R}}\nonumber\\
&-\frac{1}{2\zeta}\left(\boldsymbol{\delta}+\zeta\boldsymbol{\Omega}\right)\cdot\bm{X}
\end{align}
\begin{align}\label{eq:q_dot}
\llbracket{\dot {\bm{Q}}}\rrbracket &= \left[ \boldsymbol{\delta} - \frac{\varphi\beta}{\varphi\beta + 1}\frac{\bm{QQ}}{Q^2}\right]\cdot\Biggl(\boldsymbol{\kappa}\cdot\bm{Q}- \frac{k_BT}{\zeta}\bm{M}\cdot\frac{\partial}{\partial \bm{Q}}\ln \Psi\nonumber\\
&-\frac{1}{\zeta}\bm{M}\cdot\frac{\partial U_{\text{MS}}}{\partial \bm{Q}}-\frac{1}{2\zeta}\bm{M}\cdot\bm{Y}\Biggr)
\end{align}
where 
\begin{align*}
\bm{X} &= \bm{R}\left(H_2+H_1\right)+\bm{Q}\left(\frac{H_2-H_1}{2}\right)-\left(H_2\bm{\chi}_2+H_1\bm{\chi}_1\right)
\end{align*}
and
\begin{align*}
\bm{Y} &= \bm{R}\left(H_2-H_1\right)+\bm{Q}\left(\frac{H_2+H_1}{2}\right)-\left(H_2\bm{\chi}_2-H_1\bm{\chi}_1\right)
\end{align*}
and both $\bm{X}$ and $\bm{Y}$ have dimensions of force.

\begin{widetext}
The equation of continuity in terms of $\bm{R}$ and $\bm{Q}$ is given by~\cite{Bird1987b},
\begin{equation}
\frac{\partial \Psi}{\partial t}=-\left(\frac{\partial }{\partial \bm{R}}\cdot \llbracket{\dot {\bm{R}}}\rrbracket \Psi \right)-\left(\frac{\partial}{\partial \bm{Q}}\cdot \llbracket{\dot {\bm{Q}}}\rrbracket \Psi \right)
\end{equation}
Substituting Eqs.~(\ref{eq:r_dot}) and (\ref{eq:q_dot}) into the above expression leads to the Fokker-Planck equation that governs the configurational distribution function $\Psi\left(\bm{Q}, \bm{R},t\right)$, 
\begin{align}
\begin{split}\label{eq:fp_sdot_dim}
\frac{\partial \Psi}{\partial t} & = -\frac{\partial }{\partial \bm{R}}\cdot\Biggl\{\Biggl[\bm{v}_{0}+\boldsymbol{\kappa}\cdot\bm{R}-\frac{1}{2\zeta}\left(\boldsymbol{\delta}+\zeta\boldsymbol{\Omega}\right)\cdot\bm{X}\Biggr]\Psi\Biggr\}+\frac{k_BT}{2\zeta}\frac{\partial }{\partial \bm{R}}\cdot\left(\boldsymbol{\delta}+\zeta\boldsymbol{\Omega}\right)\cdot\frac{\partial \Psi}{\partial \bm{R}}\\
&-\frac{\partial }{\partial \bm{Q}}\cdot\Biggl\{\Biggl[ \left[ \boldsymbol{\delta} - \frac{\varphi\beta}{\varphi\beta + 1}\frac{\bm{QQ}}{Q^2}\right]\cdot\Biggl(\boldsymbol{\kappa}\cdot\bm{Q}- \frac{1}{\zeta}\bm{M}\cdot\frac{\partial U_{\text{MS}}}{\partial \bm{Q}}-\frac{1}{2\zeta}\bm{M}\cdot\bm{Y}\Biggr)\Biggr]\Psi\Biggr\}\\
&+\frac{k_BT}{\zeta}\frac{\partial }{\partial \bm{Q}}\cdot\left[ \left(\boldsymbol{\delta} - \frac{\varphi\beta}{\varphi\beta + 1}\frac{\bm{QQ}}{Q^2}\right)\cdot\bm{M}\right]\cdot\frac{\partial \Psi}{\partial \bm{Q}}
\end{split}
\end{align}
We define the following dimensionless quantities, 
\begin{equation}\label{eq:scale}
t^* = \frac{t}{\lambda_{\text{H}}};\,\bm{Q}^* = \frac{\bm{Q}}{l_{\text{H}}};\,b = \frac{Q_0^2}{l^2_{\text{H}}};\,\boldsymbol{\kappa}^* = \lambda_{\text{H}}\boldsymbol{\kappa};\,U_{\text{MS}}^* = \frac{U_{\text{MS}}}{k_BT};\, \Psi^* = \Psi {l^3_{\text{H}}};\,\bm{X}^*=\frac{\bm{X}}{\sqrt{k_BTH}}
\end{equation}
In terms of these non-dimensional variables, the Fokker-Planck equation assumes the following form,
\begin{align}\label{eq:fp_sdot_dimless}
\frac{\partial \Psi^*}{\partial t^*} & = -\frac{\partial }{\partial \bm{R^*}}\cdot\Biggl\{\Biggl[\bm{v^*}_{0}+\boldsymbol{\kappa^*}\cdot\bm{R^*}-\frac{1}{8}\left(\boldsymbol{\delta}+\zeta\boldsymbol{\hat{\Omega}}\right)\cdot\bm{X^*}\Biggr]\Psi^*\Biggr\}+\frac{1}{8}\frac{\partial }{\partial \bm{R}}\cdot\left(\boldsymbol{\delta}+\zeta\boldsymbol{\hat{\Omega}}\right)\cdot\frac{\partial \Psi^*}{\partial \bm{R^*}}\nonumber \\[5pt]
&-\frac{\partial }{\partial \bm{Q^*}}\cdot\Biggl\{\Biggl[ \left[ \boldsymbol{\delta} - \frac{\varphi\beta^*}{\varphi\beta^* + 1}\frac{\bm{Q^*Q^*}}{Q^{*2}}\right]\cdot\Biggl(\boldsymbol{\kappa^*}\cdot\bm{Q^*}- \frac{1}{4}\bm{M^*}\cdot\frac{\partial U^*_{\text{MS}}}{\partial \bm{Q^*}}-\frac{1}{8}\bm{M^*}\cdot\bm{Y^*}\Biggr)\Biggr]\Psi^*\Biggr\} \nonumber\\[5pt]
&+\frac{1}{4}\frac{\partial }{\partial \bm{Q^*}}\cdot\left[ \left(\boldsymbol{\delta} - \frac{\varphi\beta^*}{\varphi\beta^* + 1}\frac{\bm{Q^*Q^*}}{Q^{*2}}\right)\cdot\bm{M^*}\right]\cdot\frac{\partial \Psi^*}{\partial \bm{Q^*}}
\end{align}
\end{widetext}
where
\begin{equation}
\boldsymbol{\hat{\Omega}}\left(\bm{Q^*}\right)=\frac{\alpha}{\zeta Q^*}\left(A^*\boldsymbol{\delta}+B^*\frac{\bm{Q^*Q^*}}{Q^{*2}}\right)
\end{equation}
with $\alpha=\left(3/4\right)\sqrt{\pi}h^*$ and $h^*=a/\left(\sqrt{\pi}l_{\text{H}}\right)$. The quantity $\beta^*$ is dimensionless and is defined as,
\begin{equation}
\beta^*=1-\frac{\alpha}{Q^*}\left(A^*+B^*\right)
\end{equation}
Using the following identity,
\begin{equation*}\label{eq:stod}
\frac{\partial}{\partial \bm{x}}\cdot\left[\bm{L}\cdot\frac{\partial f}{\partial \bm{x}}\right] =  \frac{\partial}{\partial{\bm{x}}}\frac{\partial}{\partial{\bm{x}}}:\left[\bm{L}^{\text{T}}f\right] - \frac{\partial}{\partial \bm{x}}\cdot\left[f\frac{\partial}{\partial{\bm{x}}}\cdot\bm{L}^{\text{T}}\right] 
\end{equation*}
where $\bm{L}$ is a tensor and $f$ is a scalar, the second and fourth terms on the right-hand-side of Eq.~(\ref{eq:fp_sdot_dimless}) can rewritten in a way that renders the Fokker-Planck equation amenable to It\^o's interpretation. 
\begin{widetext}
Since $\boldsymbol{\hat{\Omega}}\left(\bm{Q^*}\right)$ is independent of $\bm{R^*}$, and 
\begin{equation}
\left(\boldsymbol{\delta} - \frac{\varphi\beta^*}{\varphi\beta^* + 1}\frac{\bm{Q^*Q^*}}{{Q}^{*2}}\right)\cdot\left(\boldsymbol{\delta}-\zeta\boldsymbol{\hat{\Omega}}\right) = \left(\frac{Q^*-A^*\alpha}{Q^*}\right)
 \left(\boldsymbol{\delta}-g_1\frac{\bm{Q^*Q^*}}{{Q}^{*2}}\right)
\end{equation}
the Fokker-Planck equation can be rewritten as follows, 
\begin{align}\label{eq:fp_ddot_dimless}
\frac{\partial \Psi^*}{\partial t^*} & = -\frac{\partial }{\partial \bm{R^*}}\cdot\Biggl\{\Biggl[\bm{v^*}_{0}+\boldsymbol{\kappa^*}\cdot\bm{R^*}-\frac{1}{8}\left(\boldsymbol{\delta}+\zeta\boldsymbol{\hat{\Omega}}\right)\cdot\bm{X^*}\Biggr]\Psi^*\Biggr\}+\frac{1}{2}\frac{\partial }{\partial \bm{R^*}}\frac{\partial }{\partial \bm{R^*}}:\left[\frac{\left(\boldsymbol{\delta}+\zeta\boldsymbol{\hat{\Omega}}\right)}{4}\Psi^*\right]\nonumber\\[5pt]
&-\frac{\partial }{\partial \bm{Q^*}}\cdot\Biggl\{\Biggl[\frac{g_2}{2}\frac{\bm{Q^*}}{Q^*}+ \left[ \boldsymbol{\delta} - \frac{\varphi\beta^*}{\varphi\beta^* + 1}\frac{\bm{Q^*Q^*}}{Q^{*2}}\right]\cdot\Biggl(\boldsymbol{\kappa^*}\cdot\bm{Q^*}- \frac{1}{4}\bm{M^*}\cdot\frac{\partial U_{\text{MS}}^*}{\partial \bm{Q^*}}-\frac{1}{8}\bm{M^*}\cdot\bm{Y^*}\Biggr)\Biggr]\Psi^*\Biggr\}\nonumber\\[5pt]
&+ {\frac{1}{2}\frac{\partial}{\partial{\bm{Q^*}}}\frac{\partial}{\partial{\bm{Q^*}}}\bm{:}\left[\left(\frac{Q^*-A^*\alpha}{Q^*}\right)\left(\boldsymbol{\delta}-g_1\frac{\bm{Q^*Q^*}}{{Q}^{*2}}\right)\Psi^*\right]}
\end{align}
where
\begin{align}
g_1 & = \frac{\alpha B^*Q^*+\varphi(Q^*-A^*\alpha)[Q^*-\alpha(A^*+B^*)]}{(Q^*-A^*\alpha)\left\{Q^*+\varphi[Q^*-\alpha(A^*+B^*)]\right\}} \\[10pt]
g_2 &= \frac{2\alpha B^*}{\left\{Q^*+\varphi[Q^*-\alpha(A^*+B^*)]\right\}^2} - 2g_1\left(\frac{Q^*-A^*\alpha}{Q^{*2}}\right) \nonumber 
\end{align}
The Fokker-Planck equation (Eq.~(\ref{eq:fp_ddot_dimless})) can be written in the following compact form, 
\begin{equation}
\frac{\partial \Psi^*}{\partial t^*}  = -\frac{\partial }{\partial \bm{R^*}}\cdot\Bigl\{\bm{e}\Psi^*\Bigr\}+\frac{1}{2}\frac{\partial }{\partial \bm{R^*}}\frac{\partial }{\partial \bm{R^*}}\bm{:}\left[\tilde{\bm{e}}\Psi^*\right] -\frac{\partial }{\partial \bm{Q^*}}\cdot\Bigl\{\bm{g}\Psi^*\Bigr\}+ {\frac{1}{2}\frac{\partial}{\partial{\bm{Q^*}}}\frac{\partial}{\partial{\bm{Q^*}}}\bm{:}\left[\tilde{\bm{g}}\Psi^*\right]}\label{eq:fp_simp}
\end{equation}
\end{widetext}
where the definitions of the quantities $\bm{e}, \tilde{\bm{e}}, \bm{g}$ and $\tilde{\bm{g}}$ are clear by comparison of Eqs.~(\ref{eq:fp_ddot_dimless}) and~(\ref{eq:fp_simp}). 

It is convenient to define a collective variable, $\bm{C}$, which is a six-element vector containing the components of $\bm{R^*}$ and $\bm{Q^*}$, such that $\bm{C}\equiv\left[R^*_{x},R^*_{y},R^*_{z},Q^*_{x},Q^*_{y},Q^*_{z}\right]$. Similarly, a six-element vector $\bm{j}$ can be defined, containing the components of $\bm{e}$ and $\bm{g}$, along with the definition of a $2\times2$ block matrix $\bm{D}$, whose off-diagonal elements are $\bm{0}$, and the diagonal elements are the matrices $\tilde{\bm{e}}$ and $\tilde{\bm{g}}$ (each of size $3\times 3$). 
With these definitions, the Fokker-Planck equation in Eq.~(\ref{eq:fp_simp}) can be written as
\begin{align}\label{eq:fp_c}
\frac{\partial \Psi^*}{\partial t^*}  = &-\frac{\partial }{\partial \bm{C}}\cdot\Bigl\{\bm{j}\Psi^*\Bigr\}+{\frac{1}{2}\frac{\partial}{\partial{\bm{C}}}\frac{\partial}{\partial{\bm{C}}}\bm{:}\left[\bm{D}\Psi^*\right]}
\end{align}

The stochastic differential equation (SDE) corresponding to Eq.~(\ref{eq:fp_c}) can be obtained using the It\^o interpretation, as 
\begin{align}\label{eq:sde_collective}
d\bm{C}=\bm{j}dt^*+\tilde{\bm{b}}\cdot d\bm{w}_t
\end{align}
where $\bm{w}_t$ is a Wiener process and $\tilde{\bm{b}}\cdot\tilde{\bm{b}}^{T}=\bm{D}$.
The SDE (Eq.~(\ref{eq:sde_collective})) is solved using a semi-implicit predictor-corrector scheme~\cite{Ottinger1996}, as discussed in the following subsection.

\begin{widetext}
\subsection*{Solver details}
With reference to Eq.~(\ref{eq:fp_c}), $\bm{D}$ is a $6\times6$ matrix, and its square root, $\tilde{\bm{b}}_j$,  at any time $t^*_j$ is found using Cholesky decomposition~\cite{press2007numerical}. Although Eq.~(\ref{eq:sde_collective}) is written in terms of the collective variable $\bm{C}$, the equation for $\bm{R^*}$ is solved purely explicitly, whereas the equation in $\bm{Q^*}$ is solved by treating only the spring force term implicitly. For the sake of clarity, the predictor and corrector equations for $\bm{R^*}$ and $\bm{Q^*}$ are presented separately. It is useful to define another six-element vector, $\Delta\bm{S}_j$, as
\begin{equation}
\Delta\bm{S}_j=\tilde{\bm{b}}_j\cdot\Delta \bm{w}_j
\end{equation}
where $\bm{w}_j$ is a vector of six independent Wiener processes, each of mean zero and variance $\Delta t^*_j$. The first three elements of $\Delta\bm{S}_j$, denoted by $\Delta\bm{S}^{\,(R)}_j$, contain the noise contribution to $\bm{R}^*$, and the next three elements, denoted by $\Delta\bm{S}^{\,(Q)}_j$, contribute to the noise in $\bm{Q^*}$.
In the following discussion, Eqs.~(\ref{eq:pred_R})\textemdash (\ref{eq:cube_def}) are in their dimensionless form, but the asterisk has been dropped from these equations for the sake of notational simplicity.
\vskip10pt
\noindent \textbf{Predictor step}
\begin{flalign}\label{eq:pred_R}
\tilde{\bm{R}}\left(t_{j+1}\right)&=\bm{R}(t_j)+\left[\bm{v}_0+\boldsymbol{\kappa}(t_j)\cdot\bm{R}(t_j)-\bm{X}_a(t_j)\right]\Delta t_j + \Delta\bm{S}^{\,(R)}_j
\end{flalign}
\begin{flalign}\label{eq:pred_Q}
\tilde{\bm{Q}}\left(t_{j+1}\right)&=\bm{Q}(t_j)+\Biggl[\boldsymbol{\kappa}(t_j)\cdot\bm{Q}(t_j)-\left(\frac{\varphi\beta(t_j)}{\varphi\beta(t_j)+1}\right)\left[\boldsymbol{\kappa}(t_j):\frac{\bm{Q}(t_j)\bm{Q}(t_j)}{Q^2(t_j)}\right]\bm{Q}(t_j)\nonumber\\
&-\frac{f}{2}\left(\frac{\beta(t_j)}{\varphi\beta(t_j)+1}\right)\frac{\bm{Q}(t_j)}{Q(t_j)}+\frac{g_2(t_j)}{2}\frac{\bm{Q}(t_j)}{Q(t_j)}-\bm{Y}_a(t_j)\Biggr]\Delta t_j +\Delta\bm{S}^{\,(Q)}_j
\end{flalign}
where
\begin{align}\label{eq:pred_def}
\bm{X}_a(t_j)&=\frac{1}{8}\left(\boldsymbol{\delta}+\zeta\boldsymbol{\hat{\Omega}}(\bm{Q}_j)\right)\cdot\bm{X}\left(\bm{Q}_j,\bm{R}_j\right)\nonumber\\[5pt]
\bm{Y}_a(t_j)&=\frac{1}{4}\left(\boldsymbol{\delta}-\frac{\varphi\beta(t_j)}{\varphi\beta(t_j)+1}\frac{\bm{Q}(t_j)\bm{Q}(t_j)}{Q^2(t_j)}\right)\cdot\left(\boldsymbol{\delta}-\zeta\boldsymbol{\hat{\Omega}}(\bm{Q}_j)\right)\cdot\bm{Y}\left(\bm{Q}_j,\bm{R}_j\right)\\[5pt]
f(t_j)&=\frac{\sqrt{b}}{3}\left[\frac{1}{2\left(1-Q(t_j)/\sqrt{b}\right)^2}-\frac{1}{2}+2\left(\frac{Q(t_j)}{\sqrt{b}}\right)\right]\nonumber
\end{align}
and the notations $\bm{Q}_j$ and $\bm{Q}(t_j)$ have been used interchangeably to refer to the same quantity.\\

\noindent \textbf{Corrector step}
\begin{flalign}\label{eq:corr_R}
\bm{R}(t_{j+1})&=\tilde{\bm{R}}\left(t_{j+1}\right)+\frac{1}{2}\left[\boldsymbol{\kappa}(t_{j+1})\cdot \tilde{\bm{R}}_{j+1}-\boldsymbol{\kappa}(t_{j})\cdot \tilde{\bm{R}}_{j}-\tilde{\bm{X}}_a(t_{j+1})+\bm{X}_a(t_j)\right]\Delta t_j
\end{flalign}
\begin{equation}
\begin{split}\label{eq:corr_Q}
\Biggl[1 & +\frac{f(t_{j+1})}{4Q(t_{j+1})}\left(\frac{\tilde{\beta}(t_{j+1})}{\varphi\tilde{\beta}(t_{j+1})+1}\right)\Delta t_j\Biggr]\bm{Q}(t_{j+1})=\tilde{\bm{Q}}\left(t_{j+1}\right)\\
&+\frac{1}{2}\Biggl[\boldsymbol{\kappa}(t_{j+1})\cdot \tilde{\bm{Q}}_{j+1}-\boldsymbol{\kappa}(t_{j})\cdot \tilde{\bm{Q}}_{j}+\tilde{\bm{q}}(t_{j+1})-\bm{q}(t_j)-\tilde{\bm{Y}}_a(t_{j+1})+\bm{Y}_a(t_j)\Biggr]\Delta t_j
\end{split}
\end{equation}
where
\begin{align}\label{eq:corr_def}
\tilde{\bm{X}}_a(t_{j+1})&=\frac{1}{8}\left(\boldsymbol{\delta}+\zeta\boldsymbol{\hat{\Omega}}(\tilde{\bm{Q}}_{j+1})\right)\cdot\tilde{\bm{X}}\left(\tilde{\bm{Q}}_{j+1},\tilde{\bm{R}}_{j+1}\right)\nonumber\\[5pt]
\tilde{\bm{Y}}_a(t_{j+1})&=\frac{1}{4}\left(\boldsymbol{\delta}-\frac{\varphi\tilde{\beta}(t_{j+1})}{\varphi\tilde{\beta}(t_{j+1})+1}\frac{\tilde{\bm{Q}}(t_{j+1})\tilde{\bm{Q}}(t_{j+1})}{\tilde{Q}^2(t_{j+1})}\right)\cdot\left(\boldsymbol{\delta}-\zeta\boldsymbol{\hat{\Omega}}(\tilde{\bm{Q}}_{j+1})\right)\cdot\bm{Y}\left(\tilde{\bm{Q}}_{j+1},\tilde{\bm{R}}_{j+1}\right)\nonumber\\[5pt]
f(t_{j+1})&=\frac{\sqrt{b}}{3}\left[\frac{1}{2\left(1-Q(t_{j+1})/\sqrt{b}\right)^2}-\frac{1}{2}+2\left(\frac{Q(t_{j+1})}{\sqrt{b}}\right)\right]\\[5pt]
{\bm{q}}(t_{j})&=\frac{g_2(t_j)}{2}\frac{\bm{Q}(t_j)}{Q(t_j)}-\left(\frac{\varphi\beta(t_j)}{\varphi\beta(t_j)+1}\right)\left[\boldsymbol{\kappa}(t_j):\frac{\bm{Q}(t_j)\bm{Q}(t_j)}{Q^2(t_j)}\right]\bm{Q}(t_j)\nonumber\\[5pt]
\tilde{\bm{q}}(t_{j+1})&=\frac{g_2(t_{j+1})}{2}\frac{\tilde{\bm{Q}}(t_{j+1})}{\tilde{Q}(t_{j+1})}-\left(\frac{\varphi\tilde{\beta}(t_{j+1})}{\varphi\tilde{\beta}(t_{j+1})+1}\right)\left[\boldsymbol{\kappa}(t_{j+1}):\frac{\tilde{\bm{Q}}(t_{j+1})\tilde{\bm{Q}}(t_{j+1})}{\tilde{Q}^2(t_{j+1})}\right]\tilde{\bm{Q}}(t_{j+1})\nonumber
\end{align}

By setting the length of the vector on the RHS of Eq.~(\ref{eq:corr_Q}) to be $L$, and the length of $\bm{Q}(t_{j+1})$ to be $\Phi$, the following cubic equation is obtained:
\begin{equation}\label{eq:cubic}
\mathcal{V}^3-\mathcal{V}^2\left[\frac{3(3\Gamma+4+2\ell)}{2(2\Gamma+3)}\right]+\mathcal{V}\left[\frac{3(1+\Gamma+2\ell)}{2\Gamma+3}\right]-\frac{3\ell}{2\Gamma+3}=0
\end{equation}
where
\begin{equation}\label{eq:cube_def}
\Gamma =\left(\frac{\tilde{\beta}(t_{j+1})}{\varphi\tilde{\beta}(t_{j+1})+1}\right)\frac{\Delta t_j}{4} \, ; \quad \mathcal{V} =\frac{\Phi}{\sqrt{b}} \, ; \quad  \ell =\frac{L}{\sqrt{b}}
\end{equation}

Eq.~(\ref{eq:cubic}) has three roots\textemdash two complex and one real\textemdash and the real root is obtained using the Newton-Raphson scheme~\cite{press2007numerical}.
Note that the equations are solved in their dimensionless form, and the dimensional quantities are obtained by a suitable multiplication with the scaling factors, as explained in the discussion surrounding Eq.~(\ref{eq:scale}).

\section{\label{sec:valid_db} Code validation for the dumbbell case}

The total Hamiltonian of the dumbbell and trap system is written as 
\begin{align}\label{eq:tot_energy}
\mathcal{H}^*\equiv\frac{\mathcal{H}}{k_BT}&=U^*_{\text{MS}}+\frac{c_1}{2}\left(\bm{r^*}_1-\bm{\chi^*}_1\right)^2+\frac{c_2}{2}\left(\bm{r^*}_2-\bm{\chi^*}_2\right)^2
\end{align}
The expression for $\mathcal{H}^*$ can be rewritten in terms of $\bm{Q^*}$ and $\bm{R^*}$ as,
\begin{align}
\begin{split}\label{eq:U_QR}
\mathcal{H}^*&=\frac{b}{3}\left[\frac{1}{2\left(1-Q^*/\sqrt{b}\right)}-\frac{1}{2}\left(\frac{Q^*}{\sqrt{b}}\right)+\left(\frac{Q^*}{\sqrt{b}}\right)^2\right] -\bm{R^*}\cdot\left(c_1\bm{\chi^*}_1+c_2\bm{\chi^*}_2\right)+\frac{c_1\chi^{*2}_1+c_2\chi^{*2}_2}{2}\\[5pt]
& +\frac{\bm{Q^*}}{2}\cdot\left(c_1\bm{\chi^*}_1-c_2\bm{\chi^*}_2\right)-\left(\frac{c_1-c_2}{2}\right)\bm{Q^*}\cdot\bm{R^*}+R^{*2}\left(\frac{c_1+c_2}{2}\right)+\frac{Q^{*2}}{4}\left(\frac{c_1+c_2}{2}\right)
\end{split}
\end{align}
The steady-state configurational distribution function can be written as
\begin{equation}
\Psi^*(\bm{Q^*},\bm{R^*})=\frac{1}{\mathcal{Z}}\exp[-\mathcal{H}^*]
\end{equation}
where $\mathcal{Z}$ is the partition function of the system, given by
\begin{equation}\label{eq:z_def}
\mathcal{Z}=\int\int\exp[-\mathcal{H}^*]d\bm{R^*}d\bm{Q^*}
\end{equation}
Substituiting the definition of $\mathcal{H}^*$ from Eq.~(\ref{eq:U_QR}) into Eq.~(\ref{eq:z_def}) yields the following equation, 
\begin{align}\label{eq:z_simp}
\mathcal{Z}=\int&\Biggl[\int\exp\biggl[-\overline{c}\left(\bm{R^*}\cdot\bm{R^*}\right)-m \left(\bm{R^*}\cdot\bm{l}\right)\biggr]d\bm{R^*}\Biggr]\exp\left[\Xi\right]d\bm{Q^*}
\end{align} 
where
\begin{align}
\begin{split}
\overline{c}&=\frac{c_1+c_2}{2} \, ; \quad m =1 \, ;\,\bm{l}=-\left[c_1\bm{\chi^*}_1+c_2\bm{\chi^*}_2+\left(\frac{c_1-c_2}{2}\right)\bm{Q}^*\right];\\
\Xi &=-\frac{Q^{*2}}{4}\left(\frac{c_1+c_2}{2}\right)-\frac{\bm{Q^{*}}}{2}\cdot\left(c_1\bm{\chi^*}_1-c_2\bm{\chi^*}_2\right)-\frac{c_1\chi^{*2}_1+c_2\chi^{*2}_2}{2}-U^{*}_{\text{MS}}
\end{split}
\end{align}
The inner integral in Eq.~(\ref{eq:z_simp}) can be evaluated using the following identity~\cite{Bird1987b} for Gaussian integrals, 
\begin{equation}\label{eq:id_integr}
\int\exp\left[-\overline{c}\left(\bm{u}\cdot\bm{u}\right)-m\left(\bm{u}\cdot\bm{j}\right)\right]d\bm{u}=\left(\frac{\pi}{\overline{c}}\right)^{3/2}\exp\left[\frac{m^2}{4\overline{c}}\left(\bm{j}\cdot\bm{j}\right)\right]
\end{equation}
resulting in
\begin{equation}
\mathcal{Z}=\left(\frac{2\pi}{c_1+c_2}\right)^{3/2}\int\exp\left[\frac{\bm{l}\cdot\bm{l}}{4\overline{c}}+\Xi\right]d\bm{Q^*}
\end{equation}
Upon simplification, one obtains
\begin{equation}\label{eq:z_anlyt}
\mathcal{Z}=\left(\frac{2\pi}{c_1+c_2}\right)^{3/2}\int\exp\left\{-k\left[\bm{Q^*}-\bm{s^*}\right]^2-U^{*}_{\text{MS}}\right\}d\bm{Q^*}
\end{equation}
where $k=(c_1c_2)/2(c_1+c_2)$, and $\bm{s^*}=\bm{\chi^*}_2-\bm{\chi^*}_1$.
The integral in Eq.~(\ref{eq:z_anlyt}) can be evaluated by converting to spherical co-ordinates, recognising that $Q^*_x=Q^*\sin\theta\cos\phi$, $Q^*_y=Q^*\sin\theta\sin\phi$, $Q^*_z=Q^*\cos\theta$.
Therefore,
\begin{align}
\begin{split}\label{eq:norm_spher}
&\mathcal{Z}=\left(\frac{2\pi}{c_1+c_2}\right)^{3/2}
 \int_{Q^*=0}^{\sqrt{b}}\int_{\theta=0}^{\pi}\int_{\phi=0}^{2\pi}\Biggl[\exp\Bigl(-k\left[Q^*_x-s^*_x\right]^2\Bigr)\exp\Bigl(-k\left[Q^*_y-s^*_y\right]^2\Bigr)\exp\Bigl(-k\left[Q^*_z-s^*_z\right]^2\Bigr)\\[10pt]
&\qquad\qquad\qquad\,\,\, \times \exp\biggl\{\frac{b}{3}\biggl[\frac{1}{2}\left(\frac{Q^*}{\sqrt{b}}\right)-\frac{1}{2\left(1-Q^*/\sqrt{b}\right)}-\left(\frac{Q^*}{\sqrt{b}}\right)^2\biggr]\biggr\}\Biggr]Q^{*2}dQ^*\sin\theta d\theta d\phi
\end{split}
\end{align}
\end{widetext}
The integral in Eq.~(\ref{eq:norm_spher}) does not have an analytically closed-form solution, and is evaluated numerically using MATLAB. The free-energy difference in going from the initial state to the final state is then given by,
\begin{equation}\label{eq:qa}
\Delta A^*_{\text{num}}=\ln\left[\frac{\mathcal{Z}(\bm{\chi^*}_2=\bm{\chi}^{\text{(i)}*}_2)}{\mathcal{Z}(\bm{\chi^*}_2=\bm{\chi}^{\text{(f)}*}_2)}\right]
\end{equation}
where the subscript `num' indicates that the free energy difference has been calculated numerically.

\begin{figure}[tb]
\centering
\includegraphics[width=0.75\linewidth]{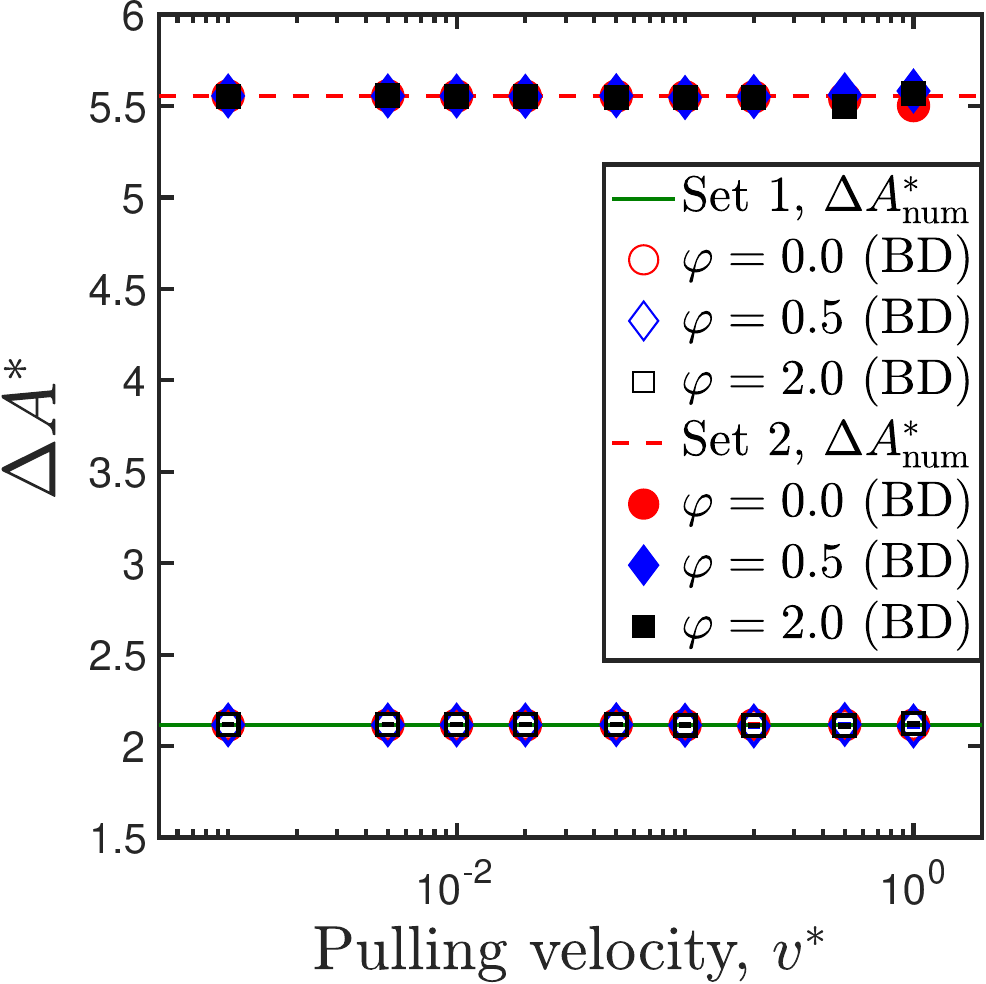}
\caption{\small \textbf{Validation of the code for pulling a single-mode spring-dashpot.} Comparison of the numerically calculated free energy differences (indicated by horizontal lines) against that calculated using the JE, for two different parameter sets shown in Table~\ref{param}.}
\label{fig:delf_validn}
\end{figure}

\begin{table}[tb]
\centering
\caption{\label{param} Parameter values for the two representative cases for which the free energy differences are evaluated using Jarzynski's equality and numerical integration.}
\begin{center}
\begin{ruledtabular}
\begin{tabular}{c c c}
{} &{Parameter sets} & {} \\
\cline{1-3}
\hline  & 1 & 2\\
\hline
$b$ & $50$ & $80$\\[5pt]
$c_1$ & $20$ & $15$ \\[5pt]
$c_2$ & $1$ & $15$\\[5pt]
$\bm{\chi^*}_1$ & $\left(0,0,0\right)$ & $\left(0,0,0\right)$ \\[10pt]
$\bm{\chi}^{\text{(i)}*}_2$ & $\left(1,0,0\right)$ & $\left(4,0,0\right)$\\[10pt]
$\bm{\chi}^{\text{(f)}*}_2$ &  $\left(3,0,0\right)$ & $\left(5,0,0\right)$\\
\end{tabular}
\end{ruledtabular}
\end{center}
\end{table} 

\begin{table}[tb]
\centering
\caption{\label{comparison} A comparison of the free-energy differences calculated using numerical integration [Eq.~(\ref{eq:qa})], and BD simulations using Jarzynski's equality [Eq.~(\ref{eq:je_use})], over $N=1\times10^5$ trajectories. Simulation data reported for freely-draining dumbbells with no internal friction $[h^*=0.0,\,\varphi=0.0]$. The error is quantified as, $\%\,\text{error}=100\times\left\vert\left(\Delta A^*-\Delta A^*_{\text{num}}\right)/\Delta A^*_{\text{num}}\right\vert$.}
\begin{center}
\begin{ruledtabular}
\begin{tabular}{c c c c}
{} &{Parameter set 1 :}& {$\Delta A^*_{\text{num}}=2.11504$} & {} \\
\hline $v^*$ & $\Delta A^*$ & $\%$ error & $\left<W^{*}_{\text{dis}}\right>$\\
\hline
$0.001$ & $2.1151\pm0.0002$ & $0.0006$& $0.0017\pm0.0003$ \\
$0.005$ & $2.1149\pm0.0004$ & $0.008$& $0.0084\pm0.0006$ \\
$0.01$ & $2.1147\pm0.0006$ & $0.02$& $0.0169\pm0.008$ \\
$0.02$ & $2.1144\pm0.0008$ & $0.03$ & $0.033\pm0.001$ \\
$0.05$ & $2.116\pm0.001$ & $0.04$ & $0.081\pm0.002$ \\
$0.1$ &  $2.114\pm0.002$ & $0.06$ & $0.155\pm0.003$ \\
$0.2$ &  $2.116\pm0.003$ & $0.05$ & $0.281\pm0.004$ \\
$0.5$ &  $2.116\pm0.004$ & $0.04$ & $0.508\pm0.005$ \\
$1.0$ &  $2.115\pm0.005$ & $0.01$ & $0.673\pm0.006$ \\
\hline
{} &{Parameter set 2 :}& {$\Delta A^*_{\text{num}}=5.55479$} & {} \\
\hline $v^*$ & $\Delta A^*$ & $\%$ error & $\left<W^{*}_{\text{dis}}\right>$\\
\hline
$0.001$ & $5.5551\pm0.0002$ & $0.005$& $0.0031\pm0.0004$ \\
$0.005$ & $5.5559\pm0.0006$ & $0.02$& $0.0156\pm0.0008$ \\
$0.01$ & $5.5546\pm0.0008$ & $0.003$& $0.031\pm0.001$ \\
$0.02$ & $5.553\pm0.001$ & $0.02$ & $0.063\pm0.002$ \\
$0.05$ & $5.554\pm0.002$ & $0.01$ & $0.155\pm0.003$ \\
$0.1$ &  $5.551\pm0.003$ & $0.07$ & $0.306\pm0.004$ \\
$0.2$ &  $5.549\pm0.005$ & $0.08$ & $0.593\pm0.006$ \\
$0.5$ &  $5.54\pm0.01$ & $0.26$ & $1.38\pm0.01$ \\
$1.0$ &  $5.50\pm0.03$ & $0.95$ & $2.42\pm0.03$ \\
\end{tabular}
\end{ruledtabular}
\end{center}
\end{table} 

Fig.~\ref{fig:delf_validn} shows a comparison between the free energy difference obtained from Brownian dynamics simulations of $N=1\times10^{5}$ trajectories using Jarzynski's equality, and that obtained from Eq.~(\ref{eq:qa}), for the two parameter sets indicated in  Table~\ref{param}. 

\section{\label{sec:valid_sc} Code validation for the single chain case}
A chain of $N_{\text{b}}$ beads connected by $N_{\text{s}}$ springs is considered, with the stiffness of each spring denoted by $H$. The first bead is held fixed at the origin, and the last bead is subjected to a harmonic trap of stiffness $c_2H$. The trap is moved from an initial position of $\chi^{\text{(i)}}$ to a final position of $\chi^{\text{(f)}}$, over a time $\tau$. The distance traveled by the trap is denoted by $d\equiv \chi^{\text{(f)}} - \chi^{\text{(i)}}$, and the pulling velocity, $v$, given by $v=d/\tau$.   
\begin{figure}[tb]
\centering
\includegraphics[width=0.85\linewidth]{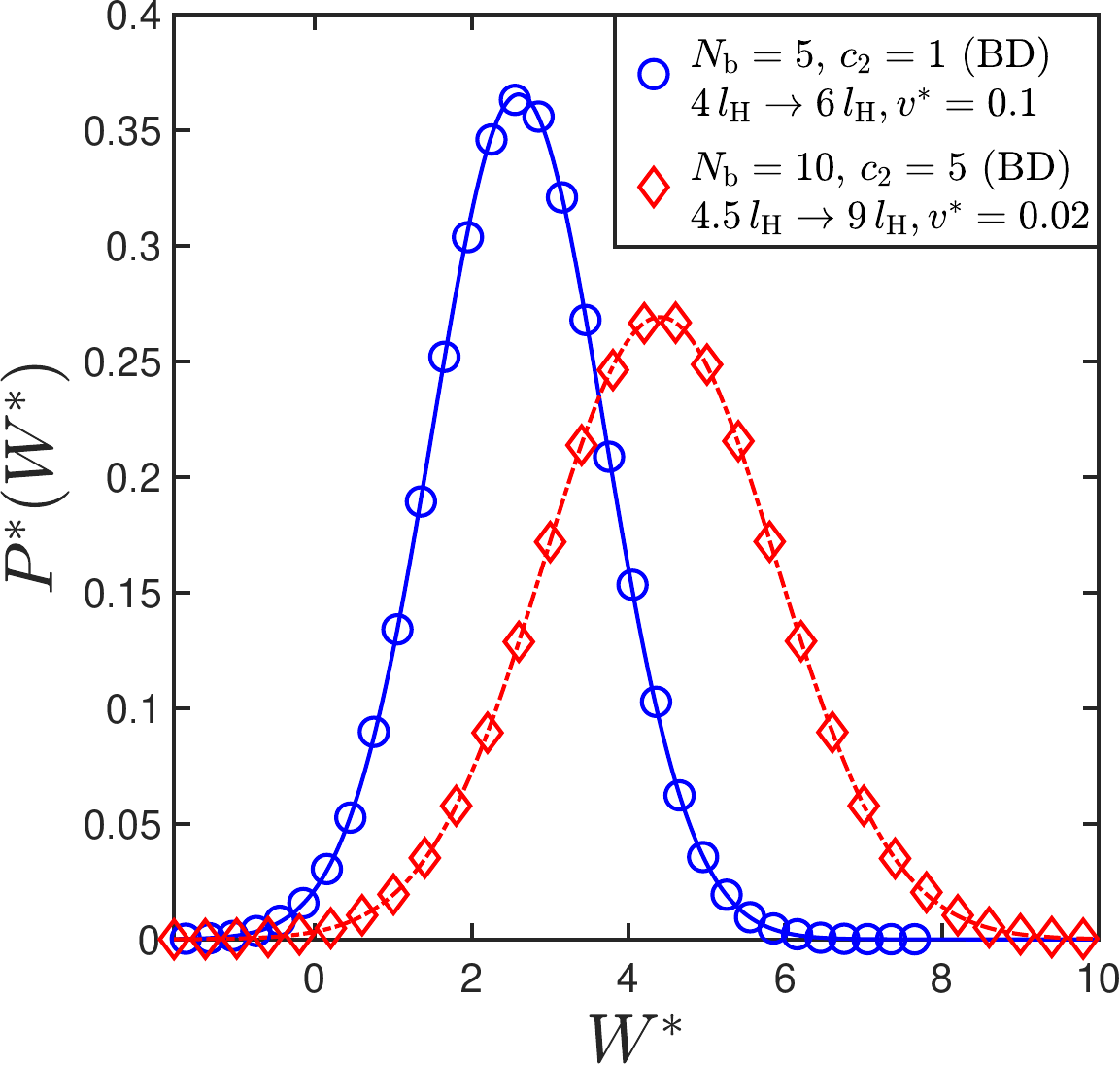} 
\caption{\small  \textbf{Validation of the code for single chain pulling.} Comparison of the analytical probability distribution function of the work trajectories, against that computed by binning the work trajectories obtained in pulling simulations on Hookean chains. The lines correspond to the Gaussian probability distribution given by Eq.~(\ref{eq:prob_dist_dhar_sc}), with variance and mean defined in Eqs.~(\ref{eq:var_sc_pull}) and (\ref{eq:mean_sc_pull}), respectively.}
\label{fig:sc_validn}
\end{figure}

By following the procedure proposed by Dhar~\cite{Dhar2005}, and using the non-dimensionalization scheme introduced in this paper [see Eq.~(\ref{eq:scale}), for example], the free energy difference associated with the stretching process is evaluated to be
\begin{align}
\Delta A^*=\dfrac{c_2}{2\left(c_2N_{\text{s}}+1\right)}\left[\left(\chi^{\text{(f)}*}\right)^2 - \left(\chi^{\text{(i)}*}\right)^2\right],
\end{align}
and $P^{*}\left(W^*\right)$ is found to be a Gaussian of the form,
\begin{align}\label{eq:prob_dist_dhar_sc} 
P^{*}\left(W^*\right)=\dfrac{1}{\sqrt{2\pi\sigma^2}}\exp\left[-\dfrac{\left(W^*-\left<W^*\right>\right)^2}{2\sigma^2}\right],
\end{align}
whose variance and mean are given by
\begin{flalign}\label{eq:var_sc_pull}
\sigma^2&=\dfrac{c_{2}^{2}v^*d^*}{2\tau^*}\left\{\bm{E}^{-2}+\dfrac{1}{\tau^*}\bm{E}^{-3}\left(e^{-\bm{E}\tau^*}-1\right)\right\}_{N_{\text{s}}N_{\text{s}}}\\
&\equiv2\left<W^{*}_{\text{dis}}\right>\nonumber 
\end{flalign}
\begin{flalign}\label{eq:mean_sc_pull}
\left<W^*\right>=&\Delta A^* + \dfrac{\sigma^2}{2}&
\end{flalign}
where the notation $\left\{...\right\}_{ij}$ refers to the $\left(ij\right)^{\text{th}}$ matrix element, and $\bm{E}$ is a symmetric, tridiagonal $N_{\text{s}}\times N_{\text{s}}$ matrix of the following form
\begin{align}\label{eq:e_matrix_def}
E_{ij}= \left\{
\begin{array}{ll}
       \dfrac{1}{2}; &  i=j\neq N_{\text{s}} \\[15pt]
      -\dfrac{1}{4}; & |i-j|=1 \\[15pt]
       \dfrac{\left(c_2+1\right)}{4}; & i=j=N_{\text{s}} \\[15pt]
       0 ; & \text{otherwise} \\[5pt]
\end{array} 
\right. 
\end{align}

The pulling of Hookean chains was simulated using Brownian dynamics, with an ensemble size of $\mathcal{O}(10^5)$. The first bead was held fixed at the origin by means of a stiff harmonic trap of strength $c_1=1000$.  
In Fig.~\ref{fig:sc_validn}, the probability distribution of work values obtained from BD simulations is compared against the analytical solution for two sample cases. The good agreement between the two suggests the validity of the code.

%

\end{document}